\documentstyle[12pt,epsf,a4]{report}

\begin{document}
\def\vpu{\mathbf{p_1}}
\def\vpd{\mathbf{p_2}}
\def\vpt{\mathbf{p_3}}
\def\vpq{\mathbf{p_4}}
\def\tpu{\mathbf{\widetilde{p_1}}}
\def\tpd{\mathbf{\widetilde{p_2}}}
\def\tpt{\mathbf{\widetilde{p_3}}}
\def\tpq{\mathbf{\widetilde{p_4}}}
\def\so#1{$\underline{\smash{\hbox{#1}}}$}
\def\rap{ {1\over 2} }
\def\rapt{\scriptstyle {1\over 2} }  
\def\raps{\scriptstyle {1/2} }
\def\gms{$G_{Ms}~$}
\def\gmp{$G_{Mp}~$}
\def\gmn{$G_{Mn}~$}
\def\ges{$G_{Es}~$}
\def\gep{$G_{Ep}~$}
\def\gen{$G_{En}~$}
\def\geN{$G_{EN}~$}
\def\gmN{$G_{MN}~$}
\def\cg#1#2#3#4#5#6{ <{#1}~{#2}~{#3}~{#4}\mid {#5}~{#6}>}

\title{Relativistically invariant analysis of 
polarization effects in exclusive deuteron
electrodisintegration process}
\author{ G.I. Gakh$^1$, A.P. Rekalo\footnote{ Permanent address:
\it NSC Kharkov Physical Technical Institute, 61108 Kharkov, Ukraine} and Egle Tomasi-Gustafsson\\
\it  DAPNIA/SPhN, CEA/Saclay, 91191 Gif-sur-Yvette Cedex, 
France}

\maketitle
\tableofcontents

\begin{abstract}                         
A general formalism for the calculation of the differential cross 
section and  polarization observables, for the process of deuteron 
electrodisintegration, is developed in the framework of relativistic impulse approximation. 
A detailed analysis of the general structure of the differential cross section 
and polarization observables for the $e^-+d\rightarrow e^-+n+p$ reaction 
is derived, using the formalism of the structure functions. The obtained expressions 
have a general nature and they hold in the one--photon--exchange 
mechanism, assuming P--invariance of the hadron electromagnetic interaction. 
The model of relativistic impulse approximation is introduced and 
the final state interaction is taken into account by means of the unitarization 
of the helicity amplitudes. A detailed description of the unitarization 
procedure is also presented. 

Using different parametrizations of the deuteron wave functions, the following polarization observables are calculated in the kinematical region of quasi--elastic deuteron electrodisintegration: the asymmetry for the 
scattering of longitudinally polarized electrons on a polarized deuteron target, the proton and neutron 
polarizations (for longitudinally polarized electron beam or vector--polarized 
deuteron target). The sensitivity to the neutron electric form factor  
is also thorougly investigated.

The predictions of the model are compared with the results of recent 
polarization measurements and a good agreement with the existing experimental 
data has been obtained.

\vspace{2 true cm}
{\it This paper is dedicated to the memory of Professor Michail P. REKALO}

\end{abstract}

\chapter{Introduction}

The deuteron, a unique two--nucleon bound system, is one of the fundamental
objects in nuclear physics. 
It has been studied for decades theoretically as well as experimentally. The 
processes of elastic 
$$\ell +d\rightarrow \ell +d,~\ell=e,~ \mu ,~ \nu$$
and inelastic 
$$\ell +d\rightarrow \ell +n+p,$$
$$ \nu _{\mu} +d\rightarrow \mu ^-+p+p,$$
$$\bar\nu _e+d\rightarrow  e^+ +n+n $$ 
lepton--deuteron scattering are the 
simplest  nuclear processes where the weak and electromagnetic characteristics 
of the two--body nuclear system appear. The fundamental properties of the 
deuteron, with zero isotopic spin and unit spin, lead to the unique possibility of 
selecting specific  reaction mechanisms and to study interesting polarization 
phenomena \cite{Ak77,Ra89,Bo96}. 

Today a large interest in the deuteron studies arises 
from the possibility  to access the kinematical region of large momentum 
transfer squared, $Q^2=-k^2$ (from the incident lepton to the hadron system),  
i.e., very short internucleon distances. In the space-like region\footnote{We use the Feynman metric where 
$k^2_{\mu}=k_0^2 -\vec k^2$.}, that we 
consider here,  $Q^2>0$.

Measurements of even very small cross 
sections and of different polarization observables can be actually realized at 
high-duty cycle electron accelerators, due to the recent 
developments of polarized electron sources, polarized targets ($p$, $d$ and $^3He$) and polarimeters for nucleons and deuterons (for a recent 
review, see, for instance, \cite{Gi02}). 

When addressing, more specifically, the electromagnetic properties of 
the deuteron, the main question concerns the description of the three elastic 
deuteron form factors in terms of the calculated deuteron wave function and of the electromagnetic isoscalar nucleon form factors. 

At low momentum transfer 
the deuteron can be understood as a system of two nucleons interacting via the 
known nucleon--nucleon interaction. The predictions agree quite well with the 
data within the impulse approximation (IA) when accounting for the one--body current only. At higher 
momentum transfers, the two--body contributions are known to be important 
(meson--exchange currents). If the  quark degrees of freedom do need to be taken explicitly into account is still under investigation.

From the experimental point of view the determination 
of the three elastic electromagnetic deuteron form factors (FF): charge $G_C$, 
quadrupole $G_Q$, and magnetic $G_M$, requires the measurement of the 
differential cross section and of one polarization observable. Two structure 
functions $A(Q^2)$, and $B(Q^2)$ are related to the forward and backward 
differential cross section respectively. Together with the measurement of the 
tensor polarization of the deuteron, $t_{20}$, in the elastic scattering of unpolarized electron by unpolarized deuteron target, they allow the full 
determination of the deuteron FFs. 

Recent experiments at the Jefferson Laboratory, JLab, (USA), measured the forward cross section of elastic $ed$--scattering up to $Q^2\simeq$ 6 GeV$^2$, with a luminosity of $4.7\cdot 10^{38}$ cm$^{-2}$ s$^{-1}$ \cite{Al99}. The highest $Q^2$ data point corresponds to a cross section of $\simeq 10^{-41}$ cm$^2$/sr. The 
tensor polarization could be measured up to $Q^2\simeq$ 1.9 GeV$^2$ \cite{Ab00}. The limitation is  due to the fact that such polarization measurement  
necessitates a double scattering and the efficiency of a tensor polarimeter, 
based on the charge exchange $d(p,pp)n$ reaction, is of the order of $10^{-3}$ 
\cite{Kox94}. 

This method, for disentangling the deuteron FFs, has been known 
and applied since the 60's \cite{Go74} and  holds in frame of the one-photon 
exchange mechanism. Going to large values of momentum transfer, it has been 
pointed out \cite{Gu73} that, at some point, the two-photon exchange mechanism may become important. The necessity of verifying the presence and the importance of this 
mechanism from the data itself has been recently discussed \cite{Re99}. In 
particular, such mechanism would induce vector polarization of the scattered 
deuteron, even in the scattering of unpolarized particles. A 
non-vanishing vector polarization of the recoil deuteron  could be measured using a 
deuteron  polarimeter based on elastic $dp$-scattering \cite{Hypom}.

The existing data definitely show that the 
two-nucleon structure of the deuteron holds in the measured range of momentum 
transfer. It has been shown that the simple picture of  a deuteron as a bound two 
nucleon system is compatible with the existing data, using definite 
prescriptions for the nucleon FFs \cite{ETG01}. However, a systematic 
difference appears between the last JLab data \cite{Ab00} and previous MIT data 
\cite{The91}, in the overlapping range, which makes the determination of the 
zero of $G_C$ less precise. Data from Novosibirsk, using a polarized deuteron 
target \cite{Ni03}, seem to favor the last JLab data.

Indeed, fundamental elements 
for the description of the deuteron structure are the nucleon electric 
\geN$(Q^2)$ and  magnetic \gmN$(Q^2)$ FFs, ($N=n$, $p$). The 
deuteron has been studied not only to check our understanding of the 
two--nucleon system, but also to get information on the neutron FFs. As 
a pure neutron target is not available,  much of our knowledge of 
the neutron charge FF \gen$(Q^2)$ came, in the past, from precision studies of 
the deuteron structure function (SF) $A(Q^2)$ \cite{Ga71,Pl90}. Only very 
recently, 
experiments involving both polarized electrons and polarized 
target (recoil--nuclei) have allowed us to get access to \gen $(Q^2)$ via 
other polarization observables. At large $Q^2$, however, \gen $(Q^2)$ is 
still unknown, a fact that 
represents a serious handicap for the quantitative understanding of the 
deuteron charge FF. Data on \gen $(Q^2)$ are also very important for our 
understanding of the electromagnetic structure of the nucleon and are 
essential for the interpretation of electromagnetic multipoles 
of nuclei. 

Experiments using polarized electrons and polarized target nuclei or recoil 
polarimetry lead to higher precision, because one has direct access to the 
interference 
terms involving the product \gen $(Q^2)$ \gmn $(Q^2)$ \cite{Re68,Do69,Ar81}. 
In the measurement of the neutron FFs,  the lack of a free neutron target is a  major difficulty, which is only partially overcome due to the progress in the construction of $^2H$ and $^3He$-targets.

Two kinds of  polarization experiments for the $e^-+d\to e^-+n+p$ reaction are especially 
interesting for this aim: the scattering of longitudinally polarized electrons 
by a vector polarized deuteron target, $\vec d (\vec e,e'n)p$, with detection of 
the scattered electron and neutron (or proton) \cite{Pa99,Zh01a,Zh01,Day}, and the 
scattering of longitudinally polarized electrons by an unpolarized deuteron 
target, with measurement of the neutron polarization, $ d (\vec e,e'\vec n)p$ 
\cite{Ed94,He99,Os99,Madey,Mainz}. Such experiments have been performed or are 
planned at NIKHEF (Amsterdam), ELSA (Bonn), MAMI (Mainz), MIT (Bates) and JLab, at momentum transfer squared up to $2$ GeV$^2$. 
 
Let us only mention the possibility to extract nucleon FFs in experiments with polarized $^3He$ target. Such experiments have been performed in  
Mainz \cite{Me94} where the ratio of the $x$- and $z$-components of the 
$^3{He}$ polarization has been measured. However, the structure of $^3He$, being a 3-nucleon system, is more complicated in 
comparison with deuteron. 

Concerning the proton FFs, very precise data 
\cite{Jo00,Ga02} obtained with the recoil proton polarization method 
\cite{Re68}, show a large deviation from the dipole-like behavior generally assumed.
These data require a revision of the models of nucleon and deuteron 
structure and show that  
polarization phenomena represent a unique tool in order to reach high precision 
measurements at large momentum transfer. 
 
In the case of polarized nuclear 
targets, such as $\vec d$ or $\vec{^3He}$, the extraction of the 
electric and magnetic FFs or their ratios from polarization observables, 
requires the careful study of various nuclear effects, as well as the final state nucleon-nucleon interaction (FSI). 

In particular, in the case of deuteron electrodisintegration in quasi-elastic 
kinematics, a correct and effective extraction of  \gen needs 
an  adequate theoretical interpretation of the reaction mechanism in $e^-+d \to 
e^-+n+p$. First 
of all, the main symmetry properties of electromagnetic hadron interaction have 
to be taken into account, such as the conservation of the hadronic 
electromagnetic current (for the subprocess $\gamma^*+d\to n+p$, where 
$\gamma^*$ is a virtual photon), i.e., the gauge invariance of the 
electromagnetic interaction and the relativistic invariance. 

The problem of relativistic corrections 
to the standard nonrelativistic approach has been widely discussed in the 
literature. For deuteron electrodisintegration, it has been shown that  
relativistic effects appear at rather low energies and lead to a
substantial modification of the observables \cite{Mo90,Mo93,Ca82,Mo92,Hu89}. 
Nucleons produced in $\gamma ^*d\rightarrow np$ are relativistic at relatively 
small $E_{np}$. Indeed, the  proton momentum at the pion 
production threshold, $E_{np}=140$ MeV, in the center of mass system (CMS), is $p$=370 MeV/c; that is, $p/m\approx 0.4$, where $m$ is the nucleon mass. 

Relativistic 
effects cannot be included as corrections, in the region of relative large $Q^2$,  and a fully  relativistic approach is 
required.  The description of FSI has also to 
be properly taken into account. In order to decrease the model dependence of 
FSI, one should avoid approaches based on the nonrelativistic concept of 
NN-potentials. Instead of NN-potentials, a model independent description of FSI 
in $\gamma^*+d\to n+p$ can be derived from the phases of NN-scattering, which 
are available from the phase-shift analysis of the huge amount of data about the 
NN-interaction. The relativistically invariant impulse approximation (RIA) 
\cite{RGR89}, with subsequent unitarization of the corresponding multipole 
amplitudes \cite{Ga66},  seems the most appropriate model for the description 
of $e^-+d\to e^-+n+p$.  
 
The description of the 
deuteron structure can be done in terms of wave functions.  Note, in this 
respect, that only the kinematical region for $e^-+d\to e^-+n+p$, which 
corresponds to quasi-elastic $e^-+n^*\to e^-+n$ scattering ($ n^*$ is a virtual 
neutron), is especially sensitive to neutron FFs. This region 
corresponds to the emission of the neutron along the three-momentum of the 
virtual photon, when $Q^2\simeq W^2-M^2$, where $M$ is the deuteron mass and 
$W$ is the invariant mass of the produced $np$-system. In such conditions the 
virtuality of the neutron is small, therefore the argument of the deuteron 
wave function (in impulse representation) is also small. In conditions of 
evidently nonrelativistic momentum, the standard S- and D-components of the 
deuteron wave function (DWF), derived from the existing NN-potentials, can be 
safely used. Therefore,  the  four components of the relativistic DWF 
can be related with good accuracy to the nonrelativistic S- and D-components 
$u(p)$ and $w(p)$ \cite{Bu79}.

The considered kinematical regime in $\gamma^*+d\to n+p$,  which is the most convenient for the determination of \gen, 
corresponds to nonperturbative QCD at any value of $Q^2$, so that all the prescriptions of pQCD such as helicity conservation, quark 
counting rules, formalism of reduced deuteron FFs or reduced nuclear 
matrix elements cannot be applied here \cite{Re03}. Moreover, all existing 
experimental data, including polarization effects, concerning different 
processes with deuteron target: $e^-+d\to e^-+d$ \cite{Ab00}, $\gamma+d\to 
d+\pi^0$ \cite{Sc01}, and $\gamma+d\to n+p$ \cite{Me99} do not confirm the pQCD 
predictions at JLab energies.

The isoscalar  nature of the deuteron allows to get a simple expression for 
the P-even asymmetry in the scattering of longitudinally polarized electrons by 
unpolarized deuterons which does not depend on the details of the deuteron 
structure, in the framework of the Standard Model \cite{Fe75,Do79,RGR87}. 
The generalization of this result has been recently done in the case of 
inelastic electroproduction of pion on nucleon \cite{Podd}. Such consideration is 
very important for the interpretation of running experiments concerning the 
study of strange quark components in the nucleon. 

On the other hand, the vector nature of the deuteron (spin $S=1$) leads to the complex spin 
structure of the deuteron electromagnetic and weak currents.  For spin-one 
particles, P-odd and T-even electromagnetic FFs 
are  in principle possible, although their determination requires a large number 
of polarization measurements.

In this paper, we develop a general formalism, based on relativistic impulse 
approximation, firstly derived in Refs. \cite{RGR86,RGR88,Re65}. 
We calculate, in particular, three types of 
observables in the kinematical region of quasi-elastic deuteron 
electrodisintegration: the asymmetry for the scattering of longitudinally polarized electrons, the proton and the neutron polarization (for longitudinally polarized electrons and 
vector polarized deuteron target) and compare the predictions with the results of recent  polarization measurements.

\chapter{General formalism}

The general structure of the 
differential cross section for the $e^-+d\rightarrow e^- +n +p$ reaction can be 
determined in the framework of the one-photon-exchange mechanism. The 
formalism in this section  is based on the most general symmetry properties of 
the hadron electromagnetic interaction, such as gauge invariance (the 
conservation of the hadronic and leptonic electromagnetic currents) and 
P-invariance (invariance with respect to space reflections) and does not depend 
on the deuteron structure and on details of the reaction mechanism for 
$\gamma^*+d\to n+p$. 
In the one--photon--exchange approximation, the matrix element of the deuteron electrodisintegration reaction
\begin{equation}\label{eq:eq1}
e^-(k_1)+d(P)\rightarrow e^-(k_2) + p(p_1)+n(p_2),
\end{equation}
(the 4-momenta of the corresponding particles are indicated in the brackets)  can be written 
as
\begin{equation}
M_{fi}=\displaystyle\frac{e^2}{k^2}j_{\mu}J_{\mu}, ~
j_{\mu}=\bar u(k_2)\gamma _{\mu}u(k_1),
\label{eq:eq2}
\end{equation}
where $k_1(k_2)$ is the 
4-momentum of the initial (final) electron, $k=k_1-k_2, $ $J_{\mu}$ is the 
electromagnetic current describing the transition $\gamma ^*+d\rightarrow n+p$. 

The electromagnetic structure of nuclei, as probed by elastic and
inelastic electron scattering, can be characterized by a set of response
functions or SFs \cite{Ra89,Bo96}. Each of these SFs is determined by different combinations of the longitudinal and
transverse components of the electromagnetic current $J_{\mu }$, thus
providing different pieces of information about the nuclear structure or possible mechanisms
of the reaction under consideration. Those ones which are determined by the
real parts of the bilinear combinations of the reaction amplitudes  are
nonzero in IA, other ones which originate from the imaginary part of SFs, 
vanish if FSI are absent. 

The formalism of SFs is especially convenient for the investigation of 
polarization phenomena in the reaction (\ref{eq:eq1}). As a starting point, let 
us write the general structure for the cross section of the reaction  
(\ref{eq:eq1}), when the 
scattered electron and one of the nucleons are detected in coincidence, and 
the electron beam is longitudinally polarized 
\footnote{The polarization states of the deuteron target and of the final nucleons can be any} \cite{Re77}:
\begin{eqnarray}
\displaystyle\frac{d^3\sigma}{dE'd\Omega_ed\Omega_N}
&=&{\cal N}\biggl[H_{xx}+H_{yy} 
+\varepsilon 
\cos(2\phi)(H_{xx}-H_{yy})
+\varepsilon 
\sin(2\phi)(H_{xy}+H_{yx}) \nonumber\\
&&-2\varepsilon\displaystyle\frac{k^2}{k_0^2}H_{zz}-\displaystyle\frac{\sqrt{-k^2}}
{k_0}\sqrt{2\varepsilon 
(1+\varepsilon)}\cos\phi(H_{xz}+H_{zx})\nonumber\\
&&-\displaystyle\frac{\sqrt{-k^2}}{k_0}\sqrt{2
\varepsilon (1+\varepsilon)}\sin\phi(H_{yz}+H_{zy})
\mp i\lambda\sqrt{(1-\varepsilon 
^2)}(H_{xy}-H_{yx}) \nonumber\\
&& \mp i\lambda 
\displaystyle\frac{\sqrt{-k^2}}{k_0}\sqrt{2\varepsilon 
(1-\varepsilon)}\cos\phi(H_{yz}-H_{zy})\nonumber\\
&&\pm i\lambda 
\displaystyle\frac{\sqrt{-k^2}}{k_0}\sqrt{2\varepsilon 
(1-\varepsilon)}\sin\phi(H_{xz}-H_{zx})\biggr],
\label{eq:eq3}
\end{eqnarray}

$$
{\cal N}= \displaystyle\frac{\alpha^2}{64\pi^3}\displaystyle\frac{E'}{E}\displaystyle\frac{p}{MW}\displaystyle\frac{1}{1-\varepsilon}\displaystyle\frac{1}
{(-k^2)},~{\vec 
|k|}=\sqrt{(W^2+M^2-k^2)^2-4M^2W^2}/2W,$$
$$~\varepsilon^{-1}=1-2\displaystyle\frac{{\vec k^2_{Lab}}}{k^2}\tan^2(\displaystyle\frac{\vartheta_e}{2}), \
\ H_{\mu\nu}=J_{\mu}J_{\nu}^*.$$
The $z$ axis is directed along the virtual-photon momentum ${\vec 
k}$, the momentum of the detected nucleon ${\vec p}$ lies in the $xz$ plane 
(reaction plane); $E (E')$ is the energy of the initial 
(scattered) electron in the deuteron rest frame (laboratory system), 
$d\Omega_e$ is the solid angle of the 
scattered electron in the laboratory (Lab) system, $d\Omega_N (p)$ is the solid 
angle (value 
of the three-momentum) of the detected nucleon in $np$--pair CMS, $\phi$ is the 
azimuthal angle between the electron scattering plane and the reaction plane, 
$k_0=(W^2+k^2-M^2)/2W$ is the virtual-photon 
energy in the $np$--pair CMS, $W$ is the invariant mass
 of the final nucleons $W^2=M^2+k^2+2M(E-E')$, $\lambda$ is the degree of the 
 electron longitudinal polarization, 
$\varepsilon$ is the degree of the linear polarization of the virtual photon.  
The upper (bottom) sign corresponds to electron 
(positron) scattering. This expression holds for zero electron mass. The 
electron mass will be neglected hereafter  wherever possible.

If we single out the 
nucleon Dirac spinors, we can write the following expression for the 
electromagnetic current $J_{\mu}$ which determines the hadronic tensor 
$H_{\mu\nu}$
\begin{equation}
\label{eq:eq4}
J_{\mu}=\bar 
u_{\alpha}(p_1)T^{\mu}_{\alpha\beta}(p_1, p_2; k, P)O_{\beta\beta'}\tilde {\bar 
u}_{\beta'}(p_2),
\end{equation}
where $O=i\tau _2C, $ $\tau _2$ is the isospin 
matrix, $C$ is the Dirac matrix of charge conjugation. The matrix $T^{\mu} (p_1, 
p_2; k, P)$ must be P-invariant  and satisfy the gauge invariance 
condition, $k_{\mu}T^{\mu}=0$. The generalized Pauli principle requires
$T^{\mu}(p_1, p_2; k, P)O=-$$[T^{\mu}(p_2, p_1; k, P)O]^T, $ where the upper 
index $T$ indicates the transpose operation. Under these conditions, the matrix 
$T^{\mu}$ can be represented as
\begin{equation}
\label{eq:eq5}
T^{\mu}=\sum_{i=1}^{18} H_i(s, t, u)I^{\mu}_i ,
\end{equation}
where $H_i$ are the invariant amplitudes depending on the invariant variables 
$s=(p_1+p_2)^2, $ $t=(k-p_1)^2, $ $u=(k-p_2)^2.$ The invariant forms 
$I^{\mu}_i (i=1-18) $ can be chosen as \cite{Re65}
\begin{eqnarray}
&I^{\mu}_1&=\displaystyle\frac{1}{2m^2}({\cal K}_{\mu}U\cdot k-U_{\mu}k\cdot {\cal K}),~
I^{\mu}_2=\displaystyle\frac{1}{2m^2}(U_{\mu}{\cal Q}\cdot k-{\cal Q}_{\mu}k\cdot U),~ 
\nonumber\\
&I^{\mu}_3&=\displaystyle\frac{1}{2m^2}(U_{\mu}k^2-k_{\mu}k\cdot U),~ I^{\mu}_4=\displaystyle\frac{U\cdot {\cal K}}{4m^4}[{\cal K}_{\mu}(k^2-4{\cal Q}\cdot k)-{\cal K}\cdot k(k_{\mu}-4{\cal Q}_{\mu})],~ 
\nonumber\\
&I^{\mu}_5&=\displaystyle\frac{U\cdot {\cal K}}{4m^4}(k_{\mu}{\cal Q}\cdot k-{\cal Q}_{\mu}k^2),~
I^{\mu}_6=\displaystyle\frac{\hat 
U}{4m^3}[{\cal K}_{\mu}(k^2-4{\cal Q}\cdot k)-{\cal K}\cdot k(k_{\mu}-4{\cal Q}_{\mu})],
\nonumber\\
&I^{\mu}_7&=\displaystyle\frac{\hat U}{4m^3}(k_{\mu}{\cal Q}\cdot k-{\cal Q}_{\mu}k^2),~
I^{\mu}_8 =\displaystyle\frac{1}{2m}(U_{\mu}\hat k-\gamma_{\mu}k\cdot U), 
\nonumber\\
&
I^{\mu}_9&=\displaystyle\frac{U\cdot k}{2m^3}(\gamma_{\mu}k^2-k_{\mu}\hat k),
~I^{\mu}_{10}=\displaystyle\frac{U\cdot {\cal K}}{m^3}(\gamma_{\mu}k^2-k_{\mu}\hat k),
\nonumber\\
&I^{\mu}_{11}&=\displaystyle\frac{1}{2m^3}[\hat k({\cal Q}_{\mu}U\cdot k+2{\cal K}_{\mu}U\cdot {\cal K})-
\gamma_{\mu}({\cal Q}\cdot k U\cdot k+2{\cal K}\cdot k U\cdot {\cal K})],  \nonumber\\
&I^{\mu}_{12}&=\displaystyle\frac{1}{2m^3}[\hat 
k({\cal K}_{\mu}U\cdot k+2{\cal Q}_{\mu}U\cdot {\cal K})-\gamma_{\mu}({\cal K}\cdot k U\cdot k+
2{\cal Q}\cdot k U\cdot {\cal K})],
\nonumber\\
&I^{\mu}_{13} &=\displaystyle\frac{U\cdot k}{4m^2}(\hat k\gamma_{\mu}-\gamma_{\mu}\hat k), 
~I^{\mu}_{14}=\displaystyle\frac{U\cdot {\cal K}}{2m^2}(\hat k\gamma_{\mu}-\gamma_{\mu}\hat k),
\nonumber\\
& I^{\mu}_{15}&=
\displaystyle\frac{1}{2m^2}[(\hat k \hat U-\hat U \hat 
k){\cal K}_{\mu}-(\gamma_{\mu}\hat U-\hat U\gamma_{\mu}){\cal K}\cdot k],
\nonumber\\ 
& I^{\mu}_{16}&=
\displaystyle\frac{1}{2m^2}[(\hat k \hat U-\hat U \hat 
k){\cal Q}_{\mu}-(\gamma_{\mu}\hat U-\hat U\gamma_{\mu}){\cal Q}\cdot k],
\nonumber\\
&I^{\mu}_{17}&=\displaystyle\frac{1}{2m^2}[(\hat k \hat U-\hat U \hat 
k)k_{\mu}-(\gamma_{\mu}\hat U-\hat U\gamma_{\mu})k^2], 
~I^{\mu}_{18}=-\displaystyle\frac{i}{2m}\varepsilon_{\mu\alpha\beta\gamma}
U_{\alpha}k_{\beta}\gamma_{\gamma}\gamma_{5},  
\label{eq:eq6}
\end{eqnarray}
where ${\cal K}=(p_1-p_2)/2, $  ${\cal Q}=(p_1+p_2)/2$, $U_{\mu}$ is the deuteron polarization 
four-vector, $P\cdot U=0.$ Assuming the conservation of the leptonic $j_{\mu}$ and 
hadronic $J_{\mu}$ electromagnetic currents the matrix element can be written 
as
\begin{equation}
\label{eq:eq7}
M_{fi}=ee_{\mu}J_{\mu}=e{\vec l} \cdot {\vec J},~e_{\mu}=\displaystyle\frac{e}{k^2}j_{\mu},~{\vec l} =\displaystyle\frac{{\vec e}\cdot {\vec 
k}}{k_0^2}{\vec k}-{\vec e}.
\end{equation}

In the final nucleon CMS we get 
$M_{fi}=e\varphi _1^{\dagger} F\varphi _2^c, $ where $\varphi _1^{\dagger}$ and $ \varphi _2^c=\sigma _2 
\tilde\varphi _2^{\dagger} $ are the proton and neutron spinors, respectively. The 
amplitude $F$ can be chosen 
as 
\begin{eqnarray}
&F=&i\varepsilon_{\alpha\beta\gamma}l_{\alpha}U_{\beta}k_{\gamma}F_1+i
\varepsilon_{\alpha\beta\gamma}l_{\alpha}U_{\beta}p_{\gamma}F_2+
i{\vec l}\cdot {\vec 
k}\varepsilon_{\alpha\beta\gamma}U_{\alpha}k_{\beta}p_{\gamma}F_3+i{\vec 
l}\cdot {\vec p}
\varepsilon_{\alpha\beta\gamma}U_{\alpha}k_{\beta}p_{\gamma}F_4\nonumber\\
&&
+{\vec 
U}\cdot {\vec k}{\vec \sigma}\cdot {\vec l}F_5+{\vec U}\cdot {\vec p}{\vec \sigma}\cdot {\vec 
l}F_6+{\vec l}\cdot {\vec k}{\vec \sigma}\cdot {\vec U}F_7+{\vec l}\cdot {\vec p}{\vec 
\sigma}\cdot {\vec U}F_8+{\vec l}\cdot{\vec U}{\vec \sigma}\cdot{\vec 
k}F_9\nonumber\\
&& 
+{\vec l}\cdot {\vec U}{\vec 
\sigma}\cdot {\vec p}F_{10}+{\vec l}\cdot {\vec k}{\vec U}\cdot {\vec k}{\vec \sigma}\cdot {\vec 
k}F_{11}+{\vec l}\cdot {\vec k}{\vec U}\cdot {\vec p}{\vec \sigma}\cdot {\vec 
k}F_{12}+{\vec l}\cdot {\vec p}{\vec U}\cdot {\vec k}{\vec \sigma}\cdot {\vec 
k}F_{13}\label{eq:eq8} \nonumber\\
&& 
+{\vec l}\cdot {\vec p}{\vec U}\cdot {\vec p}{\vec 
\sigma}\cdot {\vec k}F_{14}+{\vec l}\cdot {\vec k}{\vec U}\cdot {\vec k}{\vec \sigma}\cdot {\vec 
p}F_{15}+{\vec l}\cdot {\vec k}{\vec U}\cdot {\vec p}{\vec \sigma}\cdot {\vec 
p}F_{16}+{\vec l}\cdot {\vec p}{\vec U}\cdot {\vec k}{\vec \sigma}\cdot {\vec p}F_{17} 
\nonumber\\
&& +{\vec l}\cdot {\vec p}{\vec U}\cdot {\vec p}{\vec \sigma}\cdot {\vec p}F_{18},
\label{eq:eq8}
\end{eqnarray}
where 
$F_i \ (i=1-18)$ are the scalar amplitudes which completely determine the 
reaction dynamics. The formulae connecting the invariant $(H_i)$ and scalar 
$(F_i)$ amplitudes are given in Appendix 1. The above structures arise 
naturally in the transition from four- to the two--component spinors. As it is 
seen from Eq. (\ref{eq:eq3}), the cross section and the polarization 
characteristics of the process under consideration are determined only by the 
space components of the hadronic tensor $H_{\mu\nu}.$ Let us consider 
a polarized deuteron target. Then, the deuteron spin--density matrix can be 
written in the form \cite{Sc65}
\begin{equation}\label{eq:eq9}
U_{\mu}U_{\nu}^*=-\displaystyle\frac{1}{3}
\left (g_{\mu\nu}-\displaystyle\frac{P_{\mu}P_{\nu}}{M^2}\right )
+\displaystyle\frac{i}{2M}\varepsilon 
_{\mu\nu\alpha\beta}s_{\alpha}P_{\beta}+S_{\mu\nu},
\end{equation}
where $s_{\alpha}$ is the 4-vector of the deuteron vector polarization, 
$s^2=-1,$ $s\cdot P=0; $ $S_{\mu\nu}$ is the deuteron quadrupole polarization 
tensor,$S_{\mu\nu}=S_{\nu\mu},$ $P_{\mu}S_{\mu\nu}=0, $ $S_{\mu\mu}=0; $ the 
four-vector $s_{\alpha}$  is related to the unit vector ${\vec \xi}$ of the 
deuteron vector polarization in its rest system: $s_0=-{\vec k}{\vec \xi}/M, \ 
$${\vec s}={\vec \xi}+{\vec k}({\vec k}{\vec \xi})/M(M+\omega), \omega $ 
is the deuteron energy in the $\gamma^* d\rightarrow np$ CMS. The hadronic 
tensor $H_{ij} \ (i, j=1, 2, 3)$ depends linearly on the target polarization 
and it can be represented as follows
\begin{equation}\label{10, hadron 
tensor}
H_{ij}=H_{ij}(0)+H_{ij}(\xi)+H_{ij}(S),
\end{equation}
where the term $H_{ij}(0)$ corresponds to the case of the unpolarized deuteron 
target, and the term $H_{ij}(\xi) (H_{ij}(S))$ corresponds to the case of vector 
(tensor-)-polarized target. Let us introduce, for convenience and simplifying of
following calculations of polarization observables, the orthonormal 
system of basic unit ${\vec m}, {\vec n}$ and $\hat {\vec k}$
vectors which are built from the momenta of the particles 
participating in the reaction under consideration
$$\hat {\vec k}=
\displaystyle\frac{{\vec k}}{|{\vec k}|}, \vec n=\displaystyle\frac{{\vec k}\times{\vec p}}{|{\vec k}\times{\vec 
p}|}, {\vec m}={\vec n}\times\hat {\vec k}. $$
The unit vectors ${\vec m}$ and $\hat {\vec k}$ define the 
$\gamma^* +d\rightarrow n+p$ reaction $xz-$plane ($z$ axis is directed along 
3-momentum of the virtual photon ${\vec k}$, and $x$ axis is directed along 
the unit vector ${\vec m}$), and the unit vector ${\vec n}$ is 
perpendicular to the reaction plane. The general structure of the part of the 
hadronic tensor which corresponds to the unpolarized deuteron target has 
following form
\begin{equation}\label{eq:eq11}
H_{ij}(0)=\alpha_1\hat 
k_i\hat k_j+\alpha_2n_in_j+ \alpha_3m_im_j+\alpha_4(\hat k_im_j+\hat 
k_jm_i)+i\alpha_5(\hat k_im_j-\hat k_jm_i).
\end{equation}
The real structure 
functions $\alpha_i$ depend on three invariant variables $s, t$ and $k^2.$ Let 
us emphasize that the structure function $\alpha_5$ is determined by the strong 
interaction effects of the final--state nucleons and vanishes for  
the pole diagrams contribution in all kinematic range (independently on the 
particular parametrization of the $\gamma^* NN$-- and $dnp$-- vertices). 
This is true for the nonrelativistic approach and for the relativistic one as 
well, in describing the $\gamma^* +d\rightarrow n+p$ reaction. The scattering 
of polarized electrons by unpolarized deuteron target  allows to determine the 
$\alpha_5$ contribution. Then the corresponding asymmetry is determined only 
by the strong interaction effects. More exactly, it is determined by the 
effects arising from nonpole contributions of various nature (meson exchange 
currents can also induce nonzero asymmetry). The dibaryon resonances, if any, 
lead also to nonzero asymmetry. 

In the chosen coordinate system, the different hadron tensor components, entering in the expression of the cross section (\ref{eq:eq3}), are related to the functions $\alpha_i (i=1-5)$  by:
$$H_{xx}\pm H_{yy}=\alpha_3\pm \alpha_2,~H_{zz}=\alpha_1, ~H_{xz}+ H_{zx}=2\alpha_4, 
$$
$$H_{xz}- H_{zx}=-2i\alpha_5,~H_{xy}\pm 
H_{yx}=0,~H_{yz}\pm H_{zy}=0. 
$$
The tensor describing the deuteron vector polarization has the following general 
structure:
\begin{eqnarray}
\hspace*{-1 true cm}H_{ij}(\xi )&=&{\vec\xi }{\vec n}(\beta_1\hat k_i\hat k_j+\beta_2m_im_j+ 
\beta_3n_in_j+\beta_4\{\hat 
k,m\}_{ij}+i\beta_5[\hat k,m]_{ij})\nonumber\\
&&
+{\vec\xi }\hat {\vec 
k}(\beta_6\{\hat k,n\}_{ij}+\beta_7\{m,n\}_{ij}+i\beta_8[\hat 
k,n]_{ij}+i\beta_9[m,n]_{ij})\label{eq:eq12}\\
&&
+{\vec\xi }{\vec m}(\beta_{10}\{\hat 
k,n\}_{ij}+\beta_{11}\{m,n\}_{ij}+i\beta_{12}[\hat 
k,n]_{ij}+i\beta_{13}[m,n]_{ij}), \nonumber
\end{eqnarray}
where $\{a,b\}_{ij}=a_ib_j+a_jb_i,~[a,b]_{ij}=a_ib_j-a_jb_i. $ 

Therefore, the dependence of the polarization observables 
on the deuteron vector polarization is determined by 13 structure functions. 
On the basis of Eq. (\ref{eq:eq12}) one can make the following conclusions:
\begin{enumerate}
\item If the deuteron is vector-polarized and the  vector of polarization is 
perpendicular to the $\gamma ^* +d\rightarrow n+p$ reaction plane, then the 
dependence of the differential cross section of the $e^-d\rightarrow e^-np$ 
reaction on the $\varepsilon$ and $\phi$ variables is the same as in the case 
of the unpolarized target, and the non vanishing components of the $H_{ij}(\xi )$ tensor  are:  
$$H_{xx}(\xi ) \pm H_{yy}(\xi ),~H_{zz}(\xi ),~H_{xz}(\xi )\pm H_{zx}(\xi ). $$
\item If the deuteron target is 
polarized in the $\gamma ^* +d\rightarrow n+p$ reaction plane (along the direction of the vector ${\vec k}$ or ${\vec m}$), then the dependence of the differential cross section of the $e^-d\rightarrow e^-np$ reaction on the 
$\varepsilon$ and $\phi$ variables is:
\begin{itemize}
\item for deuteron disintegration by unpolarized electron:
$$\varepsilon \sin(2\phi), ~\sqrt{2\varepsilon(1+\varepsilon )}\sin\phi, $$
\item for deuteron disintegration by longitudinally--polarized electron:
$$\pm i\lambda\sqrt{1-\varepsilon ^2},~ \mp 
i\lambda\sqrt{2\varepsilon(1-\varepsilon )}\cos\phi. $$
\end{itemize}
\end{enumerate}
The  $H_{ij}(S)$ tensor, which depends on the deuteron tensor polarization, has the following general structure:
\begin{eqnarray}
\vspace*{-1.5truecm}&&H_{ij}(S)=S_{ab}\hat k_a\hat k_b(\gamma_1\hat k_i\hat k_j+\gamma_2m_im_j+ \gamma_3n_in_j+\gamma_4\{\hat k,m\}_{ij}+i\gamma_5[\hat 
k,m]_{ij})
\nonumber\\
&&\vspace*{.5truecm}
+S_{ab}m_am_b(\gamma_6\hat k_i\hat 
k_j+\gamma_7m_im_j+ \gamma_8n_in_j+\gamma_9\{\hat k,m\}_{ij}+i\gamma_{10}[\hat 
k,m]_{ij}) \nonumber\\
&& \vspace*{.5truecm}
+S_{ab}\{\hat k,m\}_{ab}(\gamma_{11}\hat k_i\hat 
k_j+\gamma_{12}m_im_j+ \gamma_{13}n_in_j+\gamma_{14}\{\hat 
k,m\}_{ij}+i\gamma_{15}[\hat k,m]_{ij}) \nonumber\\
&&\vspace*{.5truecm}
+S_{ab}\{\hat 
k,n\}_{ab}(\gamma_{16}\{\hat 
k,n\}_{ij}+\gamma_{17}\{m,n\}_{ij}+i\gamma_{18}[\hat 
k,n]_{ij}+i\gamma_{19}[m,n]_{ij})+\ \nonumber\\
&&\vspace*{.5truecm}
+S_{ab}\{m,n\}_{ab}(\gamma_{20}\{\hat 
k,n\}_{ij}+\gamma_{21}\{m,n\}_{ij}+i\gamma_{22}[\hat 
k,n]_{ij}+i\gamma_{23}[m,n]_{ij}). 
\label{eq:eq13}
\end{eqnarray}
In this case, the dependence of the polarization 
observables on the deuteron tensor polarization is determined by 23 structure 
functions. From Eq. (\ref{eq:eq13}) one can conclude that:
\begin{enumerate}
\item If the deuteron is tensor polarized so 
that only the $S_{zz}, ~S_{yy}$ and $(S_{yz}+S_{zy})$ components are nonzero, then 
the dependence of the differential cross section of the $e^-d\rightarrow e^-np$ 
reaction on the parameter $\varepsilon$ and on the angle $\phi$ must be the 
same as in the case of the unpolarized target (more exactly, with similar 
$\varepsilon -$ and $\phi -$ dependent terms).
\item 
If the deuteron is polarized so that only the $(S_{xy}+S_{yx})$ 
and $(S_{yz}+S_{zy})$ components are nonzero, then the typical terms follow 
 $\sin\phi$ and  $\sin(2\phi )$ dependencies - for deuteron disintegration by 
unpolarized electron, and terms which do not depend on $\varepsilon $, $\phi $ and 
$\cos\phi$ - for deuteron disintegration by longitudinally--polarized electrons.
\end{enumerate}
In polarization experiments it is possible to prepare the deuteron target with polarization 
along (opposite) its momentum (the target deuteron with definite helicity). The corresponding asymmetry is usually  defined as: 
$$A=\displaystyle\frac{d\sigma (\lambda _d=+1)-d\sigma (\lambda _d=-1)}{d\sigma 
(\lambda _d=+1)+d\sigma (\lambda _d=-1)}, $$
where $d\sigma (\lambda _d)$ is the 
differential cross section of the $e^-d\rightarrow e^-np$ reaction
when the deuteron has helicity $\lambda _d$ (in the $np-$pair CMS). 
From an experimental point of view, the measurement of an asymmetry is more convenient than a cross section, as most of systematic errors and other multiplicative factors cancel in the  ratio. 

The general form of the hadron tensor $H_{ij}(\lambda 
_d)$, which determines the differential cross section of the process under 
consideration, can be written as
\begin{eqnarray}
H_{ij}(\lambda _d=\pm 1)&=&\delta_1\hat k_i\hat k_j+\delta_2m_im_j+ 
\delta_3n_in_j+\delta_4\{\hat k,m\}_{ij}+i\delta_5[\hat 
k,m]_{ij}\nonumber\\
&&
\pm\delta_6\{\hat k,n\}_{ij}\pm i\delta_7[\hat 
k,n]_{ij}\pm\delta_8\{m,n\}_{ij}\pm i\delta_9[m,n]_{ij}. 
\label{eq:eq14}
\end{eqnarray}
The amplitude  is real in the Born approximation. So, assuming the T-invariance 
of the hadron electromagnetic interactions, we can do the following statements, according to  the deuteron polarizations:
\begin{enumerate}
\item {\bf The deuteron is unpolarized:} since the hadronic tensor $H_{ij}(0)$ 
has to be  symmetric, the asymmetry in the scattering of 
longitudinally-polarized electrons vanishes.
\item {\bf The deuteron is vector-polarized:} 
since the hadronic tensor $H_{ij}(\xi )$ has to  be antisymmetric, 
then the deuteron vector polarization can manifest itself in the 
scattering of longitudinally-polarized electrons. 

The transverse polarization of the target (lying in the $\gamma ^*d\rightarrow
np$ reaction plane) leads to a correlation of the following type 
$\mp i\lambda\sqrt{2\varepsilon (1-\varepsilon )}\cos\phi .$  
The longitudinal polarization of the target, which is perpendicular to this 
plane, leads to two correlations: $\mp 
i\lambda\sqrt{1-\varepsilon ^2}$ and $\pm 
i\lambda\sqrt{2\varepsilon(1-\varepsilon )}\sin\phi .$
\item {\bf The deuteron is tensor-polarized:} the hadronic tensor $H_{ij}(S)$ 
is symmetric. In the scattering of longitudinally polarized 
electrons the contribution proportional to $\lambda S_{ab},$ vanishes. 
If the target is polarized so that only the $(S_{xy}+S_{yx})$ or 
$(S_{yz}+S_{zy})$ components are nonzero, then in the differential cross 
section only the following two terms are present: $\varepsilon \sin(2\phi)$ and 
$\sqrt{2\varepsilon(1+\varepsilon )}\sin\phi$. For all other target 
polarizations the following structures are present: a term which does not 
depend on $\varepsilon $ and $\phi $, $2\varepsilon$, $\varepsilon \cos(2\phi ), $ 
$\sqrt{2\varepsilon(1+\varepsilon )}\cos\phi $.
\end{enumerate}
In the analysis of polarization phenomena, it is convenient to use the following  form for the amplitude $F$ of the process $\gamma^*+d\to n+p$:
\begin{eqnarray}
F&=&{\vec l}\cdot {\vec m}{\vec U}\cdot \hat {\vec k}
(f_1{\vec \sigma }\cdot \hat {\vec k}+f_2{\vec \sigma }\cdot {\vec m})+
{\vec l}\cdot {\vec m}{\vec U}\cdot {\vec m}
(f_3{\vec \sigma }\cdot \hat {\vec k}+f_4{\vec \sigma }\cdot {\vec m})\nonumber \\
&&+{\vec l}\cdot {\vec m}{\vec U}\cdot {\vec n}
(if_5+f_6{\vec \sigma }\cdot {\vec n}) 
+{\vec l}\cdot {\vec n}{\vec U}\cdot \hat {\vec k}
(if_7+f_8{\vec \sigma }\cdot {\vec n})\nonumber \\
&&+{\vec l}\cdot {\vec n}{\vec U}\cdot {\vec m}
(if_9+f_{10}{\vec \sigma }\cdot {\vec n})+
{\vec l}\cdot {\vec n}{\vec U}\cdot {\vec n}
(f_{11}{\vec \sigma }\cdot \hat {\vec k}+f_{12}{\vec \sigma }\cdot {\vec m})
\nonumber \\
&&+{\vec l}\cdot \hat {\vec k}{\vec U}\cdot \hat {\vec k}
(f_{13}{\vec \sigma }\cdot \hat {\vec k}+
f_{14}{\vec \sigma }\cdot {\vec m})+{\vec l}\cdot \hat {\vec k}{\vec U}\cdot {\vec m}
(f_{15}{\vec \sigma }\cdot \hat {\vec k}+f_{16}{\vec \sigma }\cdot {\vec m})
\nonumber \\
&&+{\vec l}\cdot \hat {\vec k}{\vec U}\cdot {\vec n}
(if_{17}+f_{18}{\vec \sigma }\cdot {\vec n}).\label{eq:eq17}
\end{eqnarray}
If the polarization of the nucleons in the final state is not measured, then all the observables are determined by bilinear 
combinations of the amplitudes $f_i$, that we can order in four groups: 
$(f_1,~ f_3,~ f_{11},~ f_{13},~f_{15})$, $(f_5,~ f_7,~ f_9,~ f_{17}), $ 
$(f_2,~ f_4,~ f_{12},~ f_{14},~ f_{16}), $ $(f_6,~ f_8,~ f_{10},~f_{18})$. 
The amplitudes $f_i$ (2.15) and $F_j$ (2.8) are related by a linear 
transformation of the form
\begin{equation}\label{eq:eq18}
f_i=M_{ij}F_j,
\end{equation}
where the nonzero elements of the $M_{ij}$ matrix 
are given in  Appendix 2. Using the explicit form for the amplitude of the 
reaction under consideration, Eq. (\ref{eq:eq17}), it is easy to obtain the 
expression for the hadronic tensor $H_{ij}$ in terms of the scalar amplitudes 
$f_i~(i=1-18)$. Appendix 3 contains the formulae for the structure functions 
$\alpha _i,\beta _i, \gamma _i, \delta _i$ in terms of the scalar amplitudes, 
which describe the polarization properties of the $e^-d\rightarrow e^-np$ reaction. 

Let us stress  again that the results listed above have a general 
nature and are not related to a particular reaction mechanism. They are valid 
for the one--photon--exchange mechanism assuming P-invariance of the hadron 
electromagnetic interaction. Their general 
nature is due to the fact that derivation of these results 
requires only the hadron electromagnetic current conservation and the fact 
that the photon has spin one.

The problems related to the deuteron structure such as the 
calculation of the meson--exchange current contribution, the determination of 
the $D-$ wave admixture in the deuteron ground state, the presence of the 
$\Delta\Delta-$ component and the six-quark configuration in the deuteron etc. 
will be shortly discussed below and they do not affect general results of this 
chapter. 

\chapter{Relativistic Impulse Approximation}
\begin{figure}
\hspace*{3true cm}\mbox{\epsfxsize=8.cm\leavevmode \epsffile{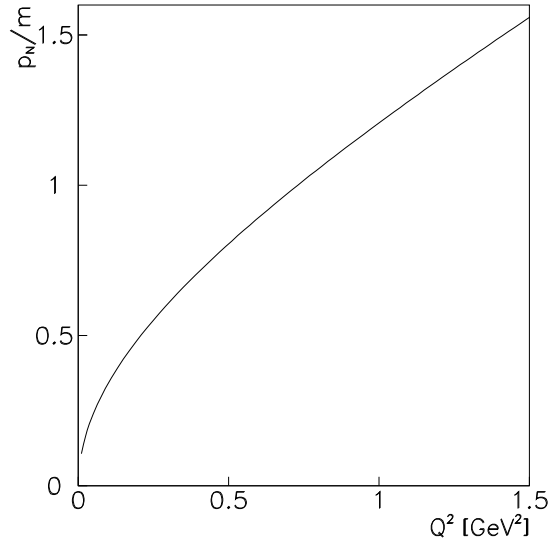}}
\caption{$Q^2$-dependence of the nucleon momentum in $d(e,e'p)n$, 
in quasi-elastic kinematics (Lab system).}
\label{fig:fig1}
\end{figure}

As already mentioned, a realistic model of the high-energy deuteron 
electrodisintegration reaction should be relativistic, due to the fast rising 
of the emitted-nucleon three-momentum, $p_N$, with increasing $Q^2$, which can 
be written in the Lab system as:
$$\displaystyle\frac{p_N}{m}=2\sqrt{\tau(1+\tau)},~
\tau=\displaystyle\frac{Q^2}{4m^2},$$
in quasi-elastic kinematics (see Fig. \ref{fig:fig1}). Therefore, the Feynman 
diagram technique,  where all particles 
are treated relativistically (the nucleons are described by the 
four-components Dirac spinors, and the deuteron  by the polarization 4-vector 
$U_{\mu}$), is a very good tool for all calculations in IA.

\begin{figure}[h]
\hspace*{2true cm}\mbox{\epsfxsize=12.cm\leavevmode \epsffile{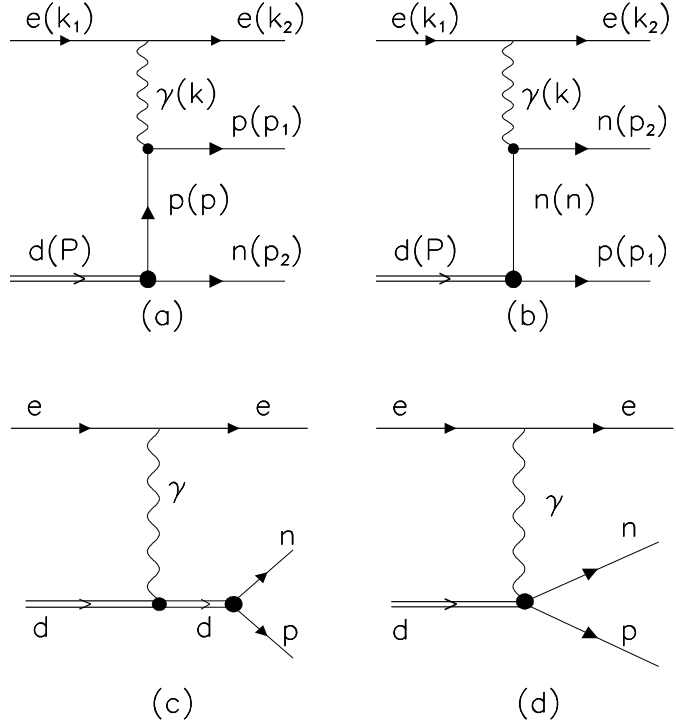}}
\vspace*{-1 true cm}
\caption{The Feynman diagrams describing RIA for the process $e^-+d\to e^-+n+p$.}
\label{fig:fig2}
\end{figure}

The diagrams illustrated in  Fig. \ref{fig:fig2} determine the amplitude 
of the reaction (\ref{eq:eq1}) in RIA.  Comparing with a  nonrelativistic 
approach, the diagrams (a) and (b) represent the relativized description of the 
one--nucleon--exchange mechanism (which are equally important in both 
approaches). The deuteron--exchange diagram, Fig. 3.2(c), as well as the 
contact diagram, Fig. 3.2(d), insure the electromagnetic current conservation 
in $\gamma ^*+d\rightarrow n+p$. The contact diagram can be related to the 
contribution of the meson--exchange currents. Of course, 
this diagram is not comprehensive of all variety of these currents,
but for its structure and origin it falls into this class. The deuteron diagram 
can be also related to the $np$-interaction  in definite states, with the 
quantum numbers of the deuteron. The one-nucleon-exchange diagrams give the 
largest contribution in the quasi-elastic region, due to the small deuteron 
binding energy compared to the hadron masses. Therefore, the pole
singularities lie near the physical region.

The deuteron structure (which, in the nonrelativistic approach, corresponds to 
DWF) is described here by the relativistic form 
factors of the $dnp-$ vertex with one virtual nucleon \cite{Bu79}. In order to 
calculate the dependence of these form factors on the nucleon virtuality we 
use their relations to both relativistic DWF Buck and Gross \cite{Bu79} and 
the nonrelativistic DWF for various $NN$ potentials.

\vspace*{.5 true cm}
The RIA amplitude for the $\gamma^*+d\rightarrow n+p$ reaction can be 
unambiguously calculated for any values of the kinematical variables $Q^2, s$ 
and $t$.

In the case of the nonrelativistic description of the deuteron
electrodisintegration the situation is somewhat different. The standard
nuclear approach to the investigation of the electron scattering by deuterons
and other nuclei requires the knowledge of the operator of the electron--nucleon
interaction. This operator cannot contain the antinucleon contribution
which is inevitably present in the covariant description of the $eN$--
scattering. The method based on the Foldy-Wouthoysen transformation is usually
applied. It allows to obtain this operator as an expansion over the powers of 
$\sqrt{Q^2}/m$. Let us note, in this connection, that this series seems to 
converge very badly \cite{Gi80}. Naturally, this method is not valid at 
$Q^2\ge m^2$, but such values of momentum transfer are currently accessible. 
Therefore, a relativistic description of the 
$e^-+d\rightarrow e^-+n+p$ reaction is more appropriate. Moreover, it allows 
to include directly the gauge invariance of the hadron electromagnetic 
interaction. 

In the relativistic description one has to face two aspects: the kinematical 
and the dynamical ones. The kinematical one, as applied to the 
$e^-+d\rightarrow e^-+n+p$ reaction, proceeds from the most general properties 
such as relativistic invariance, conservation 
of the hadron electromagnetic current as well as symmetry properties of the 
hadron electromagnetic interaction relative to $P-,C- $ and $T-$ transformations. 
The knowledge of the deuteron structure is not required in this case. This 
does not mean that the deuteron is described in an approximative way. On the 
contrary, such method leads to an adequate description of the polarization 
effects in the $e^-+d\rightarrow e^-+n+p$ process \cite{RGR87}, independently 
on the model for the deuteron structure. 

Concerning the dynamical description of the $e^-+d\rightarrow e^-+n+p$ 
process, only the RIA amplitude can be consistently calculated in 
the framework of the relativistic approach. The electromagnetic current 
$J_{\mu}$, corresponding to the diagrams of Fig. \ref{fig:fig2}, can be 
written as the sum of five contributions
$$
J_{\mu}=J_{\mu}^{(p)}+J_{\mu}^{(n)}+J_{\mu}^{(d)}+J_{\mu}^{(c)}+J_{\mu}^{(g)
}, 
$$
\begin{eqnarray}
J_{\mu}^{(p)}&=&\bar 
u(p_1)\left [F_1^p(k^2)\gamma_{\mu}-\frac{1}{2m}F_2^p(k^2)\sigma_{\mu\nu}k_{\nu}
\right ]\frac{\hat p+m}{p^2-m^2}\left \{ \hat UA(p^2,M^2)+\right .\nonumber\\
&&
\left .\frac{1}{2}U\cdot 
(p-p_2)B(p^2,M^2)+
\frac{m-\hat p}{2m}\left [\hat Ua(p^2,M^2)+U\cdot 
p_2b(p^2,M^2)\right ]\right \}u^c(p_2),\nonumber\\
J_{\mu}^{(n)}&=&\bar u(p_1) \left \{\hat 
UA(n^2,M^2)-\frac{1}{2}U\cdot (n-p_1)B(n^2,M^2)+[\hat Ua(n^2,M^2)\right .
\nonumber\\
&&\left . -U\cdot 
p_1b(n^2,M^2)]\frac{m+\hat n}{2m}\right\}\frac{\hat 
n+m}{n^2-m^2}\biggl[F_1^n(k^2)\gamma_{\mu}-\frac{1}{2m}F_2^n(k^2)\sigma_{\mu\nu}
k_{\nu}\biggr]u^c(p_2), 
\nonumber\\
J_{\mu}^{(d)}&=&(s-M^2)^{-1}\bar u(p_1)\biggl[\gamma 
_{\nu}A(s)+\frac{1}{2}(p_1-p_2)_{\nu}B(s)\biggr]u^c(p_2)
\nonumber\\
&&
\biggl\{F_1^d(k^2)U_{\nu
}(p_1+p_2+P)_{\mu}+
\frac{F_2^d(k^2)}{2M^2}[(s-M^2)U_{\mu}k_{\nu}-U\cdot kk_{\nu}(p_1+p_2+P
)_{\mu}]\nonumber\\
&&
+F_3^d(k^2)[U_{\mu}k_{\nu}-g_{\mu\nu}U\cdot Q]\biggr\}, 
\nonumber\\
J_{\mu}^{(c)}&=&-\bar 
u(p_1)\biggl\{\frac{1}{2}U_{\mu}[F_1^p(k^2)B(p^2)-F_1^n(k^2)B(n^2)]-
\frac{1}{2m}
F_1^p(k^2)\gamma_{\mu}[\hat Ua(p^2)+\nonumber\\
&&+U\cdot p_2b(p^2)]+\frac{1}{2m}F_1^n(k^2)[\hat Ua(n^2)-U\cdot p_1b(n^2)]\gamma _{\mu}\biggr\}u^c(p_2), 
\label{eq:eq19}
\end{eqnarray}
where $p=P-p_2,$ $n=P-p_1,$ 
$F_{1,2}^{p,n}$ are the Dirac and Pauli form factors of the proton and neutron; 
$A$, $B$, $a$ and $b$ are the form factors of the $dnp-$ vertex with one 
virtual nucleon \cite{Bu79}; $F_i^d (i=1,2,3)$  are the deuteron 
electromagnetic form factors and $J_{\mu}^{(g)}$ is the contribution 
(proportional to $k_{\mu}$) which insures the conservation of the hadron 
electromagnetic current $J_{\mu}$.  The four-momenta are indicated in Fig. \ref{fig:fig2}. 

Let us note 
that the contribution of the RIA diagrams (Fig. \ref{fig:fig2}) is not gauge 
invariant if the form factors of the $dnp$ and $\gamma ^*NN$ vertices are not 
correlated. Calculating the corresponding divergence $k\cdot J$ one obtains
$$k\cdot J=\bar u(p_1)\left \{ \left [\hat U(F_1^d(k^2)A(s)-F_1^p(k^2)A(p^2, M^2)-
F_1^n(k^2)A(n^2, M^2)\right ] \right .+ $$
$$\left . +\frac{1}{2}U\cdot (p_1-p_2)\left [F_1^d(k^2)B(s)-F_1^p(k^2)B(p^2, M^2)-
F_1^n(k^2)B(n^2, M^2)\right ]\right \} u^c(p_2). $$
To ensure the conservation of the electromagnetic current in the reaction 
$\gamma ^*+d\rightarrow n+p$ in RIA, we, therefore, replace the current given by
(\ref{eq:eq19}) with the current $J^{'}_{\mu }$ \cite{Fe72}:
$$J_{\mu } \ \rightarrow \ J^{'}_{\mu }=J_{\mu }-k_{\mu }(k\cdot J)/k^2. $$
Obviously, the modified current $J^{'}_{\mu}$ satisfies the continuity equation
$k_{\alpha }J^{'}_{\alpha }=0$ (the pole at the photon point $k^2=0$ does not play any role, because the second term is longitudinal and does not contribute
to the amplitude of the process involving real photons, i.e., in the limit
$k^2 \to 0.$)

The nucleon electromagnetic structure enters in the $\gamma ^*+d\rightarrow n+p$ 
reaction amplitude by means of the vertices $\gamma_{\mu}$ and 
$\sigma_{\mu\nu}k_{\nu}$, i.e., through the Dirac and Pauli form factors $F_1$ 
and $F_2$. Other parametrizations of the $\gamma^*NN$ vertex are also possible. 
They are completely equivalent for both real nucleons, but lead to different 
expressions in the case of one virtual nucleon. 

The effects of the nucleon virtuality in the $\gamma ^*NN$ vertex are, generally 
speaking, not distinguishable from the dependence on the nucleon virtuality of 
the $dnp$--vertex FFs. In this connection it is necessary to note 
that different parametrizations of the $\gamma ^*NN^*-$ vertex can affect the 
spin structure of the $\gamma^*+d\rightarrow n+p$ reaction amplitude. 
Thus, the investigation of polarization effects in $e^-+d\rightarrow e^-+n+p$ 
is sensitive to the  parametrization of the nucleon electromagnetic current. 

In principle, it is necessary to account also 
for the ambiguity of the parametrization of the deuteron electromagnetic current 
in the case when one of the deuterons is virtual (Fig. \ref{fig:fig2}c). Of course, this question is especially important near the reaction threshold where the deuteron--exchange diagram contribution is large, but the deuteron virtuality is very small. For 
the calculation of the observables it is necessary to know the $dnp$-vertex FFs,
which  can be expressed in terms of the usual nonrelativistic DWFs $u$ (S-wave)
and $w$ (D-wave). P-wave DWFs $v_t$ (triplet) and 
$v_s$(singlet) can also arise due to the fact that the virtual nucleon is out 
of mass shell. Buck and Gross determined \cite{Bu79} the set of the 
relativistic DWFs in terms of a parameter $\lambda$ which defines the 
off--mass--shell $\pi NN-$vertex
\begin{equation}
\label{eq:eq20}g_{\pi}\bar u(p_2)\biggl[\lambda 
+\frac{1}{2m}(1-\lambda )(\hat p_2-\hat 
p_1)\biggr]\gamma_5u(p_1),
\end{equation}
where $g_{\pi}$ is the coupling constant 
of the pseudoscalar $\pi NN$ interaction ($g_{\pi}^2/4\pi =14.48$).

For the electromagnetic FFs of the proton (charge $G_{Ep}$ and magnetic $G_{Mp}$) and the neutron magnetic form factor, $G_{Mn}$, the following scaling law and  the dipole parametrization are usually assumed \cite{Du83}
$$
G_{Ep}(k^2)=G_{Mp}(k^2)/\mu _p=G_{Mn}(k^2)/\mu _n=(1-k^2/0.71 
GeV^2)^{-2}, $$
with the following relations with the Pauli and Dirac nucleon (N) FFs: 
$$G_{EN}=F_{1N}-\tau F_{2N}, ~G_{MN}=F_{1N}+F_{2N},~ 
\mu_p=2.793, ~\mu _n=-1.913. $$
However, recent data \cite{Jo00,Ga02} suggest a large deviation from the dipole 
behavior for \gep, and consequently, the world data on \gmp have been 
reevaluated and a new parametrization is available \cite{Br02}. We will discuss below the effects of the new 
parametrizations.

The neutron charge form factor is less known at 
present. Experiments, based on the polarization method, have been recently performed and extend the $Q^2$ range to $\simeq$ 2 GeV$^2$ \cite{Mainz,Day,Madey}. The usual parametrizations for \gen , are 
$G_{En}=-\mu _n \tau G_{Ep}/(1+5.6\tau)$, \cite{Ga71}, or \gen=0. 
For the deuteron electromagnetic form factors, necessary for the 
calculation of the deuteron-exchange diagram, we take the parametrizations I 
and II \cite{www}, which are a phenomenological global fits of the world data. 

Before going into detailed predictions of the model for 
the $e^-+d\rightarrow e^-+n+p$ process, let us summarize the basic features of 
the suggested RIA approach:
\begin{itemize}
\item We take into account the complete set of diagrams that 
describes the $e^-+d\rightarrow e^-+n+p$ process in the Born approximation.

Besides the pole diagrams, (which, together with the diagram  of the deuteron 
structure in the $t$- and $u$-channels, constitute 
the content of the Born approximation of the model \cite{Re65}) we take into 
account also the contact or "catastrophic" diagram. We pointed out above, that 
the presence of this diagram follows from the requirement of 
the gauge invariance.

\item In Ref. \cite{Re65} the $dnp$-vertex in the $s$-channel (Fig. \ref{fig:fig2}c) (the deuteron pole contribution) was not described by the FFs, but by the constants which are related to the values of the form factors on--mass-shell of the virtual deuteron. As a result, the model \cite{Re65} predicts very large contribution of the deuteron pole diagram (approximately two 
order of magnitude larger than found by  \cite{Og80} in the 
region of the quasi-elastic peak). Taking FFs which depend on the 
intermediate deuteron virtuality, i.e., on the variable $s$, one can get a 
reasonable suppression of the $s$--channel contribution, in agreement with Ref. 
\cite{Og80}.

\item Several parametrizations of DWFs have been compared: relativistic 
\cite{Bu79}, Reid \cite{Reid}, Paris DWF \cite{La80} and charge-dependent 
Bonn DWF \cite{Ma01} which reproduces well the known low--energy properties 
of the deuteron.

\item We take into account relativistic effects such as the $P-$wave 
contribution in DWF \cite{Bu79}, which can play a role, due to the fact that the intermediate nucleons are off--mass--shell. 
\end{itemize}

\section{Unitarization procedure for $\gamma ^*d\rightarrow np$}

The scalar amplitudes $f_i,$ which correspond to 
the RIA matrix element, are  real functions. T-odd polarization observables, 
such as the asymmetry in 
the scattering of longitudinally polarized electrons by an unpolarized deuteron 
target, $d(\vec e, e'p)n$; the asymmetries in $\vec d(e, e'p)n$ which  are due to 
the vector polarization of a deuteron target, and the polarization of the proton 
(neutron) in $d(e, e'\vec p)n$ or $d(e, e'\vec n)p$ are induced by the 
imaginary parts in the $\gamma ^*d\rightarrow np$ amplitudes. Therefore, in RIA
all T-odd polarization observables  vanish.

\begin{figure}[h]
\hspace*{1truecm}\mbox{\epsfxsize=12.cm\leavevmode \epsffile{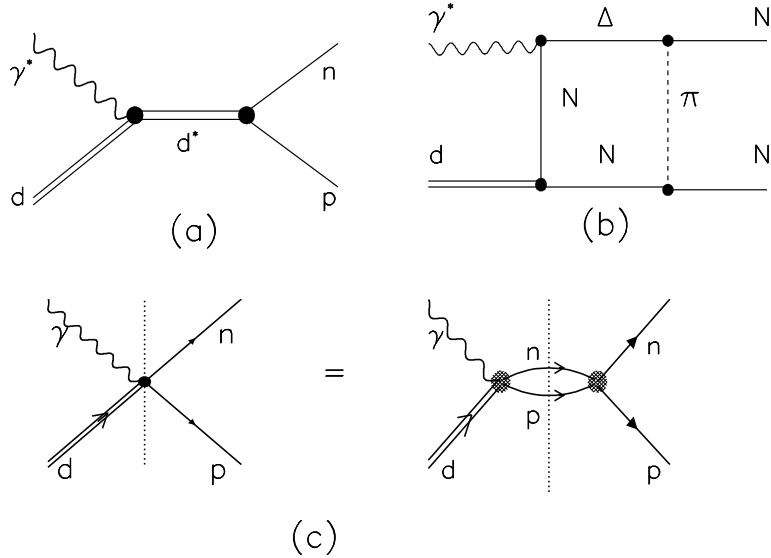}}
\vspace*{-1truecm}
\caption{Diagrams leading to complex amplitudes in $\gamma^*+d\to np$ 
reaction: (a) dibaryon resonance excitation, (b) excitation of a 
$\Delta$-isobar in an intermediate state, and (c) unitarity condition.}
\label{Fig:uni}
\end{figure}

However, additional mechanisms can be sources of complex amplitudes (Fig. \ref{Fig:uni}).
For example, the excitation of large--width dibaryon resonances in 
the $s-$ channel of $\gamma ^*d\rightarrow np$ reaction with various values of mass, spin, isospin, and space parity \cite{RGRN}.

Above the pion
production threshold, the mechanism of $\Delta -$ 
isobar excitation of one of the nucleons, with subsequent emission of a 
pion absorbed by the other nucleon (Fig. \ref{Fig:uni}b) can bring a sizeable 
complex contribution, due, on one side, to the $\Delta$--isobar propagator and, 
on the other hand, to the process where a nucleon and a $\Delta$--isobar can 
be created as free particles in the intermediate states (Fig. \ref{Fig:uni}b). 

Moreover, a universal and natural mechanism can produce complex amplitudes in 
$\gamma ^*d\rightarrow np$ from threshold: the $np\rightarrow 
np$ scattering, which is also included in nonrelativistic models of deuteron 
electrodisintegration in terms of FSI effects. In relativistic physics, 
this process is required by the  fulfilment of the unitarity condition 
(Fig. \ref{Fig:uni}c). 

\vspace*{1truecm}
In the energy range $2m\leq W\leq (2m+m_{\pi})$, the Fermi--Watson theorem, which holds at the lowest order in the electromagnetic coupling constant, follows from the 
unitarity condition \cite{Wa52}. According to this theorem, the phase of any multipole amplitude of the process $\gamma ^*d\rightarrow np$ coincides 
with the corresponding NN phase shift. We recall that a multipole amplitude describes the transition $\gamma ^*d\rightarrow np$ for a specific value of the total angular 
momentum $J$ when a virtual photon of particular multipolarity (an electric 
photon with transverse or longitudinal polarization or a magnetic one with 
transverse polarization only) is absorbed. In this case, the outgoing $np$ 
pairs are in a state  with specific values of the orbital angular momentum $L$ 
and total spin $S$. We have
\begin{equation}
\label{eq:eq21}
f_{[A]}(k^2, W)=
|f_{[A]}(k^2, W)|exp(i\delta_{[A]}(W)),
\end{equation}
where $f_{[A]}(k^2, W)$ is an RIA multipole amplitude for the process 
$\gamma ^*d\rightarrow np$ leading to the production of the np system in the
state with quantum numbers $[A]=J$, $L$, and $S$ , while $\delta _{[A]}(W)$ is the
NN phase shift in the state with quantum numbers $[A]$. This expression 
is strictly valid in the energy range $2m\leq W\leq (2m+m_{\pi})$ at any value of 
$k^2$ in the region of space-like momentum transfers. Since the $\gamma 
^*d\rightarrow np$ amplitudes are real in any version of IA (and so are all multipole amplitudes), they obviously do not 
satisfy the unitarity condition at nonzero NN phase shifts. This is true in 
both relativistic and nonrelativistic impulse approximations. 
 
Including meson--exchange currents, isobar configurations in the deuteron, 
and quark degrees of freedom does not solve this problem. Of course, the 
unitarity condition for the $\gamma ^*d\rightarrow np$  amplitude is violated in IA for any parametrization of the deuteron and 
nucleon electromagnetic FFs. A similar violation occurs for any form of 
the deuteron wave function. Furthermore, by allowing various off-mass--shell effects 
associated with the $\gamma ^*N^*N$ or $\gamma ^*d^*d$ vertices [$N^*$  ($d^*$)
is a virtual nucleon (virtual deuteron) in an intermediate state], it 
is also impossible to solve the unitarity problem for $\gamma ^*d\rightarrow np$. 

The unitarity condition is a serious constraint for any reliable model aiming to the description of the process  $\gamma ^*d\rightarrow np$. Its violation has far--reaching consequences for analysis of polarization 
effects - and above all, for the analysis of the T-odd 
polarization features mentioned above.

Of course, we do not consider here either 
actual violation of the T invariance of fundamental interactions, in which 
case elementary particles would have nonzero electric dipole moments, or 
CP-violation in the decays of neutral kaons. We discuss T-odd 
polarization correlations of the type ${\vec s\cdot (\vec k_1\times \vec k_2)}$ (${\vec s}$ 
is the pseudovector of the spin of one of the interacting particles, and ${\vec 
k_1}$ and ${\vec k_2}$ are their three-momenta). Such correlations 
are largely due to the strong interaction of the produced nucleons. The 
multipole amplitudes for  $ed\rightarrow enp$ get nonzero imaginary parts 
due to the effects of these interactions. All these imaginary parts determine 
T-odd effects in  $ed\rightarrow enp$. These effects occur not only for the 
vector polarization of one of the particles (in the initial or final states) 
but also for the tensor polarization of the target; the complex correlations 
of particle polarizations in the initial and final states may also be 
T-odd. Therefore, we can say that several polarization observables for  $e^-+d\rightarrow e^-+n+p$ have a T-odd character. Then the question arises of the correspondence between FSI effects and unitarity condition and one must specify the procedure for taking into account FSI effects. 

In framework of the nonrelativistic approach, the FSI effects in the 
reaction  $ed\rightarrow enp$ are evaluated by solving the Schrodinger equation (more 
precisely, the set of equations) in the continuous spectrum. The resulting wave functions are then  used to calculate the  $\gamma ^*d\rightarrow np$ 
multipole amplitudes following an appropriate choice of the reaction 
mechanisms and of the operator for electron--nucleon interaction. This procedure may induce a problem of electromagnetic--current conservation for the reaction 
 $\gamma ^*d\rightarrow np$. Various methods can be used to ensure the fulfilment of the unitarity 
condition (a procedure called  "unitarization") for the amplitude of the process  $\gamma 
^*d\rightarrow np$. In our opinion, the most consistent way is the one based on dispersion relations (in the late 1950s and in the early 1960s, this approach was 
very popular in the theory of low--energy strong and electromagnetic processes). 

The analyticity of multipole amplitudes allows one to obtain integral relations 
between the real and imaginary parts of the amplitudes. In a number of cases (for 
processes like  $\gamma ^*N\rightarrow N\pi $ and  $\gamma ^*d\rightarrow np$), 
the two--body unitarity condition allows to derive a set of simultaneous linear integral equations (of the Omnes--Muskhelishvili type). The existing well--developed methods for solving 
such sets of equations give answers in the form of definite integrals of the 
$\pi N$ phase shifts (for  $\gamma ^*+N\rightarrow N+\pi$) or NN phase shifts 
(for $\gamma ^*d\rightarrow np$). In this analysis one has to introduce some 
approximations, the most questionable of which is that only the two--body 
unitarity condition is used over the entire energy range from the reaction threshold up to infinite energies. It should be emphasized, 
however, that presently,  the dispersion--relation approach 
can include complex 
systems  of quarks and gluons and not only pions and nucleons. The earlier proofs of the dispersion 
relations worked very efficiently at that time, but they do not take into account 
these properties of hadrons. In view of this, we restrict ourselves to a  
simplified unitarization procedure consisting in multiplying the amplitudes of the 
process $\gamma ^*d\rightarrow np$ by the standard 
phase factor \cite{Ga66}; that is,
\begin{equation}
f_{[A]}^{(0)}(k^2, W)\longrightarrow f_{[A]}^{(0)}(k^2, W)exp(i\delta 
_{[A]}(W)),
\label{eq:eq22}
\end{equation}
where $f_{[A]}^{(0)}(k^2, W)$ is an IA multipole amplitude for the process $\gamma ^*d\rightarrow 
np$. In this connection, it should be noted that although isospin is not conserved in the reaction $\gamma ^*d\rightarrow np$ (electromagnetic interaction), the  isospin is conserved in elastic $n+p\rightarrow n+p$ scattering. Due to the generalized Pauli principle, the isospin $I$ of a nucleon pair 
with definite values of $L$ and $S$ is constrained by the relation $(-1)^{L+S+I}=-1$. This means that each amplitude  $f_{[A]}^{(0)}(k^2,W)$ determines the value of the isospin for the final $np$-system.

The substitution given by Eq.(3.4) is performed only for those 
multipole amplitudes that describe the production of the $np$-system with nonzero 
phase shift $\delta _{[A]}(W)$. This means that, at each energy $W$, there is 
a maximum value of the orbital angular momentum $L$ for the $np$-system and therefore a maximum value for the total angular momentum $J=J_m(W)$ that limits the number of those multipole amplitudes which undergo  unitarization. A similar  restriction on $L$ is imposed by the finite (and small) range of NN interaction. In spite of this, it is necessary to deal with a rather large number of multipole  amplitudes. It can be shown that, at each $J$ value with  $J\ge 1$, eighteen independent amplitudes are present ( their number coincides with the number of the scalar or invariant amplitudes of 
$\gamma ^*d\rightarrow np$). At $J=0 \ (J=1)$, there are 3 (14) independent transitions, so that it is
necessary to modify $18(J-1)+14+3=18J-1$ independent multipole amplitudes for
$J\leq J_m.$ For $J_m\geq 6,$ their number exceeds one hundred.

Of course, the substitution given by Eq. (\ref{eq:eq22}) leads to a unitary
$\gamma ^*d\rightarrow np$ amplitude (in some $W$ range). Evidently, this is not a general 
solution to the unitarity condition even in the range $2m\leq W\leq (2m+m_{\pi})$ 
(below the pion--production threshold). From Eq. (\ref{eq:eq22}), it follows that the 
unitarity condition determines only the phases of multipole amplitudes, but does not affect their moduli. Only the 
dispersion--relation approach, which makes use of  additional information about the 
analytic properties of the amplitudes, allows to determine the moduli 
of the multipole amplitudes as well. 

We will replace  Eq. (\ref{eq:eq22}) by a simple ansatz
\begin{equation}
\label{eq:eq23}
|f_{[A]}(k^2, W)|=\pm f_{[A]}^{(B)}(k^2, W).
\end{equation}
This is a very strong assumption. In this connection, we can mention that another unitarization approach based on IA, which consists in 
identifying the real part of a multipole amplitude with its 
Born part:
\begin{equation}
Ref_{[A]}(k^2, W)=f_{[A]}^{(B)}(k^2, 
W).
\label{eq:eq24}
\end{equation}
In this case, the unitary amplitude is restored with the help of 
the relation
\begin{equation}
Imf_{[A]}(k^2, W)=
\tan\delta 
_{[A]}f_{[A]}^{(B)}(k^2, W).
\label{eq:eq25}
\end{equation}
Keeping in mind this ambiguity, we 
will perform unitarization of the $\gamma ^*d\rightarrow np$ amplitude by means of a substitution in Eq. (\ref{eq:eq22}), 
because the relation (\ref{eq:eq25}) becomes meaningless near 
the $W$ values where the relevant phase shifts approach to $\pi /2$.

The proposed 
procedure for obtaining unitary multipole amplitudes has the following 
important features:
\begin{enumerate}

\item The unitary amplitude is determined only by physical 
observables that characterize the $NN$--interaction - specifically, by the phase shifts 
$\delta _{[A]}(W)$. This makes the economy of  nonrelativistic 
potentials and of the Schrodinger equation at an intermediate stage requiring a lot of complicated calculations. It is worth noting that the potential of the  $NN$ interaction is reconstructed on the basis of a large sample of data on elastic and inelastic $NN$ processes. It is not a quantity directly measurable, whereas  the $NN$ phase shifts are more directly related to the observables of $NN$ scattering.

\item The gauge invariance of the 
$\gamma ^*d\rightarrow np$ amplitude in the impulse approximation is not 
affected by the proposed unitarization scheme. Of course, the amplitude 
corresponding to the standard impulse--approximation diagrams in Fig. \ref{fig:fig1} does not 
satisfy the electromagnetic--current conservation in the general case of arbitrary 
nucleon and deuteron electromagnetic FFs.  For the RIA amplitude, we 
can calculate explicitly the divergence of the current, $k\cdot J^{(B)}$, where $k$ is
the photon four-momentum, and $J^{(B)}$is the electromagnetic current for the 
process $\gamma ^*d\rightarrow np$ considered in RIA. After 
that, the conserved current can be constructed with the procedure indicated above.
\item Including additional contributions to the $\gamma 
^*d\rightarrow np$ amplitude such as meson--exchange currents, isobar 
and quark configurations in the deuteron (as well as other  mechanisms in 
$\gamma^*d\rightarrow np$) is not a problem for the chosen unitarization 
scheme, because this changes only the Born part of the amplitude. The whole model (RIA together with meson--exchange currents, isobar 
and quark configurations) is object of unitarization.

\item The unitarized $\gamma 
^*d\rightarrow np$ amplitude, according to the procedure (\ref{eq:eq22}),
satisfies the requirement of the T-invariance of hadron electromagnetic 
interaction in the entire kinematical region with respect to $k^2$ and W. This is automatically ensured  by the procedure itself. Indeed, we note that, at the level of multipole amplitudes, the T-invariance requires that the difference between the 
phases of the amplitudes, corresponding to the absorption in $\gamma 
^*d\rightarrow np$ of electric and magnetic virtual photons is equal to $0$ or $\pi $  \cite{Ch66}. The fulfilment of this condition is especially important for the analysis of 
polarization effects in deuteron electrodisintegration - for example, the 
asymmetry of unpolarized electrons inclusively scattered by a vector-polarized 
target, ${\vec d}(e, e')np.$ According to the Christ-Lee theorem 
\cite{Ch66}, this asymmetry vanishes for all  $k^2$ and $W$ (within the 
one--photon--exchange mechanism). This means that it is necessary to add 
self--consistently other contributions to the $\gamma ^*d\rightarrow 
np$ amplitude. 
\end{enumerate}

The proposed unitarization procedure was carried out in the 
relativistic approach, without limitation in $k^2$. We also used a relativistic description of 
the nucleon electromagnetic current in terms of the Dirac ($F_1$) and 
Pauli ($F_2$) form factors. The $np$ phase shifts were taken from Ref. \cite{By87}.

\section{Unitarization of helicity amplitudes}
Since the spin 
structure of the matrix element is quite complicated, it is 
convenient to perform the unitarization procedure with the help of the helicity 
amplitudes (HA) formalism. As it was shown above, the reaction 
$\gamma ^*d\rightarrow np$ is described by eighteen amplitudes. 

Let us introduce the set of the helicity amplitudes 
$f_{\lambda\lambda '}(k^2, W, \vartheta )$ (where $\lambda$ and $\lambda '$ are 
the helicities of the initial 
($\gamma ^*+d$) and final ($n+p$) states) and consider the amplitudes 
$f_{\lambda\lambda '}=< \lambda _p, \lambda _n|T|\lambda 
_{\gamma},\lambda _d>$ (where  $\lambda _{\gamma}$, $\lambda _d, \lambda _p $ 
and $\lambda _n$ are the helicities of the virtual photon, deuteron, proton 
and neutron, respectively, with $\lambda =\lambda _{\gamma}- \lambda _d$ 
and $\lambda '= \lambda _p-\lambda _n$). We choose the following convention:
\begin{eqnarray}
h_1&=& <++|T|++> , ~h_2=<--|T|++> , ~ h_3=<++|T|+0> ,
\nonumber\\
h_4&=& <--|T|+0> , ~h_5=<++|T|+-> , ~ h_6=<--|T|+-> ,
\nonumber\\
h_7&=& <+-|T|++> , ~h_8=<-+|T|++> , ~ h_9=<+-|T|+0> , 
\nonumber\\ 
h_{10}&=&< -+|T|+0> , ~ h_{11}=<+-|T|+-> ,~h_{12}=<-+|T|+-> , ~ 
\nonumber\\ 
h_{13}&=&<++|T|0+> , ~h_{14}=<++|T|00> , ~ h_{15}=< ++|T|0-> , 
\nonumber\\ 
h_{16}&=&< -+|T|0+> , ~ h_{17}=< -+|T|00> , ~h_{18}=< -+|T|0-> ,
\label{eq:heli}
\end{eqnarray}
where, for the $np$--system, the sign $\pm $ 
denotes the nucleon helicities $\pm 1/2,$ and for the $\gamma d-$ 
system  the signs $\pm ,0$ denotes the helicities $\pm 1, 0$ of the photon and 
of deuteron, respectively. As it is shown above, the matrix element of the 
process (\ref{eq:eq1}) can be described in terms of the scalar amplitudes. 
The formulas relating the two sets of independent amplitudes $f_i$ and $h_i$ 
are given in Appendix 4. 

The formalism of helicity amplitudes is very convenient for the analysis of 
the polarization effects in the deuteron electrodisintegration. In particular, 
it is possible to perform an expansion over  the multipole amplitudes, 
which describe the transition in $\gamma 
^*d\rightarrow np$ for the states with definite values of the total 
angular momentum $J$ and particle helicity
\begin{equation}
<\lambda _p, \lambda _n|T|\lambda _{\gamma}, \lambda _d>=\frac{1}{\sqrt{4\pi }}\sum_J (2J+1)d^J_{\lambda '\lambda }(\vartheta )<J\lambda
'; \lambda_p, \lambda _n|T|J\lambda; \lambda _{\gamma}, \lambda_d>,
\label{eq:eql26}
\end{equation}
where $d^J_{\lambda '\lambda }(\vartheta )$ are the standard Wigner 
$d$--functions \cite{Wigner}.

To apply this formalism to FSI effects in the $\gamma ^*d\rightarrow 
np$ reaction, let us note that
the NN--scattering phase shifts correspond to the states of 
$np$-system with definite values of $L$ and $S$, but not to definite 
nucleon helicity states at given value of the total angular momentum of the 
$np$--system. Therefore, 
it is necessary to express the multipole amplitudes, Eq. (\ref{eq:heli}), in the $LS$--representation, namely:
\begin{eqnarray}
|J\lambda ; \lambda_p, \lambda _n>&=&\sum_{S, M_s}\biggl(\frac{1}{2}\lambda _p \frac{1}{2}-\lambda 
_n|SM_s\biggr)|J\lambda; SM_s>,
\nonumber\\
|J\lambda; 
SM_s>&=&\sum_{LM}(SM_sLM|J\lambda)|J\lambda; LS>. 
\label{eq:eq27}
\end{eqnarray}
which can be written as:
\begin{eqnarray}
|J0; \pm\pm >&=&\frac{1}{\sqrt{2}}\biggl(\pm 
|L=J,0>+\sqrt{\frac{J}{2J-1}}|L=J-1,1>\nonumber\\
&&-\sqrt{\frac{J+1}{2J+3}}|L=J+1,1>\biggr), 
\nonumber\\
|J\pm 1; \pm\mp >&=&\frac{1}{\sqrt{2}}\biggl(\mp 
|L=J,1>+\sqrt{\frac{J+1}{2J-1}}|L=J-1,1>\nonumber\\
&&+\sqrt{\frac{J}{2J+3}}|L=J+1,1>\biggr), 
\nonumber\\
|00;\pm\pm>&=&\frac{1}{\sqrt{2}}\biggl(\pm 
|L=0,0>-\frac{1}{\sqrt{3}}|L=1,1>\biggl),
\label{eq:eq28}
\end{eqnarray}
where we used the 
notation $|J\lambda ;(LS)>\equiv |LS>$. 

We will restrict ourselves to the description of the elastic $np$--scattering, 
which is correct for  deuteron electrodisintegration below the pion production 
threshold. Above threshold this approximation can be justified by the fact 
that even at large $Q^2(\ge 1$ GeV$^2$),  when relativistic effects are 
essential, the energy of the  $np-$system is $E_{np}\approx 500$ MeV, in the 
quasi-elastic kinematics. At such energies the cross section of inelastic 
$np$--scattering are small and  the phase shifts are real. As a result, we 
use the following formulas for the  multipole amplitudes of the 
$\gamma ^*d\rightarrow np$  reaction:
\begin{eqnarray}
<J\lambda',LS|T|\lambda_{\gamma }, \lambda _d>&=&F^Je^{i\delta^J_{LS}}, ~ J\leq 
J_m,\nonumber\\
<J\lambda ',LS|T|\lambda_{\gamma }, \lambda _d>&=&F^J,    
J>J_m,
\label{eq:eq29}
\end{eqnarray}
where $F^J=<J\lambda ',LS|T_B|J\lambda ,\lambda_{\gamma }, \lambda _d>$ is 
the corresponding RIA multipole amplitude, or, in other words, the Born 
approximation $T_B$. In order to simplify formally the FSI procedure described 
above, it is convenient to introduce a set of twenty-four amplitudes, $H_i$, 
which are related  to the eighteen $h_i$ amplitudes (\ref{eq:heli}) by:
\begin{eqnarray*}
&H_1=h_1, &~H_2=h_2, ~H_3=h_7, ~H_4=h_8, ~H_5=h_3, ~H_6=h_4, ~H_7=h_9, \\
&H_8= h_{10},&~ H_9=h_5, ~H_{10}=h_6, ~H_{11}=h_{11}, ~
H_{12}=h_{12}, ~H_{13}=h_{13},  \\
&~H_{14}=h_{15},&~H_{15}=h_{18}, ~H_{16}=h_{16}, ~ H_{17}=h_{14}, ~H_{18}=h_{14}, ~H_{19}=h_{17}, \\
&~H_{20}=h_{17}, &~H_{21}=h_{15}, H_{22}=h_{13}, ~H_{23}=h_{16}, ~H_{24}=h_{18}.
\end{eqnarray*}
For the corresponding multipole amplitudes, $V_i^{(J)}$, one can 
write
\begin{equation}
\label{eq:eq30}
V_i^{(J)}=\frac{1}{2}\int _{-1}^1dx 
H_i(k^2, W, x)d^J_{\lambda \lambda'}(x),
\end{equation}
where $x=\cos\vartheta $, and  $\vartheta $ is the 
angle of the proton emission relative to the virtual--photon momentum 
${\vec k}$ in CMS of the $\gamma ^*d\rightarrow np$ reaction.  In terms of 
these multipole amplitudes, the amplitudes of the transition process $\gamma 
^*d\rightarrow np$ to a state of the $np-$system with definite values of $JLS$: $|J\lambda ';L,S>$
have the form
\begin{eqnarray}
&<J\lambda ';J,0|T|i>&=\frac{1}{\sqrt{2}}(V_{i+1}^{(J)}-V_{i+2}^{(J)}), \nonumber\\
&<J\lambda ';J,1|T|i>&=\frac{1}{\sqrt{2}}(V_{i+4}^{(J)}-V_{i+3}^{(J)}),
\label{eq:eq314}\\
&
<J\lambda';J-1,1|T|i>&=\frac{1}{\sqrt{2}}\frac{\sqrt{2J-1}}{2J+1}
\left [ \sqrt{J}(V_{i+1}^{(J)}+V_{i+2}^{(J)}) \right .\nonumber\\
&&\left . +\sqrt{J+1}(V_{i+3}^{(J)}+V_{i+4}^{(J)})\right ], \nonumber\\
&
<J\lambda ';J+1,1|T|i>&=\frac{1}{\sqrt{2}}\frac{\sqrt{2J+3}}{2J+1}
\left [\sqrt{J}(V_{i+3}^{(J)}+V_{i+4}^{(J)})\right .\nonumber\\
&&\left .-\sqrt{J+1}(V_{i+1}^{(J)}+V_{i+2}^{(J)})\right], 
\nonumber
\end{eqnarray}
where we use the 
following notation for the initial states 
$|i>=|\lambda _{\gamma}\lambda _d>$:
$$ 
|i=0>=|++>, ~|i=4>=|+0>, ~|i=8>=|+->, 
$$
$$
|i=12>=|0+>, ~|i=16>=|00>, ~|i=20>=|0->. 
$$
For $J=0$, Eqs. (\ref{eq:eq314}) do not apply, so it 
is necessary to consider this case separately. 
We redefine the phases $\delta_{L,S}^J$ as:
$$
\delta _J\equiv \delta ^J_{J,0},~ \bar\delta _J\equiv \delta ^J_{J,1}, ~\alpha _J\equiv 
\delta ^J_{J-1,1}, ~\beta _J\equiv \delta ^J_{J+1,1}.
$$
For the numerical calculations, we took the phases from 
the energy--dependent phase--shift analysis of the 
$NN-$scattering \cite{By87},~in the energy range $10-800$ MeV,~for $J\leq 6$:
\begin{eqnarray} 
&\delta _0=\delta (^1S_{0,1}),&~ \delta _1=\delta (^1P_{1,0}),~ \delta _2=\delta (^1D_{2,1}),
~ \delta _3=\delta (^1F_{3,0}),~\delta _4=\delta (^1G_{4,1}),~\nonumber\\
&\delta _5=\delta (^1H_{5,0}),&~\delta _6=\delta (^1I_{6,1}),
~\nonumber\\
&\bar\delta _1=\delta (^3P_{1,1}),&~
\bar\delta _2=\delta (^3D_{2,0}),~\bar\delta _3=\delta (^3F_{3,0}),~
\bar\delta _4=\delta (^3G_{4,0}),~\bar\delta _5=\delta (^3H_{5,1}),\nonumber\\
& \alpha _1=\delta (^3S_{1,0}),&~\alpha _2=\delta (^3P_{2,1}),~\alpha _3=\delta 
(^3D_{3,0}),~~\alpha _4=\delta (^3F_{4,1}),\nonumber\\
&\alpha _5=\delta (^3G_{5,0}),&~\alpha _6=\delta (^3H_{6,1}),\nonumber\\
&\beta _0=\delta (^3P_{0,1}),&~\beta _1=\delta (^3D_{1,0}),~\beta _2=\delta (^3F_{2,1}),~\beta _3=\delta 
(^3G_{3,0}),~\nonumber\\
&\beta _4=\delta (^3H_{4,1}),&~\beta _5=\delta (^3I_{5,0}),
\label{eq:eq28a}
\end{eqnarray}
where we used the spectroscopic notation $^{2S+1}L_{J,T}$ for the states of 
the $NN$-system with total isospin $T$. The quantum numbers are not independent, but, due to the  
isotopic invariance and the generalized Pauli principle, they are related by  $(-1)^{L+S+T}=-1$. Finally, the FSI effects in the deuteron 
electrodisintegration are described in terms of twenty-four phases, at each energy of the 
$np$--system. The mixing effects will be discussed later. 

These phases are 
inserted in the RIA multipole amplitudes, Eq. (\ref{eq:eq314}), in the following way:
\begin{eqnarray}
&<J\lambda ';J,0|T|i>_U&=<J\lambda ';J,0|T|i> exp(i\delta _J),\nonumber\\
&<J\lambda ';J,1|T|i>_U&=<J\lambda ';J,1|T|i>exp(i\bar\delta _J),
\label{eq:eq32}\nonumber\\
&
<J\lambda ';J-1,1|T|i>_U&= <J\lambda ';J-1,1|T|i>exp(i\alpha _J) \nonumber\\
&
<J\lambda ';J+1,1|T|i>_U&=<J\lambda ';J+1,1|T|i>exp(i\beta _J). 
\nonumber
\end{eqnarray}
Let us note once more, that in order to calculate the unitarized amplitudes it is necessary to use the RIA results for 
$V_i^{(J)}.$ In this approach the $NN-$ interaction affects the multipole amplitudes up to $J=6$. The phases of the $S-$ and $P-$ scatterings, at energies up 
to 10 MeV, are approximated by the effective--radius approximation 
formula
$$p^{2L+1}\cot\delta _L(p)=-\frac{1}{a_L}+\frac{1}{2}r^2_Lp^2, $$
where 
$p$ is the nucleon momentum. The scattering length $a_L$  and effective radius 
$r_L$, describing the $NN-$ scattering in the states $^3S_{1,0} $, $^1S_{0,1} $, 
 $^3P_{0,1}$, $^3P_{1,1}$, $^3P_{2,1}$ and $^1P_{1,0},$ are taken from the data 
compilation \cite{Du83}.

\section{Mixing effects}
Let us briefly discuss the mixing 
effects in the $NN-$ scattering. The point is that in the general case of the triplet transitions, the orbital angular momentum is not conserved, since the transitions $L=J\pm 1\rightarrow L'=J\mp 1$ are allowed. The 
possibility of nondiagonal transitions is taken into account by introducing, 
at given $J$, the 2$\times$2 scattering matrix $S^J_{L'L}$. Unitary 
matrices of this type are determined, in the general case, by three real 
parameters. For $S^J_{L'L}$ we use the Stapp representation \cite{St57} which was also 
used for the phase-shift analysis of the $NN$--scattering 
\cite{By87}:
\begin{equation}\label{eq:eq33}
S^J_{L'L}=\left( 
\begin{array}{cc} a & 0 \\ 
~0 & b
\end{array} \right)
\left( \begin{array}{cc}
\cos2\varepsilon ^{(J)} & i\sin2\varepsilon ^{(J)} \\
~i\sin2\varepsilon ^{(J)} & \cos2\varepsilon ^{(J)}
\end{array} \right)
\left( \begin{array}{cc}a & 0 \\~0 & b
\end{array}\right),
\end{equation}
where 
$a=exp(i^3\delta ^J_{J+1}), b=exp(i^3\delta^J_{J-1}),$ and  $\varepsilon ^{(J)}$is 
the mixing parameter  of the states with total angular momentum $J$. In this 
case, the unitarity condition for the multipole amplitudes gets more involved 
and it differs from the standard Fermi--Watson form (even below the pion 
production threshold). 

In order to discuss the consequences of the unitarity for 
the $\gamma ^*d\rightarrow np$ reaction amplitudes taking into account the mixing, 
let us introduce the following states:
$$ |1>=|NN; J+1, 1>, ~|2>=|NN;J-1,1>, ~|3>=|\gamma d; \lambda _{\gamma}, \lambda_d>. $$
Then the 
elements of the scattering matrix $S_{ij}$ for the transitions between these 
states must satisfy the unitarity condition $SS^+=S^+S=1$ (neglecting the 
contributions of other channels). Taking into account that the $S$-matrix is 
symmetrical due to the $T$-invariance and accounting for the electromagnetic 
interaction in the lowest order on the coupling constant $e$, we can obtain, in 
the lowest order of the perturbation theory, a system of two equations using 
the unitarity condition:
$$ f_1^{(J)}+\cos2\varepsilon ^{(J)} e^{2i\delta 
_+^J}f_1^{(J)*}+i\sin 2\varepsilon ^{(J)} e^{i(\delta _+^J+\delta 
_-^J)]}f_2^{(J)*}=0, $$
$$ f_2^{(J)}+\cos2\varepsilon ^{(J)} e^{(2i\delta 
_-^J}f_2^{(J)*}+i\sin2\varepsilon ^{(J)} e^{[i(\delta _+^J+\delta 
_-^J)]}f_1^{(J)*}=0, $$
where$$ \delta _+^J\equiv ^3\delta ^J_{J+1}, ~\delta 
_-^J\equiv ^3\delta ^J_{J-1}, ~$$
$$  f_1^{(J)}=<NN; L=J+1, S=1|S|\gamma d; 
\lambda _{\gamma}, \lambda _d>, $$
$$  f_2^{(J)}=<NN; L=J-1, S=1|S|\gamma d; 
\lambda _{\gamma}, \lambda _d>. $$
If the mixing is neglected, then the solutions of these 
equations satisfy the Fermi--Watson theorem, that is:
$$  
f_1^{(J)}=iG_1exp(i\delta ^J_+), ~f_2^{(J)}=iG_2exp(i\delta ^J_-), ~ 
$$
where, in  turn, $G_1$ and $G_2$ are real functions of energy that 
are not constrained by the unitarity condition and are not 
correlated. In  presence of the mixing the amplitudes $f_1^{(J)}$ and 
$f_2^{(J)}$ cannot be independent.

\chapter{Polarization phenomena}
\section{Scattering of longitudinally-polarized electron beam}

In exclusive  deuteron electrodisintegration, all polarization effects induced by the vector polarization of a particle in the initial or the final state are
T-odd and are determined by the imaginary parts 
of bilinear combinations of the amplitudes.

We consider here the asymmetry in the scattering of longitudinally polarized 
electrons by an unpolarized deuteron target, $\Sigma _e $. As we already 
discussed in Chapter 3, only the mechanisms that generate complex 
amplitudes for the process $\gamma ^*d\rightarrow np$, such as nucleon FSI or 
the excitation of dibaryon resonances, can lead to T-odd effects.
Therefore, a non-zero value of  $\Sigma _e $ is the signature of the mechanisms 
beyond the one-photon exchange and/or IA. 

The asymmetry $\Sigma _e $
is determined by the so-called fifth SF $\alpha _5 $ \cite{RGR89,Le86}, see 
Eq. (2.11),  and has been object of several theoretical studies. In Ref.
 \cite{Le86} it was found, in the framework of a nonrelativistic approach, that
the asymmetry $\Sigma _e $ is sensitive to the form factor $G_{En}$. Later on 
other polarization observables were shown to be more appropriate for the
determination of this form factor (see discussion in section 4.2). A systematic 
study of the properties of the asymmetry $\Sigma _e $ in the framework of the 
unitarized RIA has been done in Ref. \cite{RGR96}.

In the one-photon-exchange approximation, the cross section for the $d({\vec e}, 
e'p)n$ reaction (with unpolarized target) can be written in terms of five 
independent contributions:
\begin{eqnarray}
&\displaystyle\frac{d^3\sigma}{dE'd\Omega_ed\Omega_N} =
{\cal N}& 
\left [ \sigma _T+\varepsilon \sigma _L+\varepsilon \cos(2\phi)\sigma _P+ \right.
\nonumber \\
&&\left .+\sqrt{2\varepsilon (1+\varepsilon)}\cos\phi\sigma _I+
\lambda \sqrt{2\varepsilon (1-\varepsilon)}\sin\phi\sigma ^{'}_I \right ], 
\label{eq:as3}
\end{eqnarray}
which are related to the hadronic-tensor components, Eq. (\ref{eq:eq3}), by: 
$$\sigma _T=H_{xx}+H_{yy},~ \sigma _P=H_{xx}-H_{yy},~
\sigma _L=-2\displaystyle\frac{k^2}{k_0^2} H_{zz},  $$
$$\sigma _I=-\displaystyle\frac{\sqrt{-k^2}}{k_0}(H_{xz}+H_{zx}),~
\sigma _I^{'} =i\displaystyle\frac{\sqrt{-k^2}}{k_0}(H_{xz}-H_{zx}).  $$
The asymmetry $\Sigma_e(\phi)$ is defined as:  
\begin{eqnarray}
&\Sigma_e(\phi)  &=\displaystyle\frac
{d\sigma (\lambda =+1)-d\sigma (\lambda =-1)}
{d\sigma (\lambda =+1)+d\sigma (\lambda =-1)}= 
\nonumber \\
&&=
\displaystyle\frac
{\sin\phi \sqrt {2\varepsilon (1-\varepsilon )} \sigma'_I}
{\sigma_T+\varepsilon \sigma_L+\varepsilon \cos(2\phi)\sigma_P+
\sqrt{2\varepsilon (1+ \varepsilon )} \cos\phi\sigma_I }.
\label{eq:asy3}
\end{eqnarray}
Due to the $\phi $-dependence, this asymmetry has to be measured  in noncoplanar geometry (out-of-plane kinematics). For $\phi=-90^{\circ}$ one finds:
\begin{equation}
A_{LT}^{'}=\Sigma _e(-90^{\circ}) =-\displaystyle\frac{\sqrt{2\varepsilon (1-\varepsilon )}\sigma'_I}
{\sigma_T +\varepsilon (\sigma_L -\sigma_P )}. 
\label{eq:asy4}
\end{equation}

Another asymmetry, the so-called left-right asymmetry, is defined as: 
\begin{equation}
A_{LT}=
\displaystyle\frac{d\sigma (\phi=0^0)-d\sigma (\phi=180^0)}
{d\sigma (\phi=0^0)+d\sigma(\phi=180^0)}
=\displaystyle\frac{\sqrt{2\varepsilon (1+\varepsilon )}\sigma_I}
{\sigma_T+\varepsilon (\sigma_L+\sigma_P)}. 
\label{eq:asy5}
\end{equation}

Note that in Ref. \cite{Ar89} another notation is used:
\begin{equation}
f_T=c\sigma _T, \ \ f_L=-\displaystyle\frac{c}{2}\displaystyle\frac{W^2}{M^2}
\displaystyle\frac{{\vec k}^2}{k^2}\sigma_L,~f_{TT}=-c\sigma_P,
\label{eq:asy6}
\end{equation}
$$f_{LT}=\sqrt{2}c\frac{W}{M}\displaystyle\frac{|{\vec k}|}{\sqrt{-k^2}}\sigma_I, \ \ 
f_{LT}^{'}=\sqrt{2}c\displaystyle\frac{W}{M}\displaystyle\frac{|{\vec k}|}{\sqrt{-k^2}}\sigma_I^{'},~
c=\displaystyle\frac{3\alpha }{16\pi }\displaystyle\frac{p}{MW}. $$
The subscripts $L$ and $T$ refer to longitudinally and transverse components of
the electromagnetic current, respectively, and the 
double subscripts indicate interference terms.

The structure function $\alpha _5$ is determined by
the interference of the reaction amplitudes that characterize the absorption of 
virtual photons with nonzero longitudinal and transverse polarizations. One 
finds that $\alpha_5\sim \sin\vartheta $ for any mechanism of the 
reaction $\gamma ^*d\rightarrow np$. The structure function $\alpha_5$ vanishes 
at proton emission angles $\vartheta =0^0$ and $180^0$ due to the conservation of 
the total helicity of the interacting particles in the process 
$\gamma^*d\rightarrow np$. The structure function $\alpha_5$ is nonzero only if 
the complex amplitudes of the reaction $\gamma ^*d\rightarrow np$ have nonzero 
relative phases. This is a very specific observable, which has no corresponding 
quantity in the deuteron photodisintegration $\gamma d\rightarrow np$.

The study of $\Sigma_e$ was firstly suggested (about 30 years ago) for the 
process $e+N\rightarrow e'+N+\pi$ \cite{Ge71}. Afterwards it has been discussed 
in  the processes of the type $A({\vec e}, e'h)X $  where $A$ is a target 
nucleus and  $h$ is the detected hadron \cite{Bo85,Pi85}, but it has 
only recently been measured \cite{Ma94,Ba02,Do95,Do99}.

In this section we calculate  $\Sigma_e$ in the kinematical conditions of recent experiments. The kinematical conditions of the performed experiments on measuring the fifth
SF are reported in the Table \ref{tab:asym1}. 

\vspace{0.5cm}
\begin{table}[h]
\begin{center}
\begin{tabular}{|c||c||c||c||c||c|}
\hline\hline
 $E$ [GeV] & $E'$ [GeV] & $\vartheta _e$ [deg] & 
$Q^2$ [GeV$^2]$ & $\varepsilon$ & Ref. \\
\hline\hline
 0.560 & 0.4897 & 40.0 & 0.128 & 0.784  &\protect\cite{Ma94,Ba02,Do95}\\
\hline
 0.800 & 0.645 & 31.0 & 0.147 & 0.848 & \protect\cite{Do99}\\
\hline
\end{tabular}
\vspace{0.5cm}
\caption{ Kinematical conditions of the experiments on measurement of the 
asymmetry $\Sigma_e$:
$E$ and $E'$ are the energies of the initial and final electron, $\vartheta _e$ is
the electron scattering angle, $Q^2=-k^2$ is the four-momentum transfer squared,
$\varepsilon $ is the linear-polarization degree of the virtual photon.}
\label{tab:asym1}
\end{center}
\end{table}

The first measurement of the fifth SF in a coincidence electron scattering experiment with  out-of-plane detection of the outgoing nucleon, has been performed at 
Bates for the reactions $^{12}C({\vec e},e'p)$, $d({\vec e},e'p)$ in the 
quasielastic kinematics \cite{Ma94}, but the results for the deuteron target
were not published in this paper. The conclusion of the paper is that, in the 
quasielastic kinematics, the fifth SF is an excellent tool for the study and 
separation of knockout and rescattering amplitudes of the investigated
reactions. Other cases, where the fifth SF may be used to isolate interfering
amplitudes, are the separation of resonant from competing channels in the study
of nucleon resonances. The first measurement of 
the beam-helicity asymmetry in a $p({\vec e}, e'p)\pi^0 $ reaction has been 
performed at MAMI (Mainz) \cite{Ba02}.
The importance of this new observable is proved by the fact that the results
of the experiment are in disagreement with three up-to-date model calculations. This shows the large sensitivity of this observable to the details of the models.

In Ref. \cite{Do95} the measurements of the $d({\vec e}, e'p)n$ out-of-plane electron 
helicity asymmetry $\Sigma _e $, the helicity-independent cross section, and 
the derived imaginary part of the longitudinal-transverse interference response 
(i.e., the fifth SF) were presented. The measurements were done in quasielastic
kinematics over a range of recoil momentum $p_m$ between $0$ and 180 MeV/c. 
In a more complete paper \cite{Do99} the authors present detailed description of the experiment and the analysis of the data. Finally, some future prospects for 
additional measurements are discussed. 

Measurements of the $d({\vec e}, e'p)n$ 
reaction were performed at the MIT-Bates \cite{Zh01a}. Using their notation, 
the longitudinal-transverse, $f_{LT}$ and $f^{'} _{LT}$, and the transverse-transverse, 
$f_{TT}$, interference SFs at a missing momentum of 210 MeV/c were 
simultaneously extracted in the "dip" region between quasi-elastic ridge and the
$\Delta-$ resonance.

\begin{figure}[h]
\begin{center}
\mbox{\epsfxsize=10.cm\leavevmode \epsffile{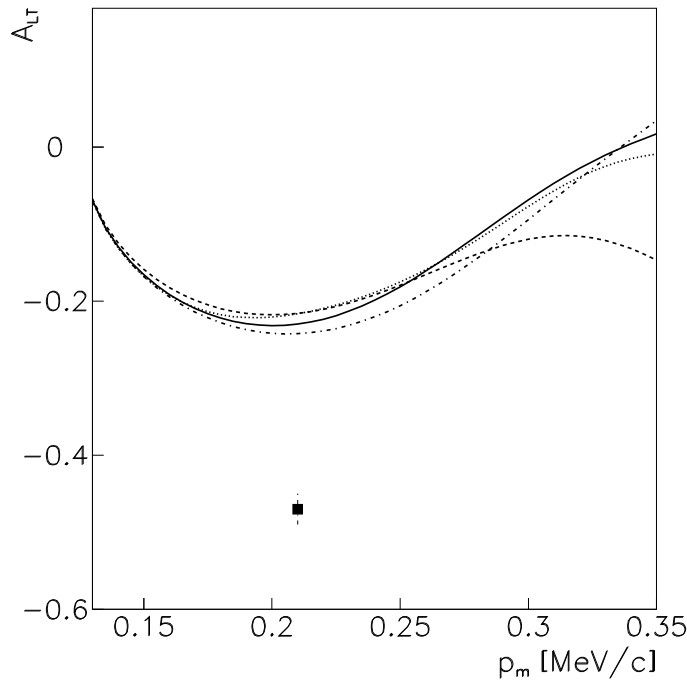}}
\caption{The asymmetry $A_{LT}$ as a function of a missing momentum $p_m$. The
asymmetry is calculated for various DWFs: the Paris (solid line), thr 
Reid soft-core (dotted line), charge-dependent Bonn (dash-dotted line). The dashed line shows
the calculation without FSI effects (for the Paris DWF). The Galster
parametrization was used for $G_{En}.$ The data are taken from Ref. 
\protect\cite{Zh01a}.}
\label{fig:asymm}
\end{center}
\end{figure}

\begin{figure}[h]
\begin{center}
\mbox{\epsfxsize=10.cm\leavevmode \epsffile{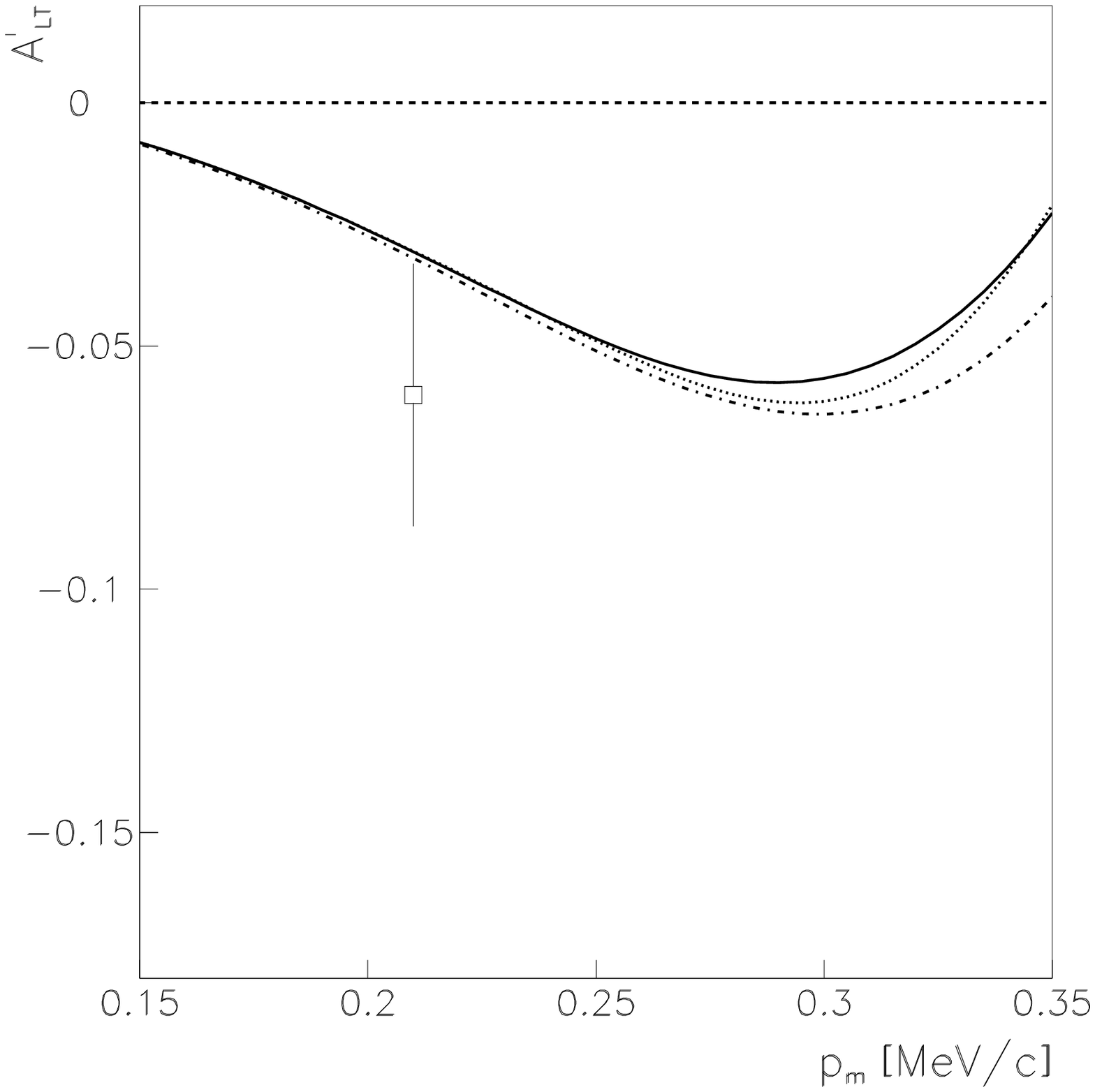}}
\caption{The asymmetry $A_{LT}^{'}$ as a function of $p_m$. The same notations 
as in Fig. \ref{fig:asymm}.}
\label{fig:asymprime}
\end{center}
\end{figure}

The results of our calculations of the asymmetries $A_{LT}$ and $A_{LT}^{'}$ 
for the  kinematics conditions of the experiment \cite{Zh01a} are
shown in Fig. \ref{fig:asymm} and \ref{fig:asymprime}, respectively. The predicted asymmetry $A_{LT}$ is
strongly underestimated in our model in contrast with the Arenhovel's prediction
\cite{Zh01a} (their calculations show very little sensitivity to the two-body
currents for the asymmetry $A_{LT}$). So, the reason of this dscrepancy is not
clear. The importance of FSI is not large in the region where $p_m\leq 280
MeV/c$. Beyond this region the role of FSI is appreciably increased. The
sensitivity of the asymmetry $A_{LT}$ to the choice of DWFs is not strong in all
investigated region of the missing momentum. The behaviour of the asymmetry $A_{LT}$
as a function of $p_m$ is similar to the Arenhovel's one and strongly differs from
the behaviour predicted by the relativistic model of the Hummel and Tjon \cite{Hu94}
and the $\sigma _{ccl}$ prescription of the de Forest \cite{Fo83}. The predicted value
for the asymmetry $A_{LT}^{'}$ is not contradicted to the experimental data. The
sensitivity of this asymmetry to the choice of DWFs appears only in the region
where $p_m\geq 280 MeV/c.$ The qualitative behaviour of this asymmetry as a function
of $p_m$, predicted in our model, is similar to the Arenhovel's prediction.

\begin{figure}[h]
\begin{center}
\mbox{\epsfxsize=10.cm\leavevmode \epsffile{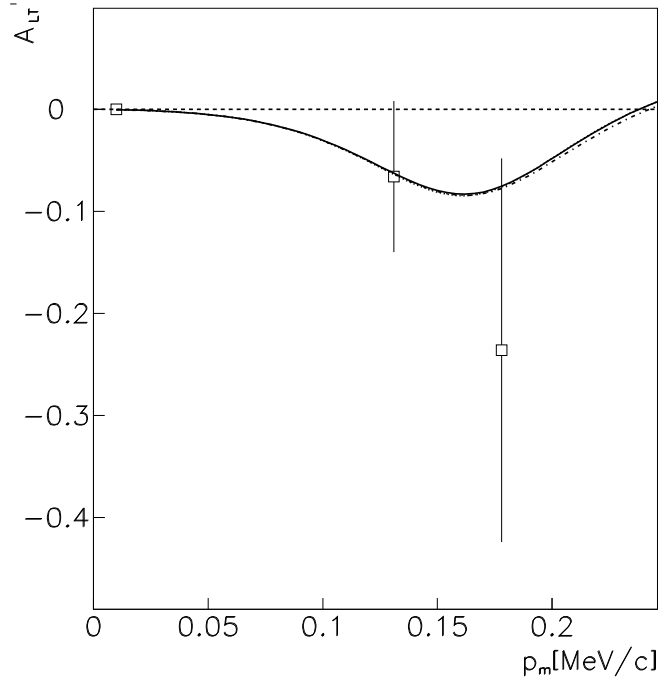}}
\caption{The asymmetry $A_{LT}^{'}$ as a function of a missing momentum $p_m$. The
asymmetry was calculated for various DWFs: the Paris (solid line), charge-dependent 
Bonn (dash-dotted line). The dashed line shows the calculation without FSI 
effects (for the Paris DWF). The Galster parametrization was used for $G_{En}.$ 
The data are the experimental results that have been approximately unfolded to
account for acceptance averaging, corresponding to the Table XVII 
from Ref. \protect\cite{Do99}.}
\label{fig:asymdolf}\end{center}
\end{figure}

The asymmetry $A_{LT}^{'}$ calculated for the kinematics of the experiment
\cite{Do99} is presented in Fig. \ref{fig:asymdolf}. This experiment was 
performed at less missing momentum ($p_m\leq 180 MeV/c$). The kinematics of 
this experiment were chosen to be in kinematic regime where the subnuclear 
degrees of freedom are not expected to contribute significantly. The predicted 
asymmetry agree relatively well with the data. The sign, the magnitude, and 
the general trend of the calculation are in agreement with the data. There is 
no  sensitivity of the asymmetry to the choice of the DWF model at such 
low values of $p_m$. The behaviour
of the asymmetry as a function of momentum $p_m$ calculated by the Arenhovel is 
similar to one calculated by us (the Arenhovel's asymmetry somewhat larger in 
magnitude at $p_m\geq 150 MeV/c$). As it was pointed out in Ref. \cite{Do99}, 
the overall out-of-plane angular acceptance has a significant effect on the
measured values and one has to take carefully into account the experimental 
conditions for a meaningful comparison. The data reported in 
Fig. \ref{fig:asymdolf} are the unfolded data (for the details see Ref. 
\cite{Do99}). It has been shown  that relativistic
contributions to the nonrelativistic approach are quite large  even at this
low momentum transfer \cite{Do99}, and that they are of the same order as the 
difference in potentials.

The data \cite{Zh01a} agree with the full calculations (improved by including
retardation diagrams)  \cite{Ri97}. The calculations  of Ref. \cite{Hu94}, which 
do not contain two-body contributions, fail to describe
the data. The conclusion of the paper \cite{Zh01a} is that the data clearly reveal
strong effects of relativity and FSI, as well as of two-body currents arising
from the meson-exchange currents and isobar configurations. The two-body currents
and relativity are extremely important to the understanding of the deuteron, and
thus, more rigorous relativistic calculations including all ingredients
discussed in this paper are needed. The authors noted the substantial
cancellations between the effects of the two-body currents and FSI. 

As a first step towards implementing a systematic program of measurements of the
response functions for nucleon resonances and few-body nuclei, measurements were
made of the fifth SF for the $^{12}C({\vec e}, e'p)$ and $d({\vec e}, e'p)n$
reactions by using a specially designed experimental apparatus \cite{Do99}.

Out-of-plane measurements with higher statistical precision have been planned in
the near future, especially in the region of higher missing momentum ($\ge$ 250 
MeV/c) and as a function of the energy transfer up to the $\Delta $-resonance
region \cite{Be97}. These data will clarify the role of the relativity and two-body
currents and provide a detailed understanding of the isobar configurations and
possible knowledge of the $\Delta - N$ interactions.

\section{Scattering of longitudinally-polarized electrons on a vector-polarized deuteron target}

The differential cross section of the reaction ${\vec d}({\vec e},e'p)n$, 
where the electron beam is longitudinally--polarized and the deuteron target 
is vector--polarized, can be written as follows
\begin{equation}\label{definition of cross--section, 3}
\displaystyle\frac{d^3\sigma }{dE'd\Omega_ ed\Omega_ N} =
\sigma _0\left [ 1+\lambda\Sigma _ e + (A^d_x +\lambda A^{ed}_x)\xi _x+
(A^d_y +\lambda A^{ed}_y)\xi _y+(A^d_z +\lambda A^{ed}_z)\xi _z\right ] ,
\end{equation}
where $\sigma _0$ is the unpolarized differential cross section, $\lambda$ is 
the polarization of the electrons, $\Sigma _e $ is the beam analyzing power (the
asymmetry induced by the electron-beam polarization), $A^d_i (i=x, y, z)$ are the analyzing 
powers due to the vector polarization of the deuteron target, and $A^{ed}_i 
(i=x, y, z)$ are the spin--correlation parameters. The direction of the deuteron 
polarization vector is defined by the angles $\vartheta^*, \phi^*$ in the frame 
where the $z$ axis is along the direction of the three-momentum transfer 
$\mathbf{ k}$ and the $y$ axis is defined by the vector product of the detected 
nucleon and virtual photon momenta (along the unit vector ${\vec n}$). The 
target analyzing powers and spin-
correlation parameters depend on the orientation of the target polarization 
vector. $\Sigma _e $ and $A^d_i$ are T--odd observables and they are completely 
determined by the reaction mechanisms beyond RIA, for example, by the FSI
effects. On the contrary, $A^{ed}_i$ are T-even observables and they do not 
vanish in the absence of the FSI effects. 

The expressions of the $A_i^d$ and $A_i^{ed}$ asymmetries can be  explicitly written as functions of the azimuthal angle $\phi$, of the 
virtual-photon linear polarization $\varepsilon $, and of contributions of the
longitudinal $(L)$ and transverse $(T)$ components of the hadronic current of
the $\gamma ^*d\rightarrow np$ reaction:
\begin{eqnarray}
A_x^{d}\sigma _0&=&N\sin\phi [\sqrt{2\varepsilon (1+\varepsilon )}A_x^{(LT)}
+\varepsilon \cos\phi A_x^{(TT)}], \nonumber\\
A_z^{d}\sigma _0&=&N\sin\phi [\sqrt{2\varepsilon (1+\varepsilon )}A_z^{(LT)}
+\varepsilon cos\phi A_z^{(TT)}], \nonumber\\
A_y^{d}\sigma _0&=&N[A_y^{(TT)}+\varepsilon A_y^{(LL)}+
\sqrt{2\varepsilon (1+\varepsilon )}\cos\phi A_y^{(LT)}
+\varepsilon \cos(2\phi )\bar A_y^{(TT)}], \nonumber  \\
A_x^{ed}\sigma _0&=&N[\sqrt{1-\varepsilon ^2}B_x^{(TT)}
+\sqrt{2\varepsilon (1-\varepsilon )} \cos\phi B_x^{(LT)}], 
\label{eq:eq47}\\
A_z^{ed}\sigma _0&=&N[\sqrt{1-\varepsilon ^2}B_z^{(TT)}
+\sqrt{2\varepsilon (1-\varepsilon )} \cos\phi B_z^{(LT)}], \nonumber\\
A_y^{ed}\sigma _0&=&N\sqrt{2\varepsilon (1-\varepsilon )} 
\sin\phi B_y^{(LT)}, \nonumber
\end{eqnarray}
where the individual contributions to the considered asymmetries in terms of 
SFs $\beta _i$, introduced in Eq. (2.12), are given by 
\begin{eqnarray}
A_x^{(TT)}&=&4\beta _{11}, \ A_y^{(TT)}=\beta _2+\beta _3, \ 
\bar A_y^{(TT)}=\beta _2-\beta _3, \  A_z^{(TT)}=4\beta _7,\nonumber\\ A_x^{(LT)}&=&-2\displaystyle\frac{\sqrt{Q^2}}{k_0}\beta _{10}, \
 A_y^{(LT)}=-2\displaystyle\frac{\sqrt{Q^2}}{k_0}\beta _4, \ 
 A_z^{(LT)}=-2\displaystyle\frac{\sqrt{Q^2}}{k_0}\beta_6, \label{eq:eq48} \\
A_y^{(LL)}&=&2\displaystyle\frac{Q^2}{k^2_0}\beta _1, \ 
B_x^{(TT)}=2\beta _{13}, \ B_z^{(TT)}=2\beta _9,  \nonumber\\
B_x^{(LT)}&=&-2\displaystyle\frac{\sqrt{Q^2}}{k_0}\beta _{12},
B_y^{(LT)}=2\displaystyle\frac{\sqrt{Q^2}}{k_0}\beta _5, 
B_z^{(LT)}=-2\displaystyle\frac{\sqrt{Q^2}}{k_0}\beta _8. \nonumber
\end{eqnarray}

Of course, the expressions for SFs $\beta _i$ in terms of the reaction
amplitudes $f_i$ (2.15) are general ones (see Appendix 3) and  they do not 
depend on any details of the reaction
mechanism. The determination of all these SFs constitutes the
complete $\gamma ^* +d\rightarrow n+p$ experiment. 

At this stage, the general model-independent analysis of the polarization
observables in the reactions ${\vec d}(e,e'p)n$ and ${\vec d}({\vec e},e'p)n$ 
is completed. To proceed further in the calculation of the observables, one 
needs a model for the reaction mechanism and for the deuteron structure.

The asymmetry $A_y^d$ has been studied in Ref. \cite{RGR02} using RIA with unitarized
multipole amplitudes. Since SFs, which define the asymmetries for the ${\vec
d}(e,e'p)n$ reaction, are determined by the imaginary parts of the bilinear
combinations of the reaction amplitudes these asymmetries are zero in IA. The
calculations of the asymmetry $A_y^d$ were performed in the coplanar kinematics 
at $\phi = 0^0$ and $180^0$. The values of $Q^2= \ $0.2 and 1.5 GeV $^2$ 
correspond to the nonrelativistic and relativistic regions beyond the 
quasielastic scattering. The conclusion is that the investigation of the T-odd 
asymmetries in the vector-polarized deuteron electrodisintegration can give important information about the reaction mechanisms, especially about the importance of the interference of various contributions to the total reaction  amplitude.

\subsection{Predictions and results for the reaction $\vec d(\vec e, e'p)n$ }

\subsubsection{The reaction $\vec d(\vec e, e'p)n$ in the NIKHEF experiment}

\begin{figure}
\mbox{\epsfxsize=14.cm\leavevmode \epsffile{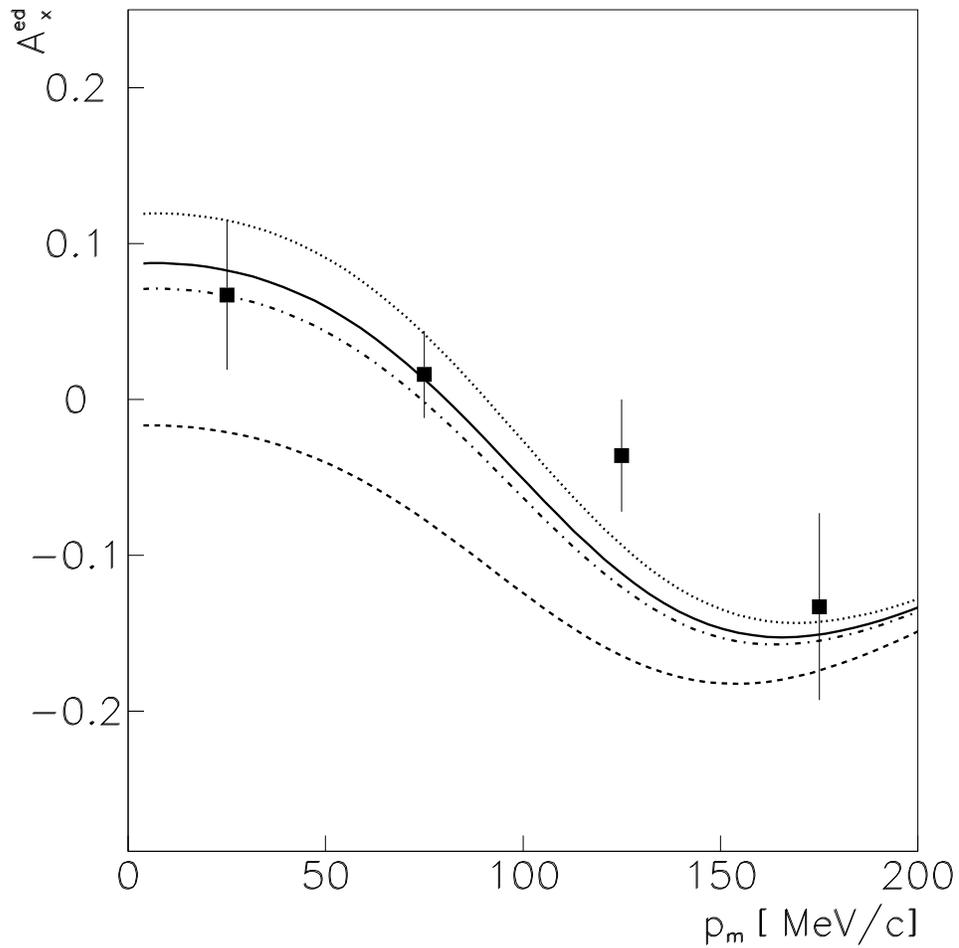}}
\caption{  Data \protect\cite{Pa99} and theoretical predictions for the 
sideways asymmetry, $A_x^{ed},$ versus the missing momentum $p_m$ for the 
${\vec d}({\vec e},e'n)p$ reaction. The asymmetry was calculated for the Paris 
DWF and different parametrizations of the form factor $G_{En}$: the Galster 
(p=5.6) - solid line, the Galster (p=10) - dash-dotted line, 0 - dashed line, 
and $G_{En}=-\tau G_{Mn}$ (or $F_{1n}=0)$ - dotted line.}
\label{fig:asymed} 
\end{figure}

The sideways asymmetry $A_x^{ed}$ has been measured at the NIKHEF accelerator
\cite{Pa99}, in the following kinematics : $E=720 MeV$,
$\vartheta _e=40^\circ $ and $E'= 610$ MeV,  $Q^2 $= 0.206 GeV$^2$.
In Fig. \ref{fig:asymed} the data are shown as function of the missing 
momentum. As already pointed out, this observable is especially sensitive to 
the form factor \gen. The different calculations correspond to the Paris DWF. The 
calculation corresponding to the Galster parametrization \cite{Ga71} with 
$p=5.6$ (solid line) best reproduces the data. The  agreement of the 
calculation with the experimental data is excellent, particularly in the 
quasi-elastic region. The calculations for the Galster parametrization with 
$p=10$ is represented as dash-dotted line, for \gen = 0 is represented as the 
dashed line and for $F_{1n}= 0$ - as the dotted line. The sensitivity of the 
asymmetry to the choice of the DWF model is small. The influence of the FSI 
effects is also insignificant. So, the measurement of this asymmetry in the 
quasi-elastic region can give a reliable value of the form factor $G_{En}.$

\subsubsection{The reaction $\vec d(\vec e, e'p)n$ in the  JLab-E93-026 experiment}

\begin{figure}
\mbox{\epsfxsize=14.cm\leavevmode \epsffile{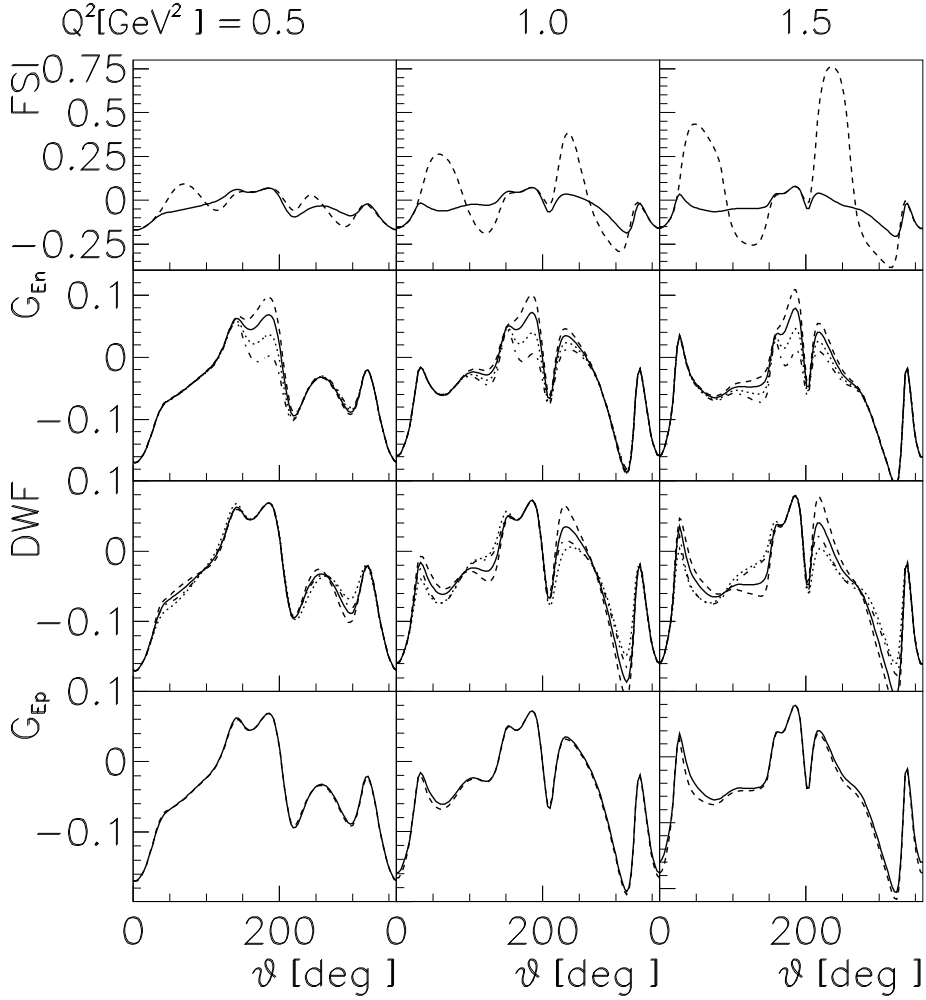}}
\caption{ $\vartheta $-dependence of $A_x$-asymmetry for $Q^2$ =0.5, 1, 
and  1.5 GeV$^2$ are drawn: from left to right and from top to bottom:
- with FSI effects (solid line) and without ones (dashed line);
- for different parametrizations of $G_{En}$: the 
Galster parametrization (solid line), the same one scaled by a 
factor 0.5 (dashed line), the same one scaled by a factor 1.5 
(dotted line), \gen=0 (dash-dotted line);
- for different DWFs: the Paris (solid line), the Reid soft-core 
(dashed line), the Bonn (dotted line), the Buck-Gross (dashed-dotted line);
-for the dipole-like formula (solid  line) and for the  
parametrization given in Ref. \protect\cite{Br02}  (dashed line) of the
\gep form factor.}
\label{fig:asymx} 
\end{figure}

\begin{figure}
\mbox{\epsfxsize=14.cm\leavevmode \epsffile{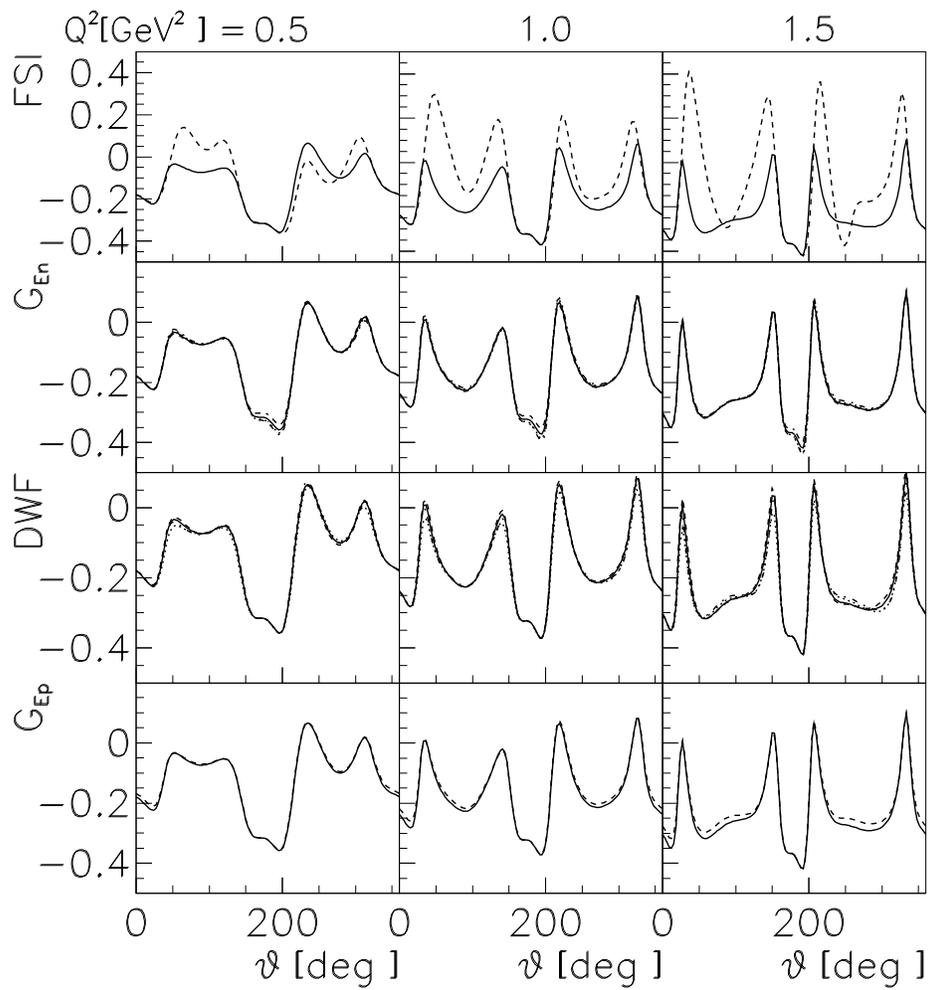}}
\caption{ $\vartheta $-dependence of the $A_z$-asymmetry. The same notations 
as in Fig. \protect\ref{fig:asymx}.}
\label{fig:asymz} 
\end{figure}

We calculated the asymmetries for the electron kinematics shown in 
Table \ref{tab:Day}, which correspond to the kinematical conditions of the 
experiment E93-026 at JLab \cite{Day} \footnote{We are grateful to D. Day for providing us with updated values of the kinematics of the experiment E93-026.}.  
Note that $\epsilon\simeq 1$ for the quasi-elastic regime.
\begin{center}
\vspace{0.5cm}
\begin{table}[h]
\hspace*{3 truecm}
\begin{tabular}{|c||c||c||c||c|}
\hline
$Q ^2$ [GeV$^2$] & $E$ [GeV] & $E'$ [GeV] & $\vartheta_e$ [deg] & 
$\varepsilon$ \\
\hline
\it 0.5 & 2.332 & 2.063 & 18.55 & 0.942\\
\hline
\it 1.0 & 3.481 & 2.948 & 17.96 & 0.940\\
\hline
\it 1.5 & 4.232 & 3.376 & 19.26 & 0.923\\
\hline
\end{tabular}
\caption{ Kinematical parameters for the JLab experiment \protect\cite{Day}.}
\label{tab:Day}
\end{table}
\end{center}
We report in Figs. \ref{fig:asymx} and \ref{fig:asymz} the results for $A_x$ 
and $A_z$ asymmetries, for coplanar kinematics,  as a function of $\vartheta $ 
variable ($\vartheta $ is the angle between the virtual photon and emitted 
nucleon momenta in CMS of the $np$--system) in the full angular range, in order 
to give a global view of the sensitivity of these observables to different 
ingredients of the calculation. The range $\vartheta\le 180^{\circ}$ corresponds
 to $\phi=0$, and the range $\vartheta \ge 180^{\circ}$ corresponds to $\phi=180$. 

From top to bottom the effects of FSI, of different choices of form factor \gen, 
 DWF and \gep form factor are shown. The results for different values of $Q^2$ 
($Q^2$ = 0.5, 1, and  1.5 GeV$^2$) are drawn from the left to the right. 

The solid line, in all graphs, corresponds to the Paris DWF \cite{La80}, to the Galster parametrization for the form factor 
\gen  (with $p=5.6$)\cite{Ga71}, and to the standard dipole parametrization for 
the form factor \gep . It includes FSI effects.

Both asymmetries show a strong $\vartheta $-dependence, for all considered 
values of momentum transfer squared. Switching off FSI (dashed line, top series of figures) modifies essentially the results at large $Q^2$, in an angular range outside the quasi-elastic region where the cross section is smaller. The same effect, although less 
apparent, applies to DWF choice, as one can see from  the the corresponding 
set of figures where the results for the Paris (solid line), the Reid soft-core 
(dashed line), the Bonn (dotted line), and  the Buck-Gross (dashed-dotted line) 
DWFs are represented. Two parametrizations of \gep form factor give similar 
results in the whole kinematical region (bottom series of figures): the 
dipole-like (solid  line) and the recent parametrization \protect\cite{Br02} 
(dashed line) slightly differ at the highest values of $Q^2$.

\begin{figure}
\mbox{\epsfxsize=14.cm\leavevmode \epsffile{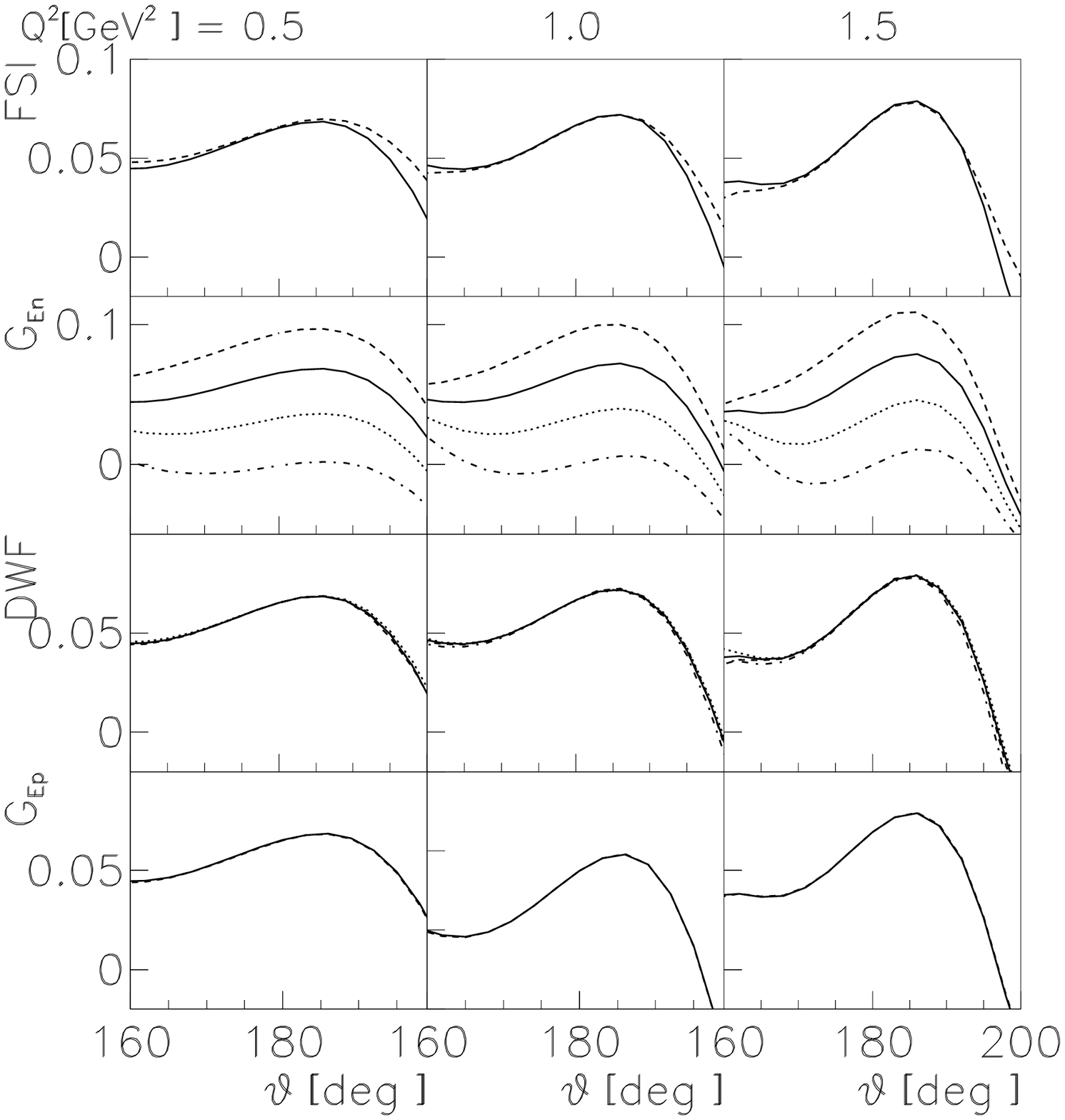}}
\caption{ $\vartheta $-dependence of the $A_x$-asymmetry in the quasi-elastic 
kinematics region. The same notations as in Fig. \protect\ref{fig:asymx}.}
\label{fig:qeasymx} 
\end{figure}

\begin{figure}
\mbox{\epsfxsize=14.cm\leavevmode \epsffile{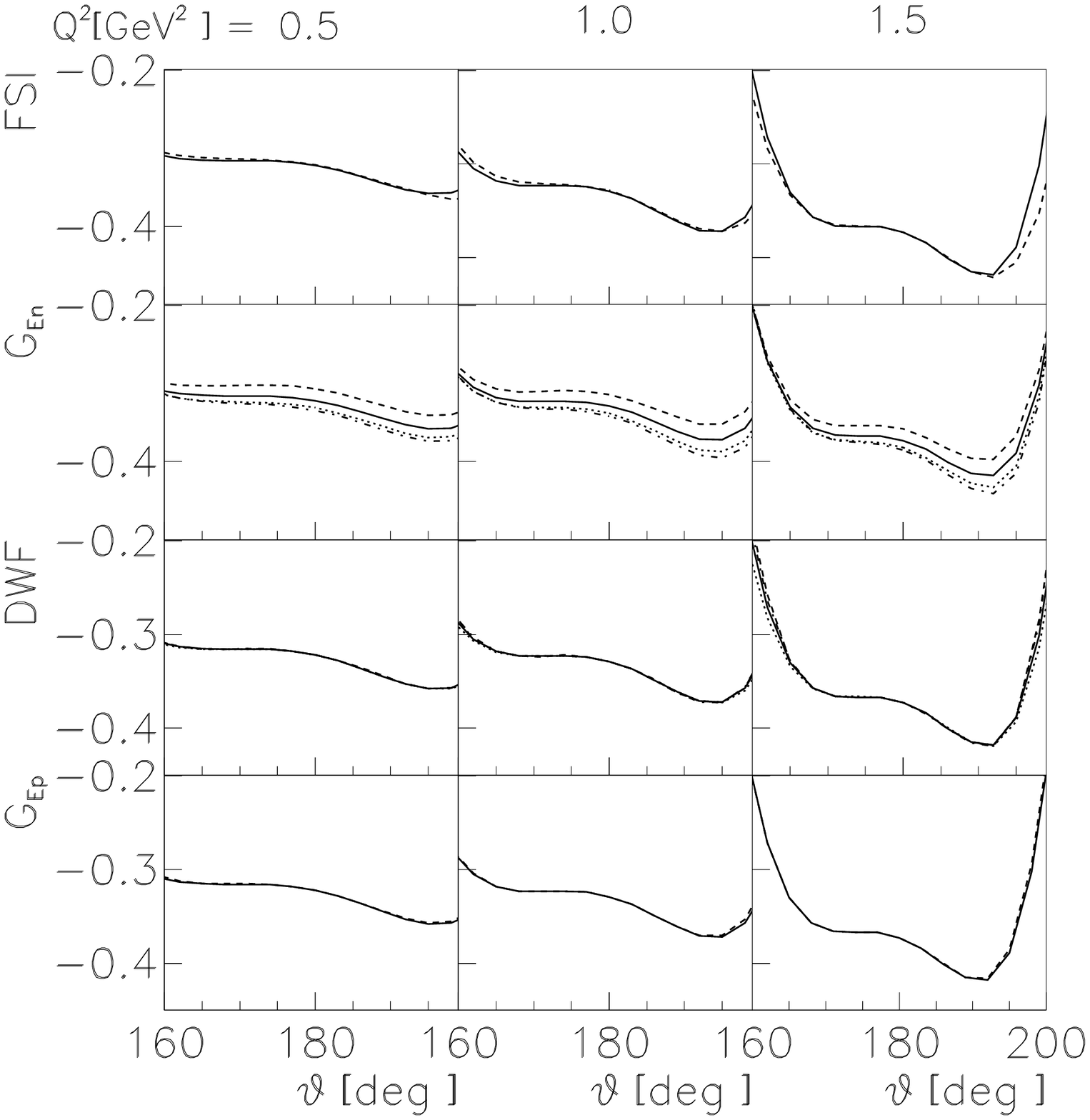}}
\caption{ $\vartheta $-dependence of the $A_z$-asymmetry in the quasi-elastic 
kinematics region. The same notations as in Fig. \protect\ref{fig:asymx}.}
\label{fig:qeasymz} 
\end{figure}

In Figs. \ref{fig:qeasymx} and \ref{fig:qeasymz} we restrict the angular 
domain to the quasi-elastic region. The main sensitivity to the form factor 
\gen appears in this angular range, as shown by  the calculations corresponding 
to the Galster parametrization scaled by 0.5 (dashed line) and 1.5 (dotted line). 
The calculation for   \gen =0 is represented as dash-dotted line.

Therefore, the reaction $\vec d(\vec e, e'p)n$, in the quasi-elastic regime, 
is a good source of information on the form factor \gen.  A better 
determination of the form factor \gen might be through the ratio $A_x/A_z$, 
which does not depend on the electron helicity. 

A similar procedure, firstly suggested in 
Ref. \cite{Re68}, has been recently realized for the processes 
$\vec e+p\to e+\vec p$ \cite{Jo00,Ga02}, $d(\vec e,e'\vec n p)$  
\cite{He99,Os99}, where the ratio of the $x$- and $z$-components of the 
nucleon polarization has been measured, and for the $\vec{^3{He}}(\vec e,e'
n)pp$ process as well (for the ratio of the $x$--  and $z$--  components of 
the $\vec{^3{He}}$ polarization \cite{Me94}).

The  above mentioned ratios (in the impulse approximation) for polarized $d$ 
and $^3He$ targets, is essentially determined by the ratio of the electric and 
magnetic form factors $G_{En}/G_{Mn}$. From the experimental point of view this  
measurement is very convenient as many systematic errors essentially cancel 
and the analysis is simplified.

It is evident that in the case of polarized nuclear targets, such as $\vec d$ 
or $\vec{^3He}$, the ratios of target asymmetries (or neutron polarizations) 
contain various nuclear effects, as well as other corrections (FSI, etc).

In the case of a deuteron target the situation is more complicated due, on one 
side, to the Fermi motion of the bound nucleons, and from another side, to the 
Wigner rotation of the nucleonic spin. The final $np-$interaction plays also a 
role. All these aspects of the deuteron physics should be carefully taken into 
account for the extraction of the form factor $G_{En}$  from the respective 
polarization observables in the $e^-+d\to e^-+n+p$ reaction.

The ratio of the asymmetries can be written as follows \cite{ERGR}:
\begin{equation}
R_{xz}=\displaystyle\frac{A_x}{A_z}=\displaystyle\frac{ \sqrt{1+\epsilon}A_x^{(0)}+\sqrt{2\epsilon}A_x^{(1)}\cos\phi}
{ \sqrt{1+\epsilon}A_z^{(0)}+\sqrt{2\epsilon}A_z^{(1)}\cos\phi}, 
\label{eq:eqR}
\end{equation}
where $A_x^{(0)}=2\beta_{13}$, $A_z^{(0)}=2\beta_{9}$, 
$A_x^{(1)}=-2(\sqrt{Q^2}/{k_0})\beta_{12}$, $A_z^{(1)}=-2(\sqrt{Q^2}/{k_0})\beta_{8}$.
The specific dependence of the ratio $R_{xz}$ on $\epsilon$ and $\phi$ is a 
model independent result, which is based on the following properties of the 
hadron electromagnetic interaction:
\begin{itemize}
\item the validity of the one-photon-exchange mechanism for $e^-+d\to e^-+n+p$;
\item the conservation of the electromagnetic current for $\gamma^*+d\to n+p$ (the gauge invariance);
\item the P-invariance of hadron electrodynamics;
\item the validity of QED for the description of the $\gamma e e$-vertex.
\end{itemize}

Note that $A_x^{(0)}$ and $A_z^{(0)}$ are determined by quadratic combinations of the transversal components of the electromagnetic current for 
$\gamma^*+d\to n+p$ reaction, whereas $A_x^{(1)}$ and $A_z^{(1)}$ are 
driven by the interference of the longitudinal and the transversal components 
of this current. Moreover,  $A_x^{(0)}$ and $A_z^{(1)}$ vanish at 
$\vartheta =0^{\circ}$ and $180^{\circ}$, due to the helicity conservation for 
collinear kinematics in  $\gamma^*+d\to n+p$ reaction. These statements are also 
model independent. 

However, considering the one-nucleon contributions of the electromagnetic 
current for the $\gamma^*+d\to n+p$ reaction, one can expect that its 
longitudinal component is determined by the $G_{En}$ form factor and the 
transversal one by the $G_{Mn}$ form factor, at least in the Breit system. 
Therefore, the contributions $A_{x,z}^{(1)}$, in Eq. (\ref{eq:eqR}), should 
contain terms proportional to the product $G_{En}G_{Mn}$.

The presence of two contributions in both asymmetries,  $A_x$ and $A_z$, 
results from the specific character of the deuteron dynamics for 
$\gamma^*+d\to n+p$ reaction and their relative role is determined by the 
corresponding model.

\begin{figure}
\mbox{\epsfxsize=14.cm\leavevmode \epsffile{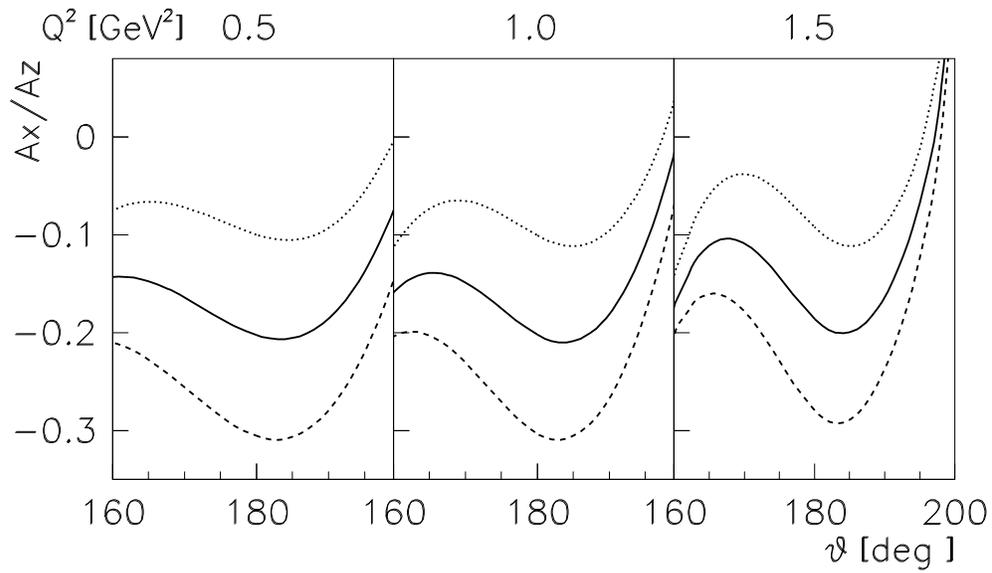}}
\caption{$\vartheta $-dependence of the ratio $R_{xz}=A_x/A_z$ for different 
parametrizations of the neutron electric form factor. The same notations as 
in Fig. \ref{fig:asymx}.}
\label{fig:str3}
\end{figure}

In Fig. \ref{fig:str3} the results for the ratio $R_{xz}$ are shown at 
$Q^2$ = 0.5, 1, and  1.5 GeV$^2$, for different parametrizations of the \gen 
form factor, with the previous notations. The sensitivity of this ratio is 
quite large in the considered $Q^2$ and $\vartheta $ range.  Again, at larger 
$Q^2$ the effects of FSI and DWF choice are negligible in the quasi-elastic 
kinematics, making simpler and less model dependent the extraction of the 
$G_{En}$ form factor from  the experimental data on the ratio $R_{xz}$ . 

\begin{figure}
\mbox{\epsfxsize=14.cm\leavevmode \epsffile{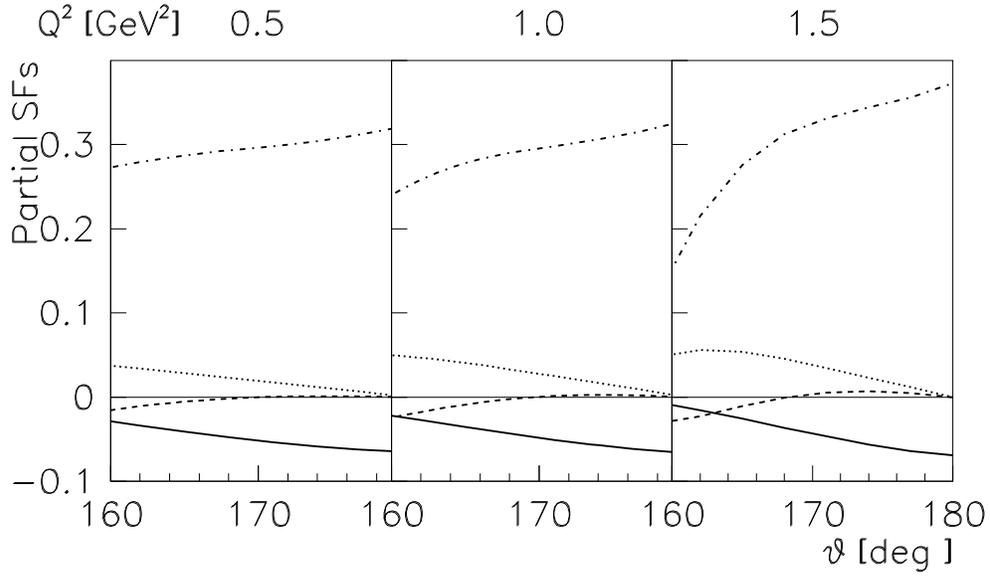}}
\caption{$\vartheta $-dependence of the partial structure functions $A_x^{(0)}$ 
 (dashed line), $A_x^{(1)}$ (solid thick line), $A_z^{(0)}$ (dash-dotted line), and 
 $A_z^{(1)}$ (dotted line) for $Q^2$= 0.5, 1, and 1.5 GeV$^2$.}
\label{fig:str4}
\end{figure}

The relative role of the two possible contributions to the asymmetries $A_x$ 
and $A_z$ is shown in Fig. \ref{fig:str4}. For the region where 
$\vartheta \ne 180^\circ$ these contributions are comparable for the asymmetry 
$A_x$, being negative in the region where $\vartheta  < 180^\circ$. For the 
asymmetry $A_z$, the transversal contribution $A_z^{(0)}$ , related to 
$G_{Mn}^2$, is essentially larger in comparison with the asymmetry $A_z^{(1)}$. 

This  method for the determination of the $G_{En}$ form factor, at relatively 
large $Q^2$, from the ratio $R_{xz}=A_x/A_z$ of the T-even asymmetries in 
$\vec d(\vec e,e'p)n$, measured in the kinematical conditions of quasi-elastic 
$en$- scattering, seems promising and may be comparable in accuracy with the 
measurements of the $G_{Ep}$ form factor through the recoil polarization method.

\section{Proton polarization}

The polarization 
properties of the proton, produced in the $d(e, e'{\vec p})n$ and 
$d({\vec e}, e'{\vec p})n$ reactions, are 
determined by the ${\vec P}_{ij} $ tensor (see Eqs. (\ref{eq:eq7}) 
and (\ref{eq:eq8})):
\begin{equation}
{\vec P}_{ij}=Tr~F_iF_j^+{\vec \sigma},~F=l_iF_i, \ \ i,j = x, y, z.
\label{eq:eq15}
\end{equation}
The proton polarization vector ${\vec P}$ (multiplied by the 
unpolarized differential cross section $d^3\sigma /dE'd\Omega _ed\Omega _p$) is expressed by a formula similar to Eq. (\ref{eq:eq13}), replacing the components of the hadron tensor $H_{ij}$  by the corresponding ${\vec P}_{ij}$ tensor components. The tensor 
${\vec P}_{ij}$ can be represented in the following general form:
$$
{\vec P}_{ij}=\hat {\vec k}P_{ij}^{(k)}+{\vec m}P_{ij}^{(m)}+{\vec n}P_{ij}^{(n)}. 
$$
Assuming the P-invariance of the hadron electromagnetic 
interaction, we can explicitate the tensor structure of the quantities 
$P_{ij}^{(a)},$ $a=k,~m,~n$, in terms of SFs $P_i$, $i=1-13$,  
which depend on three independent kinematical variables: $k^2$, $W$, and $t$. 
\begin{eqnarray}
P_{ij}^{(k)}&=&P_1\{\hat k,n\}_{ij}+P_2\{m,n\}_{ij}+iP_3[\hat 
k,n]_{ij}+iP_4[m,n]_{ij},\nonumber\\
P_{ij}^{(m)}&=&P_5\{\hat k,n\}_{ij}+P_6\{m,n\}_{ij}+iP_7[\hat 
k,n]_{ij}+iP_8[m,n]_{ij},\nonumber\\
P_{ij}^{(n)}&=&P_9\hat k_i\hat 
k_j+P_{10}m_im_j+P_{11}n_in_j+P_{12}\{\hat k,m\}_{ij}+iP_{13}[\hat k,m]_{ij}. 
\label{eq:eq16}
\end{eqnarray}

The expressions for 
SFs $P_i$, in terms of the scalar amplitudes, are given in Appendix 5.  We can 
see that the symmetric parts of the tensors in Eq. (\ref{eq:eq16}) (which 
correspond to  eight SFs $P_i, i=1,2,5,6,9,10,11,12$) determine the components 
of the polarization vector of the protons produced in collisions of unpolarized 
electrons with an unpolarized target, for the reaction $d(e,e'{\vec p})n$. The 
antisymmetric parts of the tensors in Eq. (\ref{eq:eq16}) (that is, five SFs 
$P_i,~ i=3,4,7,8,13$) determine the components of the polarization 
vector of the protons produced in collisions of longitudinally polarized 
electrons with an unpolarized target, $d({\vec e},e'{\vec p})n.$ 

It can also be shown that eight SFs $P_1$ , $ P_2$ , $ P_5$ , $ P_6$ , $ P_9-P_{12}$ 
(in the symmetric parts of the corresponding tensors) determine the T-odd 
contributions to the proton polarization $\mathbf{ P}$ (for the scattering of 
unpolarized electrons), whereas  the five SFs $P_3$ , $ P_4$ , $ P_7$, $ P_8$, $ P_{13}$ 
(in the antisymmetric parts of the corresponding tensors) determine the T-even  
contributions to the proton polarization $\mathbf{ P}$ (for  the scattering of 
longitudinally polarized electrons). 

These five T-even SFs are nonzero even when the $\gamma ^* +d\rightarrow n+p$ amplitudes 
are real functions, which is true in framework of IA. In the scattering of the longitudinally polarized electrons, they determine the proton polarization induced by the absorption of 
circularly polarized virtual photons (by unpolarized deuterons) in the 
$\gamma ^* +d\rightarrow n+p$ reaction: the polarization is 
transferred from the electron to the produced proton by the virtual photon. 
The eight T-odd SFs, defined above, are nonzero only for complex 
$\gamma ^* +d\rightarrow n+p$ amplitudes (with different relative phases). 

Due to the tensor structure of the quantities $\mathbf{P}_{ij}$, in the 
scattering of unpolarized electrons by unpolarized deuterons, the polarization 
component of the protons which is orthogonal to the
$\gamma ^* +d\rightarrow n+p$ reaction plane is characterized by the same 
$\varepsilon$ and $\phi $ dependences as in the unpolarized case. The 
polarization vector of the protons polarized in the 
$\gamma ^* +d\rightarrow n+p$ reaction plane (components $P_x$ and $P_z$) is 
characterized by two dependences: $\varepsilon \sin(2\phi ) $ and 
$\sqrt{2\varepsilon(1+\varepsilon )}\sin\phi$.  

To prove these statements, 
we explicitly single out the dependence of the proton polarization on the 
kinematic variables $\phi $ and $\varepsilon$. In the general case, the vector 
of the proton polarization can be represented as the sum of two terms: 
$\mathbf{P}^{(0)}$ and $\mathbf{P}^{(\lambda)}$, where the polarization 
$\mathbf{P}^{(0)}$ corresponds to the unpolarized electron beam (induced 
polarization) and the polarization $\mathbf{P}^{(\lambda)}$ 
corresponds to the longitudinally polarized electron beam (polarization 
transfer). So, the components of the proton polarization vector $\mathbf{P}$ in 
the reactions $d(e,e'{\vec p})n$, $d({\vec e},e'{\vec p})n$ are given 
by
\begin{eqnarray}
\mathbf{P}&=&\mathbf{P}^{(0)}+\lambda \mathbf{P}^{(\lambda )}, \nonumber\\
P_x^{(0)}\sigma _0&=&{\cal N}\sin\phi [\sqrt{2\varepsilon (1+\varepsilon 
)}P_x^{(LT)}+\varepsilon \cos\phi P_x^{(TT)}],  \nonumber\\
P_z^{(0)}\sigma _0&=&{\cal N}\sin\phi 
[\sqrt{2\varepsilon (1+\varepsilon )}P_z^{(LT)}+\varepsilon \cos\phi P_z^{(TT)}], \nonumber\\
P_y^{(0)}\sigma _0&=&{\cal N}[P_y^{(TT)}+\varepsilon P_y^{(LL)}+\sqrt{2\varepsilon 
(1+\varepsilon )}\cos\phi P_y^{(LT)}+\varepsilon \cos(2\phi )\bar P_y^{(TT)}],
\nonumber\\
P_x^{(\lambda )}\sigma _0&=&{\cal N}[\sqrt{1-\varepsilon ^2}R_x^{(TT)}+\sqrt{2\varepsilon (1-\varepsilon )} 
\cos\phi R_x^{(LT)}], \nonumber\\
P_z^{(\lambda )}\sigma 
_0&=&{\cal N}[\sqrt{1-\varepsilon ^2}R_z^{(TT)}+\sqrt{2\varepsilon (1-\varepsilon )} 
\cos\phi R_z^{(LT)}], 
\nonumber\\
P_y^{(\lambda )}\sigma _0&=&{\cal N}\sqrt{2\varepsilon 
(1-\varepsilon )} \sin\phi R_y^{(LT)},\nonumber
\label{eq:eq412}
\end{eqnarray}
where  the individual contributions to 
the polarization vector in terms of SFs $P_i$ are :
\begin{eqnarray*}
P_x^{(TT)}&=&4P_6, \ P_y^{(TT)}=P_{10}+P_{11}, ~\bar P_y^{(TT)}=P_{10}-P_{11},~ P_z^{(TT)}=4P_2, \nonumber\\
P_x^{(LT)}&=&-2\frac{\sqrt{Q^2}}{k_0}P_5,~
P_y^{(LT)}=-2\frac{\sqrt{Q^2}}{k_0}P_{12},~ 
P_z^{(LT)}=-2\frac{\sqrt{Q^2}}{k_0}P_1, \nonumber\\
P_y^{(LL)}&=&2\frac{Q^2}{k^2_0}P_9, ~
R_x^{(TT)}=2P_8, \ R_z^{(TT)}=2P_4,~
R_x^{(LT)}=-2\frac{\sqrt{Q^2}}{k_0}P_7, \nonumber\\
R_y^{(LT)}&=&2\frac{\sqrt{Q^2}}{k_0}P_{13},~
R_z^{(LT)}=-2\frac{\sqrt{Q^2}}{k_0}P_3.
\end{eqnarray*}
The expressions for SFs $P_i$ in terms of the reaction amplitudes are general 
and do not depend on the details of the reaction mechanism. As explicitly 
shown in the Appendix 5, each of the thirteen SFs $P_i(W, k^2,t), i=1-13,$ 
carries independent information about the scalar amplitudes. Therefore, 
measurement of all these SFs is in principle necessary to perform the complete 
$\gamma ^* +d\rightarrow n+p$ experiment. 

The polarization transfer to the proton in the reactions 
$H({\vec e}, e'{\vec p})$ and $d({\vec e}, e'{\vec p})n$, with a beam 
of longitudinally polarized electrons, was measured in three experiments in 
quasi-free kinematics \cite{Ey95,Mi98,Ba99}. These experiments have been done 
in conditions of in--plane kinematics, where the electron scattering plane 
coincides with the reaction plane. Moreover, the induced proton polarization 
was also measured at MIT. The kinematical conditions of these experiments are 
presented in Table \ref{tab:tab2}.
\vspace{0.5cm}
\begin{table}[ht]
\begin{center}
\begin{tabular}{|c||c||c||c||c|}
\hline \hline
$Q^2$ $[GeV^2]$  & $ E$  [GeV] &  $ E'$  [GeV] & $\vartheta_e$ [deg] & Ref. \\
\hline \hline
0.31 & 0.855 & 0.700 & 43 &\cite{Ey95} \\
\hline
0.38 & 0.580 & 0.395 & 82.7 & \cite{Mi98}\\
\hline
 0.50 & 0.580 & 0.395 & 113 &\cite{Mi98}\\
\hline
\it 0.38 & 0.579 & 0.379 & 82.7 & \cite{Ba99} \\
\hline\hline
\end{tabular}
\caption{Kinematics of the $d(e,e'{\vec p})n$, $d({\vec e},e'{\vec p})n$  experiments} 
\label{tab:tab2}
\end{center}
\end{table}
\vspace{0.5cm}

Previous calculations  \cite{Gi02} showed that some polarization
observables, in the quasi-free kinematic
region, are insensitive to specific nuclear mechanisms such as FSI,
meson exchange currents, and isobar configurations. Therefore, this 
kinematical region is very well adapted to a reliable determination of  
the form factor \gen. Since the same experimental setup can be used for the
proton and deuteron targets, it is possible to compare the measurements on 
free and bound proton in two reactions: $d(e,e'{\vec p})n$,  
$p({\vec e},e'{\vec p})$ and $d({\vec e},e'{\vec p})n$. 

On the contrary, the induced polarization (which is helicity independent, 
unlike other polarization components) is very sensitive to the reaction
mechanism. For elastic electron-proton scattering the induced polarization
vanishes in the one-photon-exchange approximation. A value different from 
zero would sign the presence of an interference contribution with the 
two-photon-exchange mechanism.

The main findings from these experiments can be summarized as follows. 

The data indicate 
that the description of quasi-elastic deuteron electrodisintegration reaction 
in terms of purely elastic electron-nucleon amplitudes is consistent with the spin 
transfer observables as well as with the cross section. 

On the contrary, the 
predictions of the induced polarization are not consistent with the data. 

The experiments 
have demonstrated that the proton transfer polarization can be precisely 
determined at intermediate energies and that this method is very powerful 
for determining \gen and \gep form factors at large $Q^2$ values. 

According to the Madison convention, the reference system, where the 
proton-polarization components are defined, is related to the outgoing 
particle. The $z^{'}$ axis is now directed along the three-momentum of the 
emitted proton, while the $y^{'}$ axis is normal (as before) to the 
$\gamma ^* +d\rightarrow n+p$ reaction plane. Let us define the proton 
polarization components in this coordinate frame: $P_{\ell}$ (longitudinal), 
$P_t$ (transverse), and $P_n$ (normal).  These components are then related 
to the components $P_x, \ P_y,$ and $P_z $ by:
\begin{equation}
P_n=P_y,~P_{\ell}=\cos\vartheta P_z+\sin\vartheta P_x,~
P_t=\cos\vartheta P_x-\sin\vartheta P_z.
\label{eq:eqp16}
\end{equation}

\begin{figure}
\mbox{\epsfxsize=14.cm\leavevmode \epsffile{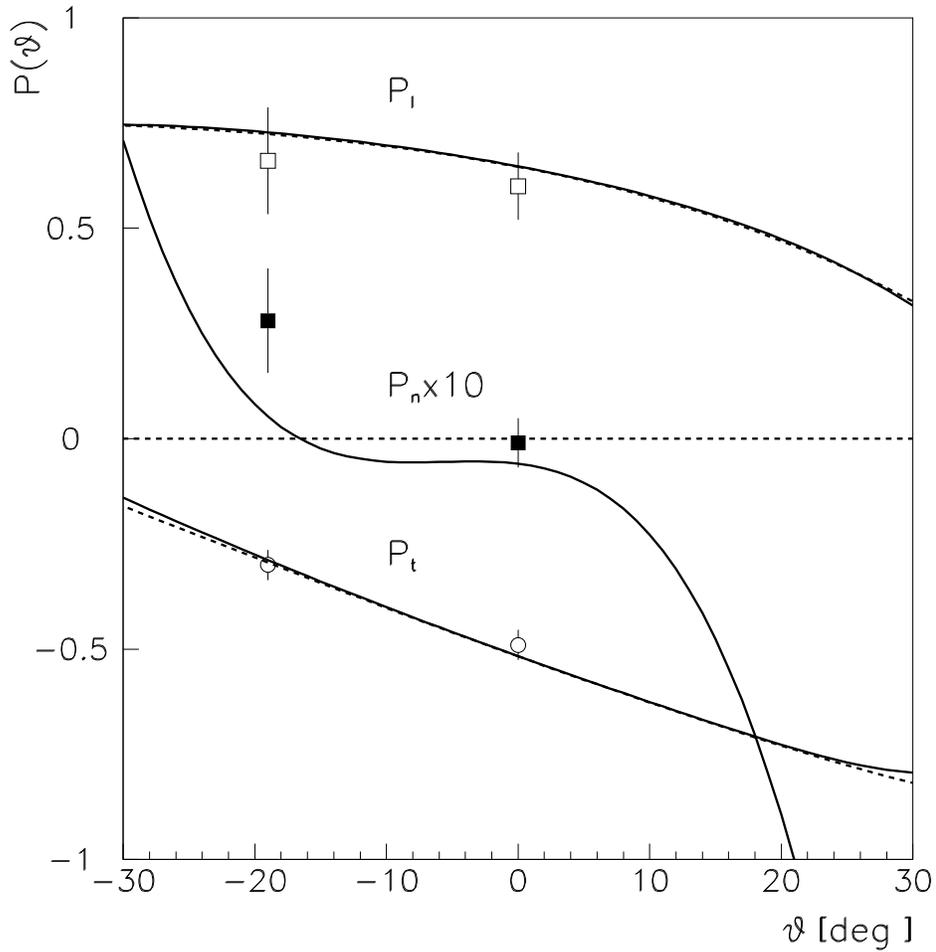}}
\caption{$\vartheta$-dependence of the recoil proton polarization components 
$P_{\ell}$ (open squares), $P_n$ (solid squares) and $P_t$ (open circles) in the  quasi--elastic region of the $d({\vec e},e'{\vec p})n$
reaction from Ref. \protect\cite{Ba99}. The corresponding predictions, for the Paris DWF are shown (solid lines). The dashed lines correspond to the calculations without FSI
effects.The results for the normal component have been magnified by a factor ten.
}
\label{fig:ppol0}
\end{figure}

The predictions of our model for the proton polarization components are
presented in Fig. \ref{fig:ppol0} for the kinematical conditions of the 
experiment at MIT \cite{Ba99}.
Protons were detected at two angles corresponding to neutron recoil momenta 
$p_m$ of 0 and 100 MeV/c. The calculations of the polarization transfer
components $P_l$ and $P_t$ describe the data well, but calculations of the
induced polarization $P_n$ at $p_m=100$ MeV/c underestimated the experimental
result. We can see that contribution of FSI (for the $P_l$ and $P_t$ components)
is noticeable only at $p_m\geq$ 100 MeV/c (which corresponds to $\vartheta \geq 20^0$). 
Our calculations show that the longitudinal and transverse components are
insensitive to the choice of the $NN$-potential.

The data in Ref. \cite{Ba99} have been compared with a  nonrelativistic calculation which 
includes FSI, meson-exchange currents, and isobar configurations, as well as leading order relativistic corrections from H. Arenhovel. The prediction from the present model is very close to the one reported in the paper \cite{Ba99}.

\begin{figure}
\mbox{\epsfxsize=14.cm\leavevmode \epsffile{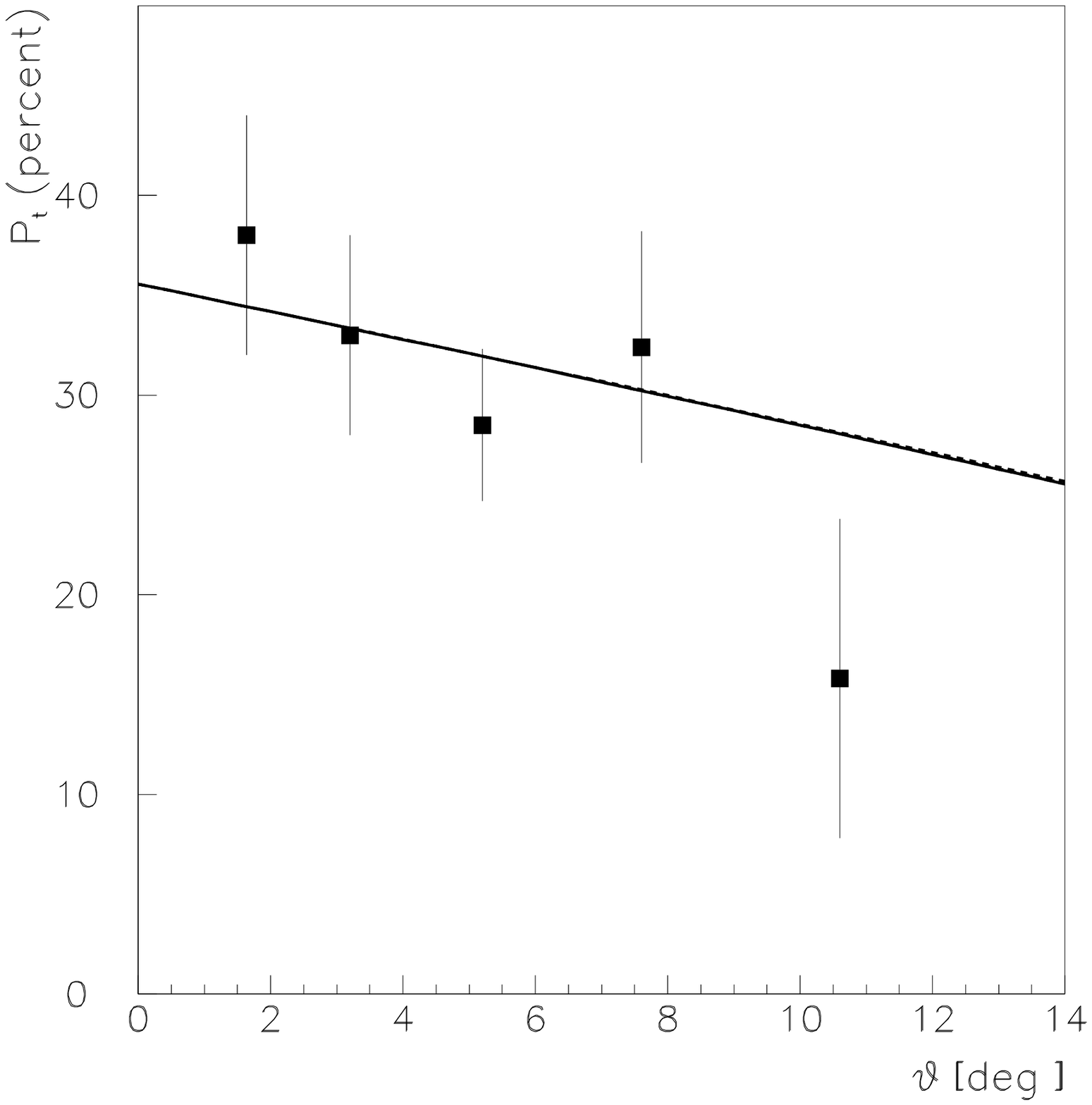}}
\caption{$\vartheta$-dependence of the proton polarization $P_t$ calculated for 
various DWFs. Predictions for the Paris 
wave function (solid line), the charge-dependent Bonn wave function (dashed 
line) are almost overlapping. Data and kinematical conditions are  from Ref. \protect\cite{Ey95}}
\label{fig:ppolx}
\end{figure}

The calculated $P_t$-component ($P_x$ according to the notation of the paper
\cite{Ey95}), for the kinematical conditions of the experiment \cite{Ey95}, is 
shown in Fig. \ref{fig:ppolx}. In this experiment the recoil-neutron momentum $p_m$  $\leq$ 
100 MeV/c. Therefore,  the polarization component is insensitive to the choice 
of the DWF model. The influence of the FSI effects on this observables is also small.

We calculated the discussed polarization observables for the kinematical 
conditions of the experiment \cite{Mi98}. The recoil momentum of the neutron was
restricted in this experiment to the range $0-60$ MeV/c. So, we averaged the
predicted values over this momentum range. Due to the low value of the missing
momentum the predicted observables are insensitive to the choice of DWF model and to
the inclusion of FSI contribution. The predicted polarization components are
presented in the Table \ref{tab:tab4}. We see that the data are consistent with our
theoretical calculations. 

A discussion concerning  the experimental evidence of the 
validity of IA in the quasi-elastic region is given in Ref. \cite{Kl98}. Disentangling medium
modification of nucleon properties from additional reaction mechanisms remains a central problem in intermediate energy nuclear physics \cite{Kl98}, for which 
polarization measurements can bring unique information. 

\vspace{0.5cm}
\begin{table}[h]
\begin{tabular}{|c||c||c||c||c|}
\hline\hline
 $Q^2$ [GeV$^2$]& $P_l$ &$P_t$ & $P_n$  &\\
\hline\hline
 0.38 & 0.60$\pm$ 0.07 & -0.48$\pm$ 0.02 & 0.001$\pm$ 0.005 & Experiment \\
\hline
 0.38 & 0.63 & -0.51 & -0.008 &Theory \\
\hline\hline
0.50& 0.77$\pm$ 0.04 & -0.41$\pm$ 0.02 & 0.005$\pm$ 0.005 & Experiment\\
\hline
0.50 & 0.85 & -0.43 & -0.002& Theory \\
\hline
\end{tabular}
\vspace{0.5cm}
\caption{The recoil proton polarization observables from the experiment 
\protect \cite{Mi98} and from the present calculation averaged over the neutron
missing-momentum range $p_m$=0-60 MeV/c. The calculation corresponds to the Paris DWF.}
\label{tab:tab4}
\end{table}

\section{Neutron polarization in case of unpolarized deuteron target}

The polarization of the neutron, emitted in the reaction $\vec e +d\to e +\vec n + p$, is determined by the ${\vec N}_{ij} $ tensor:
\begin{equation}
{\vec N}_{ij}=-Tr F_i{\vec \sigma}F_j^+, \ \ i,j = x,y,z.  
\label{eq:eqn13}
\end{equation}
As in the case of the proton polarization, the tensor ${\vec N}_{ij}$ can be represented in the 
following general form:
$${\vec N}_{ij}=\hat {\vec k}N_{ij}^{(k)}+{\vec m}N_{ij}^{(m)}+
{\vec n}N_{ij}^{(n)}$$
and  we can express the tensor structure 
of the contributions $N_{ij}^{(k)},$ $ N_{ij}^{(m)},$ and $ N_{ij}^{(n)}$ in 
terms of SFs  $N_i$, (which are determined by Eq. (\ref{eq:eq16}) after  
substituting $P_i\rightarrow N_i$).  The neutron polarization is also driven 
by thirteen real SFs $N_i$ depending on three kinematical variables: $k^2$, 
$W$, and $t.$ The formulae for SFs $N_i$ in terms of the scalar amplitudes are given in Appendix 6.    

\begin{figure}
\mbox{\epsfxsize=14.cm\leavevmode \epsffile{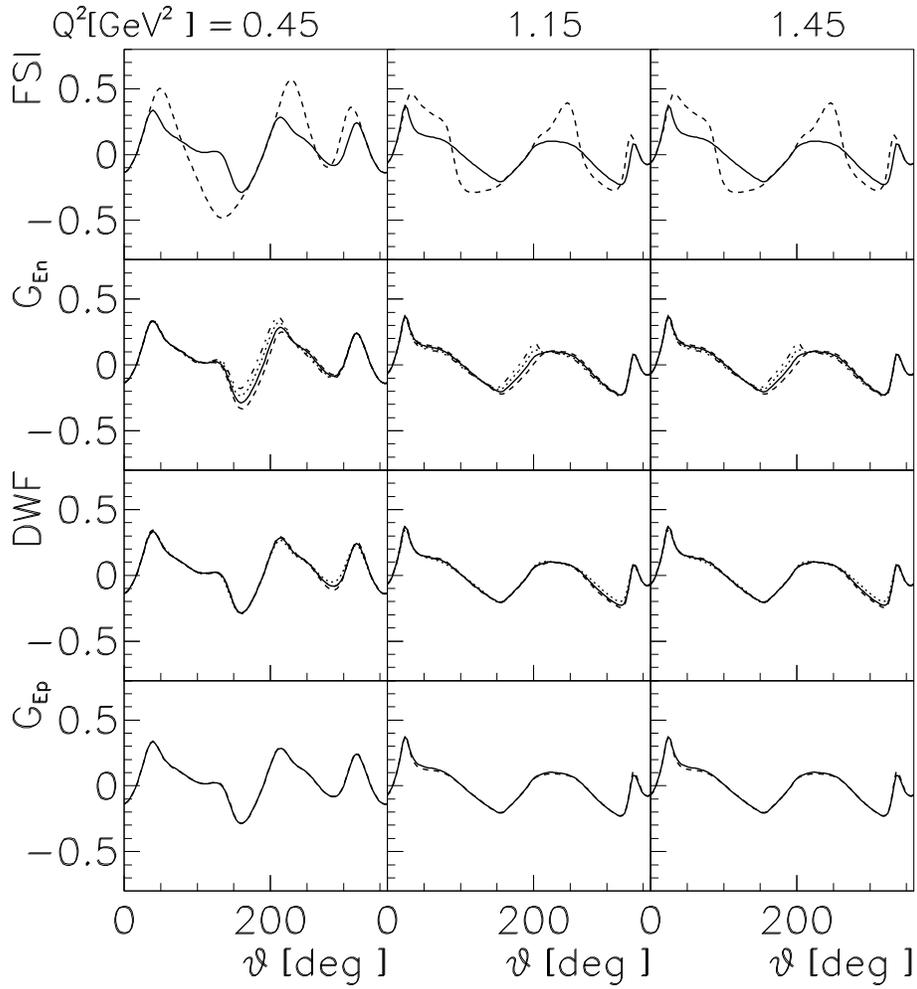}}
\caption{$\vartheta$-dependence of the neutron transversal polarization in 
$d(\vec e,e'\vec n)p$ at the kinematics in Tab. \protect\ref{tab:Madey}. The angle $\vartheta=180^\circ$ corresponds to 
quasi-elastic kinematics, i.e. the neutron and the virtual photon move in 
the same direction. Notations as in Fig. \protect\ref{fig:asymx}.}
\label{fig:npolt}
\end{figure}

\begin{figure}
\mbox{\epsfxsize=14.cm\leavevmode \epsffile{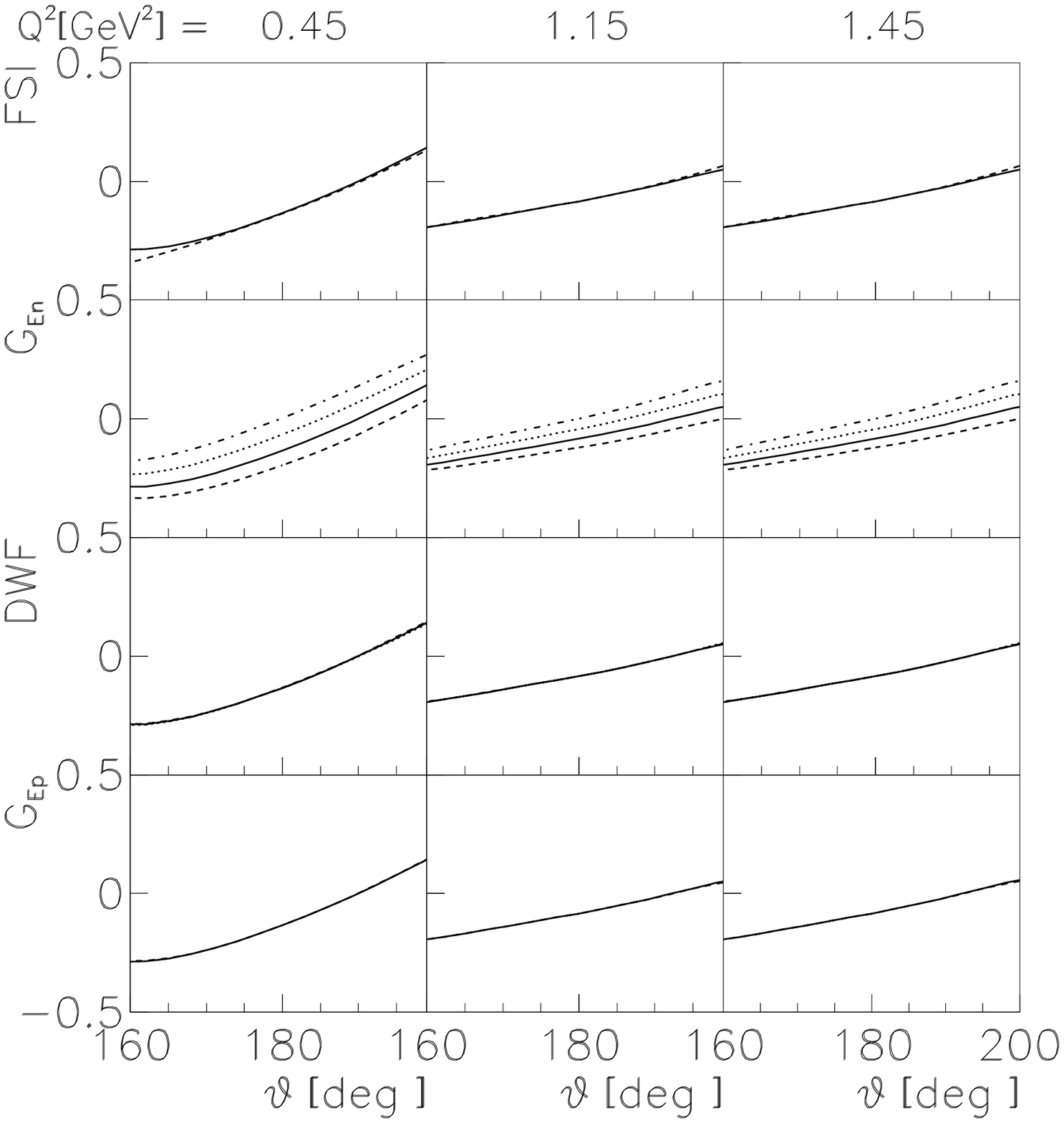}}
\caption{$\vartheta$-dependence of the neutron transversal polarization, 
in $d(\vec e,e'\vec n)p$, for quasi-elastic kinematics, corresponding to Ref. \protect\cite{Madey}. Notations as in Fig. \protect\ref{fig:asymx}.}
\label{fig:qenpolt}
\end{figure}

The measurement of the \gen form factor by means of the recoil--neutron 
polarization method in quasi-elastic $d(\vec e,e'\vec n)p$ has been recently performed at JLab  \cite{Madey} for $Q^2$ up to $\simeq$ 1.5 GeV$^2$. The kinematics of the experiment is given in 
Table \ref{tab:Madey} and the calculations are presented in 
Figs. \ref{fig:npolt} and \ref{fig:qenpolt}. 
\vspace{0.5cm}
\begin{table}[h]
\begin{center}
\begin{tabular}{|c||c||c||c||c|}
\hline
$Q^2$ [GeV$^2$] & $E$ [GeV] & $E'$ [GeV] & $\theta_e$ [deg] & 
$\varepsilon$ \\
\hline
\it 0.45 & 0.884 & 0.646 & 52.7 & 0.644\\
\hline
\it 1.15 & 2.33 & 1.750 & 30.8 & 0.836\\
\hline
\it 1.47 & 3.40 & 2.584  & 23.6 & 0.887\\
\hline
\end{tabular}
\caption{ Kinematical parameters for the JLab experiment \protect\cite{Madey}.}
\label{tab:Madey}
\end{center}
\end{table}

\section{Neutron polarization in case of  vector-polarized deuteron target}

The polarization state of the neutron, produced in the 
${\vec d}(e, e'{\vec n})p$ reaction (deuteron in this case is 
vector-polarized), is determined by the ${\vec N}_{ij}(\vec \xi ) $ tensor:
\begin{equation}\label{eq:eqn15}
{\vec N}_{ij}({\vec \xi })=-TrF_i{\vec \sigma}F_j^+, \ \ i,j = x,y,z.  
\end{equation}
The tensor ${\vec N}_{ij}(\vec \xi )$ can be represented 
in the following general form, as it was previously done for the case of the neutron 
polarization:
$${\vec N}_{ij}(\vec \xi )={\vec\xi }{\vec m}{\vec N}_{ij}^{(m)}+{\vec\xi }
{\vec n}{\vec N}_{ij}^{(n)}+
{\vec\xi }{\vec {\hat k}}{\vec N}_{ij}^{(k)}. $$
The tensors ${\vec N}_{ij}^{(l)},$ $l=m, n, k$ can be expanded over a complete set of orthonormal vectors. For these tensors we have following
general structure:
\begin{eqnarray*}
{\vec N}_{ij}^{(m)}&=&N_{ij}^{(mk)}{\vec {\hat k}}+
N_{ij}^{(mm)}{\vec m}+
N_{ij}^{(mn)}{\vec n}, \\
{\vec N}_{ij}^{(k)}&=&N_{ij}^{(kk)}{\vec{\hat k}}+
N_{ij}^{(km)}{\vec m}+
N_{ij}^{(kn)}{\vec n}, \\
{\vec N}_{ij}^{(n)}&=&N_{ij}^{(nk)}{\vec {\hat k}}+
N_{ij}^{(nm)}{\vec m}+
N_{ij}^{(nn)}{\vec n}. 
\end{eqnarray*}
The tensors $N_{ij}^{(lr)},$  $l,r=m, n, k$ can be decomposed, in turn, in the
way that was used in the analysis of previous observables. In terms of SFs, we may write these tensors as follows
\begin{eqnarray}
&N_{ij}^{(mk)}=&D_1\hat k_i\hat k_j+D_2m_im_j
+ D_3n_in_j+D_4\{\hat k,m\}_{ij}+iD_5[\hat k,m]_{ij},  \nonumber \\
&N_{ij}^{(mm)}=&D_6\hat k_i\hat k_j+D_7m_im_j
+ D_8n_in_j+D_9\{\hat k,m\}_{ij}+iD_{10}[\hat k,m]_{ij}, \nonumber \\
&N_{ij}^{(mn)}=&D_{11}\{\hat k,n\}_{ij}+D_{12}\{m,n\}_{ij}+
iD_{13}[\hat k,n]_{ij}+iD_{14}[m,n]_{ij}, \nonumber \\
&N_{ij}^{(kk)}=&D_{15}\hat k_i\hat k_j+D_{16}m_im_j
+ D_{17}n_in_j+D_{18}\{\hat k,m\}_{ij}+iD_{19}[\hat k,m]_{ij},\nonumber \\
&N_{ij}^{(km)}=&D_{20}\hat k_i\hat k_j+D_{21}m_im_j
+ D_{22}n_in_j+D_{23}\{\hat k,m\}_{ij}+iD_{24}[\hat k,m]_{ij}, \nonumber \\
&N_{ij}^{(kn)}=&D_{25}\{\hat k,n\}_{ij}+D_{26}\{m,n\}_{ij}
+iD_{27}[\hat k,n]_{ij}+iD_{28}[m,n]_{ij}, \label{eq:eqn16} \\
&N_{ij}^{(nk)}=&D_{29}\{\hat k,n\}_{ij}+D_{30}\{m,n\}_{ij}
+iD_{31}[\hat k,n]_{ij}+iD_{32}[m,n]_{ij}, \nonumber \\
&N_{ij}^{(nm)}=&D_{33}\{\hat k,n\}_{ij}+D_{34}\{m,n\}_{ij}
+iD_{35}[\hat k,n]_{ij}+iD_{36}[m,n]_{ij}, \nonumber\\
&N_{ij}^{(nn)}=&D_{37}\hat k_i\hat k_j+D_{38}m_im_j
+ D_{39}n_in_j+D_{40}\{\hat k,m\}_{ij}+iD_{41}[\hat k,m]_{ij}.\nonumber 
\end{eqnarray}
Thus, the dependence of the neutron polarization on the vector 
polarization of the deuteron is determined by forty-one SFs $D_i $, $i=1-41$. 
The formulae for SFs $D_i$ in terms of the reaction scalar amplitudes are 
given in Appendix 7.    

The components of the neutron polarization vector, $\vec P(n)$, can be written as 
\begin{equation}
P_i(n)=P_{ij}\xi _j.  
\label{eq:eqn17}
\end{equation}
 
The polarization components have a definite general structure, if we single out 
the explicit dependences on the azimuthal angle $\phi ,$ virtual-photon linear 
polarization $\varepsilon ,$ and contributions of the longitudinal $(L)$ and 
transverse $(T)$ components of the electromagnetic current of the $\gamma 
^*d\rightarrow np$ reaction. As a result, we can obtain:

\noindent\underline{For the $P_x$-component:}
\begin{eqnarray*} 
P_{xx}\sigma _0&=&{\cal N}[P_{xx}^{(TT)}+\varepsilon P_{xx}^{(LL)}+
\sqrt{2\varepsilon (1+\varepsilon )}cos\phi P_{xx}^{(LT)}
+\varepsilon cos(2\phi )\bar P_{xx}^{(TT)}], \\ 
P_{xy}\sigma _0&=&{\cal N}sin\phi [\sqrt{2\varepsilon (1+\varepsilon )}P_{xy}^{(LT)}
+\varepsilon cos\phi P_{xy}^{(TT)}],   \\
P_{xz}\sigma _0&=&{\cal N}[P_{xz}^{(TT)}+\varepsilon P_{xz}^{(LL)}+
\sqrt{2\varepsilon (1+\varepsilon )}cos\phi P_{xz}^{(LT)}
+\varepsilon cos(2\phi )\bar P_{xz}^{(TT)}], 
\end{eqnarray*}
where 
\begin{eqnarray*} 
P_{xx}^{(TT)}&=&D_7+D_8, ~ P_{xx}^{(LL)}=2\displaystyle\frac{Q^2}{k^2_0}D_6,~
\bar P_{xx}^{(TT)}= D_7-D_8, \ \ P_{xx}^{(LT)}=-2\displaystyle\frac{\sqrt{Q^2}}{k_0}D_9, \\
P_{xy}^{(TT)}&=&4D_{34}, ~ P_{xy}^{(LT)}=-2\displaystyle\frac{\sqrt{Q^2}}{k_0}D_{33}, ~
P_{xz}^{(TT)}=D_{21}+D_{22}, ~ \bar P_{xz}^{(TT)}=D_{21}-D_{22},   \\
P_{xz}^{(LL)}&=&2\displaystyle\frac{Q^2}{k^2_0}D_{20}, ~
P_{xz}^{(LT)}=-2\displaystyle\frac{\sqrt{Q^2}}{k_0}D_{23},
\end{eqnarray*}
\noindent\underline{For the $P_y$-component:} 
\begin{eqnarray*} 
P_{yx}\sigma _0&=&{\cal N}sin\phi [\sqrt{2\varepsilon (1+\varepsilon )}P_{yx}^{(LT)}
+\varepsilon cos\phi P_{yx}^{(TT)}], \\
P_{yy}\sigma _0&=&{\cal N}[P_{yy}^{(TT)}+\varepsilon P_{yy}^{(LL)}+
\sqrt{2\varepsilon (1+\varepsilon )}cos\phi P_{yy}^{(LT)}
+\varepsilon cos(2\phi )\bar P_{yy}^{(TT)}], \\
P_{yz}\sigma _0&=&{\cal N}sin\phi [\sqrt{2\varepsilon (1+\varepsilon )}P_{yz}^{(LT)}
+\varepsilon cos\phi P_{yz}^{(TT)}], \\
 \end{eqnarray*}
where 
\begin{eqnarray*} 
P_{yx}^{(TT)}&=&4D_{12},~ P_{yx}^{(LT)}=-2\displaystyle\frac{\sqrt{Q^2}}{k_0}D_{11},~
P_{yy}^{(TT)}=D_{38}+D_{39}, ~ \bar P_{yy}^{(TT)}=D_{38}-D_{39}, \\
P_{yy}^{(LL)}&=&2\displaystyle\frac{Q^2}{k^2_0}D_{37}, ~
P_{yy}^{(LT)}=-2\displaystyle\frac{\sqrt{Q^2}}{k_0}D_{40}, \\
P_{yz}^{(TT)}&=&4D_{26}, ~
P_{yz}^{(LT)}=-2\displaystyle\frac{\sqrt{Q^2}}{k_0}D_{25}, 
\end{eqnarray*}
\noindent\underline{For the $P_z$-component: }
\begin{eqnarray*} 
P_{zx}\sigma _0&=&{\cal N}[P_{zx}^{(TT)}+\varepsilon P_{zx}^{(LL)}+
\sqrt{2\varepsilon (1+\varepsilon )}cos\phi P_{zx}^{(LT)}
+\varepsilon cos(2\phi )\bar P_{zx}^{(TT)}], \\
P_{zy}\sigma _0&=&{\cal N}sin\phi [\sqrt{2\varepsilon (1+\varepsilon )}P_{zy}^{(LT)}
+\varepsilon cos\phi P_{zy}^{(TT)}], \\
P_{zz}\sigma _0&=&{\cal N}[P_{zz}^{(TT)}+\varepsilon P_{zz}^{(LL)}+
\sqrt{2\varepsilon (1+\varepsilon )}cos\phi P_{zz}^{(LT)}
+\varepsilon cos(2\phi )\bar P_{zz}^{(TT)}], 
\end{eqnarray*}
where 
\begin{eqnarray*} 
P_{zx}^{(TT)}&=& D_2+D_3, ~ P_{zx}^{(LL)}=2\displaystyle\frac{Q^2}{k^2_0}D_1, ~\bar P_{zx}^{(TT)}=D_2-D_3, ~ P_{zx}^{(LT)}=-2\displaystyle\frac{\sqrt{Q^2}}{k_0}D_4, \\
P_{zy}^{(TT)}&=&4D_{30}, ~ P_{zy}^{(LT)}=-2\displaystyle\frac{\sqrt{Q^2}}{k_0}D_{29},~
P_{zz}^{(TT)}=D_{16}+D_{17}, ~ \bar P_{zz}^{(TT)}=D_{16}-D_{17}, \\
P_{zz}^{(LL)}&=& 2\displaystyle\frac{Q^2}{k^2_0}D_{15}, ~
P_{zz}^{(LT)}=-2\displaystyle\frac{\sqrt{Q^2}}{k_0}D_{18}.
\end{eqnarray*}
As it was shown earlier, the components of the neutron polarization, containing 
the terms with the longitudinal contribution of the electromagnetic current, are the most sensitive to the form factor $G_{En}$. For  coplanar experimental
conditions ($\phi =0^0$ or $180^0$) the following components are nonzero: 
$P_{xx}$, $P_{xz}$, $P_{yy}$, $P_{zx}$, and $P_{zz}$.

\section{Proton polarization in case of vector--polarized deuteron target}

A new method for determination of the neutron electric form factor has been
suggested in Ref. \cite{G90}. This
method requires no longitudinally polarized electron beam and is based on
measurement of the polarization of the nucleons (protons or neutrons)
produced in the disintegration of the vector polarized deuteron. The
sensitivity of the spin transfer coefficients to the form factor $G_{En},$ as well as to
various parameterizations of the deuteron wave functions has been analyzed
for the case of the reaction ${\vec d}(e, e'{\vec p})n.$ The analysis was
carried out in the framework of RIA. It was shown that in the region of the
quasi--elastic peak some spin transfer coefficients are large and sensitive
to the form factor $G_{En}$ and practically do not depend on the choice of the deuteron wave function. Later, the outgoing
nucleon polarization in the case of polarized beam and target (the vector and
tensor deuteron polarizations were considered) has been investigated in Ref. 
\cite{Mos93}. In the numerical applications,
they have studied all the nonvanishing  components of proton and neutron
polarization in the coplanar kinematics and in the quasi--elastic region. The
calculations were made in the standard theory with emphasis on the effect of
nucleonic and pionic relativistic corrections. It was shown, in particular,
that the longitudinal component of neutron polarization with vector polarized
deuterons is as sensitive to the parametrization of the form factor $G_{En}$ as 
the sideways beam polarization transfer.

In this chapter we present the formulae describing the polarization state
of the proton, produced in the ${\vec d}(e, e'{\vec p})n$ reaction (where the deuteron is vector--polarized). The polarization of the proton is
determined by the ${\vec P}_{ij}({\vec \xi })$ tensor:
\begin{equation}
{\vec P}_{ij}({\vec \xi})=Tr{\vec \sigma }F_iF_j^+. 
\end{equation}
The tensor ${\vec P}_{ij}({\vec \xi })$ can be represented in the following
general form
\begin{equation}
{\vec P}_{ij}({\vec \xi })={\vec \xi}\cdot {\vec m}{\vec P}_{ij}^{(m)}+
{\vec \xi}\cdot{\vec n}{\vec P}_{ij}^{(n)}+
{\vec \xi}\cdot \hat {\vec k}{\vec P}_{ij}^{(k)}. 
\end{equation}
The tensors ${\vec P}_{ij}^{(l)}, l=m,n,k$ can be expanded over
the complete set of the orthonormal vectors. As a result, we have following
general structures for these tensors
$${\vec P}_{ij}^{(m)}=\bar P_{ij}^{(mk)}{\hat {\vec k}}+
\bar P_{ij}^{(mm)}{\vec m}+
\bar P_{ij}^{(mn)}{\vec n}, $$
\begin{equation}
{\vec P}_{ij}^{(k)}=\bar P_{ij}^{(kk)}{\hat {\vec k}}+
\bar P_{ij}^{(km)}{\vec m}+
\bar P_{ij}^{(kn)}{\vec n}, 
\end{equation}
$${\vec P}_{ij}^{(n)}=\bar P_{ij}^{(nk)}{\hat {\vec k}}+
\bar P_{ij}^{(nm)}{\vec m}+
\bar P_{ij}^{(nn)}{\vec n}. $$
The tensors $\bar P_{ij}^{(lr)}, l,r=m,n,k$ can be decomposed, in turn, in the
way that was used in previous chapter. Then these tensors have the same
form as the tensors  $N_{ij}^{(lr)}$ (see formula (4.14)), describing the
neutron polarization, where SFs $D_i$ are replaced by SFs $B_i$. Thus, the
dependence of the proton polarization on the vector polarization of the
deuteron is determined by forty-one SFs $B_i, i=1-41.$ The formulae for SFs
$B_i$ in terms of the reaction scalar amplitudes are given in Appendix 8.

The components of the proton polarization vector, $\vec P (p)$, can be written as
\begin{equation}
P_i(p)=\bar P_{ij}\xi _j. 
\end{equation}
The polarization components have definite general structure, if we single out
the explicit dependences on the azimuthal angle $\phi ,$ virtual--photon
linear polarization $\varepsilon ,$ and contributions of the longitudinal (L)
and transverse (T) components of the electromagnetic current of the
$\gamma ^*d\rightarrow np$ reaction. The expressions for the $\bar P_{ij}$
quantities can be obtained from the formulae for the $P_{ij}$ quantities by
following substitution $D_i \rightarrow B_i.$ Let us note that the
components of the proton polarization, containing the terms with the
longitudinal contribution of the electromagnetic current, will be the most
sensitive to the form factor $G_{En}.$ For coplanar experimental conditions
($\phi =0^0$ or $\phi =180^0$) the following components are nonzero:
$\bar P_{xx},$ $\bar P_{xz},$ $\bar P_{yy},$ $\bar P_{zx},$ and $\bar P_{zz}.$

\section{Disintegration of tensor--polarized
deuteron target by unpolarized electron beam}

The differential cross section of the tensor--polarized deuteron
disintegration by the unpolarized electron beam (in the coincidence
experimental setup) has the following general structure (where the $z$ axis is directed
along the virtual--photon momentum ${\vec k}$, and the $xz$ plane coincides
with the $({\vec k}, {\vec p})$ plane):
\begin{eqnarray*}
\frac{d^3\sigma }{dE'd\Omega _ed\Omega _p}&=&
N\left \{\sigma _T+A_{xz}^TQ_{xz}+A_{xx}^T(Q_{xx}-Q_{yy})+A_{zz}^TQ_{zz}+ 
\right .\\
&&
\varepsilon \left  [\sigma _L+A_{xz}^LQ_{xz}+A_{xx}^L(Q_{xx}-Q_{yy})+
A_{zz}^LQ_{zz}\right ]+ \\
&&\sqrt{2\varepsilon (1+\varepsilon )}\cos\phi
\left  [\sigma _I +A_{xz}^IQ_{xz}+A_{xx}^I(Q_{xx}-Q_{yy})+A_{zz}^IQ_{zz}\right  ]+
\\
&&\sqrt{2\varepsilon (1+\varepsilon )}\sin\phi (A_{xy}^IQ_{xy} +A_{yz}^IQ_{yz})+ \\
&&
\varepsilon \sin 2\phi (A_{xy}^PQ_{xy}+A_{yz}^PQ_{yz})+\\
&&\left .
\varepsilon \cos 2\phi \left  [\sigma _P+A_{xz}^PQ_{xz}+A_{xx}^P(Q_{xx}-Q_{yy})+A_{zz}^PQ_{zz}\right ]\right \}, 
\end{eqnarray*}
where the quantities $Q_{ij}, \ (i,j=x, y, z)$ are the components of the
quadrupole--polarization tensor of the deuteron in its rest system (the
coordinate system is specified similarly to the case of the np--pair
CMS). These components  satisfy to the following conditions: $Q_{ij}=
Q_{ji}, \ $ $Q_{ii}=0.$

The general characteristic property of all these tensor asymmetries is that 
they vanish in the region of the quasi--elastic scattering. This can be
explained as follows. All the asymmetries are determined by the
convolution $X_{\mu\nu }Q_{\mu\nu },$ where the tensor $X_{\mu\nu }$ is built with the four-momenta describing the $d\rightarrow np$ transition.
Due to the condition $P_{\mu }Q_{\mu\nu }=0,$ the most general form of this tensor is
$$X_{\mu\nu }=a_1g_{\mu\nu }+ia_2[\gamma _{\mu }, \gamma _{\nu }]+
a_3\gamma _{\mu }p_{\nu }+a_4\gamma _{\nu }p_{\mu }+a_5p_{\mu }p_{\nu }, $$
where $p_{\mu }$ is the four-momentum of the neutron--spectator. However, if we
take into account that $Q_{\mu\nu }g_{\mu\nu }=0, \ Q_{\mu\nu }=Q_{\nu\mu },$
then the convolution $X_{\mu\nu }Q_{\mu\nu }$ is determined by  $a_3$, $a_4$, and $a_5$. From the condition $P_{\mu }Q_{\mu\nu }=0$, it
follows that the time components of the $Q_{\mu\nu }$ tensor are zero in the
laboratory system. After all, the convolution $X_{\mu\nu }Q_{\mu\nu }$ turns
out proportional to the nucleon--spectator three-momentum which is zero in the
peak of the quasi--elastic scattering.

Thus, in the general case the exclusive cross section of the electrodisintegration
of the tensor--polarized deuteron is determined by 16 independent asymmetries
$A_{ij}^m(W, k^2, \vartheta ),$ where $i,j=x,y,z; m=T,P,L,I.$ The asymmetries
$A_{ij}^m$ can be related to 18 independent scalar amplitudes $f_i, (i=1-18)$
describing the $\gamma ^*+d\rightarrow n+p$ reaction. These relations are: 
\begin{eqnarray*}
A_{xz}^T&=&4\frac{\omega}{M}Re(f_{1}f_{3}^*+ f_{2}f_{4}^*
+f_{7}f_{9}^*+ f_{8}f_{10}^*), \\
A_{xx}^T&=&|f_3|^2+|f_4|^2+|f_9|^2+|f_{10}|^2-|f_5|^2-|f_6|^2-
|f_{11}|^2-|f_{12}|^2, \\
A_{zz}^T&=&2\frac{\omega ^2}{M^2}\left [
|f_1|^2+|f_2|^2+|f_7|^2+|f_{8}|^2-\frac{M^2}{\omega ^2}
(|f_5|^2+|f_6|^2+|f_{11}|^2+|f_{12}|^2)\right  ]-A_{xx}^T, \\
A_{xz}^I&=&-4\frac{\omega}{M}\frac{\sqrt{-k^2}}{k_0}
Re(f_{1}f_{15}^*+ f_{3}f_{13}^*+f_{2}f_{16}^*+ f_{4}f_{14}^*), \\
A_{xx}^I&=&-2\frac{\sqrt{-k^2}}{k_0}
Re(f_{3}f_{15}^*+ f_{4}f_{16}^*-f_{5}f_{17}^*-f_{6}f_{18}^*), \\
A_{zz}^I&=&-4\frac{\omega ^2}{M^2}\frac{\sqrt{-k^2}}{k_0}
Re\left [f_{1}f_{13}^*+ f_{2}f_{14}^*-
\frac{M^2}{\omega ^2}(f_{5}f_{17}^*+ f_{6}f_{18}^*)\right ]-A_{xx}^I, \\
A_{xy}^I&=&-4\frac{\sqrt{-k^2}}{k_0}
Re(f_{9}f_{17}^*+ f_{10}f_{18}^*+f_{12}f_{18}^*+ f_{11}f_{15}^*), \\
A_{yz}^I&=&-4\frac{\omega}{M}\frac{\sqrt{-k^2}}{k_0}
Re(f_{7}f_{17}^*+ f_{8}f_{18}^*+f_{12}f_{14}^*+ f_{11}f_{13}^*), \\
A_{xz}^L&=&-8\frac{\omega}{M}\frac{k^2}{k_0^2}
Re(f_{13}f_{15}^*+ f_{14}f_{16}^*), \\
A_{xx}^L&=&-2\frac{k^2}{k_0^2}
[|f_{15}|^2+|f_{16}|^2-|f_{17}|^2-|f_{18}|^2], \\
A_{zz}^L&=&-4\frac{\omega ^2}{M^2}\frac{k^2}{k_0^2}
\left [|f_{13}|^2+|f_{14}|^2-\frac{M^2}{\omega ^2}(|f_{17}|^2
+|f_{18}|^2)\right ]-A_{xx}^L , \\
A_{xy}^P&=&4Re(f_{5}f_{9}^*+ f_{6}f_{10}^*
+f_{4}f_{12}^*+ f_{3}f_{11}^*), \\
A_{yz}^P&=&4\frac{\omega}{M}Re(f_{5}f_{7}^*+ f_{6}f_{8}^*
+f_{2}f_{12}^*+ f_{1}f_{11}^*), \\
A_{xz}^P&=&4\frac{\omega}{M}Re(f_{1}f_{3}^*+ f_{2}f_{4}^*
-f_{7}f_{9}^*- f_{8}f_{10}^*), \\
A_{xx}^P&=&|f_3|^2+|f_4|^2-|f_9|^2-|f_{10}|^2-|f_5|^2-|f_6|^2+
|f_{11}|^2+|f_{12}|^2, \\
A_{zz}^P&=&2\frac{\omega ^2}{M^2}\left [
|f_1|^2+|f_2|^2-|f_7|^2-|f_{8}|^2-\frac{M^2}{\omega ^2}
(|f_5|^2+|f_6|^2-|f_{11}|^2-|f_{12}|^2)\right ]-A_{xx}^P. 
\end{eqnarray*}
One can see from this formula that the scattering of unpolarized
electrons by a tensor polarized  
deuteron target 
with components $Q_{xy}=Q_{yz}=0$, is characterized by the same $\phi -$ and $\varepsilon -$ dependences as in
the case of the scattering of unpolarized electrons by the unpolarized
target. If $Q_{xy}\neq 0,$ $Q_{yz}\neq 0$, then new tems of the type $\sqrt{2\varepsilon (1+\varepsilon )}\sin\phi $ and
$\varepsilon \sin2\phi $ are present in the cross section. The asymmetries
with upper indices $T, P(L)$ are determined only by the transverse
(longitudinal) components of the electromagnetic current for the $\gamma ^*+
d\rightarrow n+p$ reaction, while the asymmetries with upper index $I$ are
determined by the interference of the longitudinal and transverse components
of the electromagnetic current.

\chapter{Conclusions}
We developed a relativistic approach to the calculation of the differential 
cross section and various polarization observables for the deuteron 
electrodisintegration process, $e^-+d\rightarrow e^-+n+p.$ Our formalism is based on the most general symmetry
properties of the hadron electromagnetic interaction, such as gauge invariance 
(the conservation of the hadronic and leptonic electromagnetic currents) and 
P--invariance (invariance with respect to the space reflections) and does not 
depend on the deuteron structure and on the details of the reaction mechanism 
for $\gamma ^*+d\rightarrow n+p.$ 

We established a general formalism for the structure of the differential 
cross section and various polarization observables. This general analysis was done with the help 
of the structure function formalism which is especially convenient for the 
investigation of the polarization phenomena in this reaction.

The observables related to the cases of a polarized deuteron target, longitudinally polarized
electron beam, polarizations of the outgoing nucleons (proton and neutron),  
as well as the polarization transfer from electron to final nucleon and from 
deuteron (vector--polarized) to nucleon, and the correlation of the electron 
and deuteron polarizations were considered in detail. 

For the quantitative estimations of the observables a dynamical model of the  
$\gamma ^*+d\rightarrow n+p$ reaction was developed: the relativistic impulse 
approximation which accounts for four Feynman diagrams: two diagrams represent 
the relativized description of the one--nucleon--exchange mechanism, and the 
others are the deuteron--exchange and contact diagrams. The deuteron structure 
is described here by the relativistic form factors of the $dnp-$vertex with 
one virtual nucleon. In order to calculate the dependence of these form factors 
on the nucleon virtuality we use the approach developed by the Buck and Gross 
\cite{Bu79}. The resulting amplitude for the  $\gamma ^*+d\rightarrow n+p$ 
reaction can be unambiguously calculated for any values of the kinematics 
variables $Q^2, s$ and $t$.

The RIA amplitude does not contain the effects of the $np\rightarrow np$ 
scattering. These effects are also included in nonrelativistic models of the 
deuteron electrodisintegration in terms of the FSI effects. In relativistic 
physics, this process (the $np\rightarrow np$ scattering) is required by the 
fulfilment of the unitarity condition. This is especially important for the 
analysis of the polarization effects in $e^-+d\rightarrow e^-+n+p$ reaction - 
for example, the asymmetry of unpolarized electrons inclusively scattered by 
a vector--polarized target, ${\vec d}(e, e')np.$

The proposed unitarization procedure is carried out in the relativistic 
approach, for any value of $Q^2$. We also use a relativistic
description of the nucleon electromagnetic current in terms of the Dirac ($F_1$) 
and Pauli ($F_2$) form factors. The $np$ phase shifts were taken from Ref. 
\cite{By87}.

We calculated the asymmetry $\Sigma _e$ for the deuteron disintegration by 
longitudinally polarized electron beam. It is a T--odd polarization observable 
determined by the so--called fifth SF, $\alpha_5.$ In impulse approximation 
this asymmetry is zero.

The predicted value for the asymmetry $\Sigma _e$ for the kinematics conditions 
of the experiment performed at the MIT--Bates \cite{Zh01a} (measurements were
done at the missing momentum of 210 MeV/c) does not contradict the 
experimental data. The sensitivity of this asymmetry  to the choice of DWFs 
appears only in the region where $p_m \ge ~280~MeV/c.$ The qualitative behaviour 
of this asymmetry  as a function of $p_m$, predicted in our model, is consistent
with one in Ref. \cite{Ar89}. The predicted asymmetry for the experimental 
conditions of Ref. \cite{Do99} agree relatively well with the data.

The sideways asymmetry $A_x^{ed}$ (the vector--polarized deuteron 
disintegration by the longitudinally polarized electron beam) has been measured 
at the NIKHEF accelerator \cite{Pa99} for missing momenta less than 
200 MeV/c. The agreement of the predicted $A_x^{ed}$ with the experimental data 
is excellent, particularly in the quasi--elastic region. The sensitivity of 
the asymmetry to the choice of the DWF model is small and influence of the FSI 
effects is also insignificant. Therefore, the measurement of this asymmetry in 
the quasi--elastic region can give a reliable value of the $G_{En}$ form factor.

We suggested a method for the determination of the $G_{En}$ form factor, at 
relatively large $Q^2$, from the ratio $R=A_x/A_z$ of the T--even asymmetries 
in ${\vec d}({\vec e}, e'p)n$ reaction, measured in the kinematical conditions 
of the quasi--elastic $en$--scattering. This method seems promising and may be 
comparable in accuracy with the measurements of the $G_{Ep}$ form factor 
through the recoil polarization method.

We calculated the proton polarization in $d(e, e{\vec p})n$ and 
$d({\vec e}, e{\vec p})n$ reactions and compared the results with existing 
experimental data. The calculations of the polarization transfer components 
$P_l$ and $P_t$ describe the data, obtained at MIT in the region of the 
neutron recoil momenta lower than $100~ MeV/c,$ well, but calculations of the 
induced polarization $P_n$ at $p_m = 100~ MeV/c $ somewhat underestimate the 
experimental result. The data from Ref. \cite{Mi98} are also consistent with our 
theoretical calculations of these quantities.

In conclusion we can say that we developed a fully relativistic model for the 
description of various polarization effects in the deuteron
electrodisintegration reaction. The reliability of the suggested model was tested 
by comparing the predicted results of our model with the existing experimental 
data. The agreement is quite reasonable. This model is especially adapted to 
the description of the deuteron electrodisintegration reaction at large $Q^2$ 
where relativistic effects must be sizeable.

\chapter{Appendices}

\section{Appendix 1: relations between invariant amplitudes  and scalar amplitudes}

The scalar amplitudes $F_i$ are related to the invariant amplitudes $H_i$ in  
the following way:
\begin{eqnarray*}
F_1&=&-H_{18}-\frac{W}{\omega}\frac{{\vec p}\cdot {\vec k}}{m^2}H_{13}-
2\frac{{\vec p}^2}{m^2}H_{14}, \\
F_2&=&\frac{1}{m^2}\biggl[{\vec k}^2\frac{W}{\omega}H_{13}+
2{\vec p}\cdot {\vec k}(H_{14}+H_{15})-Wk_0H_{16}-2k^2H_{17}\biggr], \\
F_3&=&\frac{1}{m^2}\biggl(2H_{17}+\frac{W}{\omega}H_{13}\biggr), \ \ 
F_4=\frac{2}{m^2}\biggl(H_{14}+H_{15}\biggr), \\
F_5&=&\frac{1}{2m\omega}\biggl[2Wk_0H_{13}+4(k^2H_{17}+Ek_0H_{16}-
{\vec p}\cdot {\vec k}H_{15})+ \\ 
&&+W^2(H_{8}-\frac{k^2}{m^2}H_{9}+\frac{Ek_0}{m^2}H_{11}-
\frac{{\vec p}\cdot {\vec k}}{m^2}H_{12}-
2\frac{{\vec p}\cdot {\vec k}}{W^2}H_{18})\biggr], \\
F_6&=&\frac{k_0}{m}\biggl(2H_{14}-H_{18}\biggr)-
\frac{W}{m^3}\biggl(k^2H_{10}+{\vec p}\cdot {\vec k}H_{11}-
Ek_0H_{12}\biggr), \\
F_7&=&2\frac{k_0}{m}H_{17}-\frac{W}{4m^3}\biggl({\vec p}\cdot
{\vec k}H_{6}+Ek_0H_{7}\biggr), \ \ \\
F_8&=&\frac{k_0}{m}(2H_{15}+H_{18})+\frac{W}{4m^3}(2Wk_0-k^2)H_{6}, \\
F_9&=&-\frac{W}{2m}H_8, \ \   
F_{10}=\frac{1}{m^2}\biggl({\vec p}\cdot {\vec k}H_{1}+Ek_0H_{2}+
k^2H_3+{\vec p}\cdot {\vec k}\frac{m}{E+m}H_8\biggr), \\
F_{11}&=&\frac{1}{2m\omega}(4H_{17}-\frac{W^2}{m^2}H_9), \ \ 
F_{12}=-\frac{W}{m^3}H_{10}, \\
F_{13}&=&\frac{1}{2m\omega}(\frac{W^2}{m^2}H_{12}+4H_{15}+2H_{18}), \ \ 
F_{14}=\frac{W}{m^3}H_{11}, \\
F_{15}&=&\frac{1}{m^2}\frac{W}{\omega}\biggl[H_3+\frac{1}{W}
\frac{{\vec p}\cdot {\vec k}}{E+m}(\frac{W}{m}H_9+2H_{17})\biggr], \\
F_{16}&=&\frac{1}{2m^4}\biggl[-{\vec p}\cdot {\vec k}H_4-Ek_0H_5+
{\vec p}\cdot {\vec k}\frac{m}{E+m}(H_6+4H_{10})+
\frac{mk_0}{E+m}(EH_7+4mH_{17})\biggr], \\
F_{17}&=&-\frac{1}{m^2}\frac{W}{\omega}\biggl[H_{11}+
\frac{m}{E+m}(H_8-\frac{k^2}{m^2}H_{9}+\frac{Ek_0}{m^2}H_{11})- \\
&&-\frac{k_0}{E+m}(H_{13}+H_{16}+2\frac{k^2}{Wk_0}H_{17})\biggr], \\
F_{18}&=&\frac{2}{m^2}\frac{k_0}{E+m}\biggl[H_{14}+H_{15}+
\frac{E}{m}(\frac{k^2}{Ek_0}H_{10}-H_{12})+ \\
&&+\frac{2Wk_0-k^2}{4m^2}\frac{E+m}{k_0}(H_{4}-\frac{m}{E+m}H_{6})\biggr], \nonumber
\end{eqnarray*}
where $W$ is the total energy of the $np-$pair; $E, \omega, k_0$ are the
energies of the nucleon, deuteron, virtual photon in the $np-$ pair CMS,
repectively; ${\vec p}$ and ${\vec k}$ are the 3-momenta of the proton
and virtual photon in the final--hadrons CMS.

\section{Appendix 2: relations between two sets of scalar amplitudes}
Here we give the explicit expression for the nonzero elements of the matrix $M_{ij}, $ which relate two equivalent
sets of the amplitudes $f_i$ and $F_j$ :
\begin{eqnarray*}
M_{2\ 5}&=&M_{3\ 9}=M_{5\ 1}=M_{8\ 5}=M_{11\ 9}=M_{13\ 5}=M_{13\ 7}=
M_{13\ 9}=M_{16\ 7}= \\
&=&M_{18\ 7}=-M_{9\ 1}=|{\vec k}|,~M_{13\ 11}=|{\vec k}|^3,\\
M_{2\ 6}&=&M_{3\ 10}=M_{5\ 2}=M_{8\ 6}=M_{11\ 10}=M_{13\ 6}= 
M_{13\ 8}=M_{13\ 10}=M_{16\ 8}=\\
&=&M_{18\ 8}=-M_{9\ 2}=|{\vec p}|\cos\vartheta , \\
M_{1\ 8}&=&M_{4\ 6}=M_{4\ 8}=M_{4\ 10}=M_{6\ 8}= 
M_{7\ 2}=M_{10\ 6}=M_{12\ 10}=M_{14\ 10}=\\
&=&M_{15\ 6}=-M_{17\ 2}=
|{\vec p}|\sin\vartheta , \\
M_{13\ 12}&=&M_{13\ 13}=M_{13\ 15}={\vec k}\ ^2|{\vec p}|\cos\vartheta ,
M_{1\ 13}=M_{14\ 15}=M_{15\ 12}=M_{17\ 3}= \\
&=&{\vec k}\ ^2|{\vec p}|\sin\vartheta , \\
M_{13\ 14}&=&M_{13\ 16}=M_{13\ 17}={\vec p}\ ^2|{\vec k}|\cos^2\vartheta ,\\ 
M_{2\ 17}&=&M_{3\ 14}=  
M_{5\ 4}=M_{16\ 16}={\vec p}\ ^2|{\vec k}|\sin^2\vartheta , \\
M_{1\ 14}&=&M_{1\ 17} 
=M_{14\ 16}=M_{14\ 17}=M_{15\ 14}=M_{15\ 16}=
M_{17\ 4}={\vec p}\ ^2|{\vec k}|\cos\vartheta \sin\vartheta , \\
M_{13\ 18}&=&|{\vec p}|\ ^3\cos^3\theta ,~
M_{4\ 18}=|{\vec p}|\ ^3\sin^3\vartheta ,~
M_{1\ 18}=M_{14\ 18}= M_{15\ 18}=|{\vec p}|\ ^3\cos^2\vartheta \sin\vartheta , \\ 
M_{2\ 18}&=&M_{3\ 18}=M_{16\ 18}=|{\vec p}|\ ^3\cos\vartheta \sin^2\vartheta , \\
\end{eqnarray*}  
where $\vartheta $ is the angle between the 3-momenta of the proton and virtual 
photon in the final--hadrons CMS.    
  
\section{Appendix 3: expressions of SFs for  $d(e,e' p)n$ and 
$\vec d(\vec e,e' p)n$ in terms of the scalar amplitudes}

Relations between SFs $\alpha _i$ $(i=1-5),$ $\beta _i$ $(i=1-13)$,
$\gamma _i $ $(i=1-23),$ $\delta _i$ $(i=1-9)$ and the 
scalar amplitudes $f_i$ $(i=1-18)$ determining the $\gamma ^*d\rightarrow np$ 
reaction
\begin{eqnarray*}
\alpha _1&=&\frac{2}{3}\biggl[\frac{\omega ^2}{M^2}(|f_{13}|^2+|f_{14}|^2)+
|f_{15}|^2+|f_{16}|^2+|f_{17}|^2+|f_{18}|^2\biggr], \\
\alpha _2&=&\frac{2}{3}\biggl[\frac{\omega ^2}{M^2}(|f_{7}|^2+|f_{8}|^2)+
|f_{9}|^2+|f_{10}|^2+|f_{11}|^2+|f_{12}|^2\biggr], \\
\alpha _3&=&\frac{2}{3}\biggl[\frac{\omega ^2}{M^2}(|f_{1}|^2+|f_{2}|^2)+
|f_{3}|^2+|f_{4}|^2+|f_{5}|^2+|f_{6}|^2\biggr], \\
\alpha _4&=&\frac{2}{3}Re\biggl[\frac{\omega ^2}{M^2}(f_{1}f_{13}^*+
f_{2}f_{14}^*)+f_{3}f_{15}^*+f_{4}f_{16}^*+f_{5}f_{17}^*+
f_{6}f_{18}^*\biggr], \\
\alpha _5&=&\frac{2}{3}Im\biggl[\frac{\omega ^2}{M^2}(f_{13}f_{1}^*+
f_{14}f_{2}^*)+f_{15}f_{3}^*+f_{16}f_{4}^*+f_{17}f_{5}^*+
f_{18}f_{6}^*\biggr], \\
\beta_1&=&2\frac{\omega}{M}Im(f_{16}f_{14}^*+f_{15}f_{13}^*), \\ 
\beta_2&=&2\frac{\omega}{M}Im(f_{4}f_{2}^*+f_{3}f_{1}^*), \\
\beta_3&=&2\frac{\omega}{M}Im(f_{9}f_{7}^*+f_{10}f_{8}^*), \\
\beta_4&=&\frac{\omega}{M}Im(f_{4}f_{14}^*+f_{3}f_{13}^* -
f_{2}f_{16}^*-f_{1}f_{15}^*), \\
\beta_5&=&\frac{\omega}{M}Re(f_{4}f_{14}^*+f_{3}f_{13}^* -
f_{2}f_{16}^*-f_{1}f_{15}^*), \\
\beta_6&=&Im(f_{12}f_{16}^*+f_{11}f_{15}^* -
f_{9}f_{17}^*-f_{10}f_{18}^*), \\  
\beta_7&=&Im(f_{12}f_{4}^*+f_{11}f_{3}^* -
f_{9}f_{5}^*-f_{10}f_{6}^*), \\
\beta_8&=&Re(f_{12}f_{16}^*+f_{11}f_{15}^* -
f_{9}f_{17}^*-f_{10}f_{18}^*), \\
\beta_9&=&Re(f_{12}f_{4}^*+f_{11}f_{3}^* -
f_{9}f_{5}^*-f_{10}f_{6}^*), \\
\beta_{10}&=&\frac{\omega}{M}Im(f_{7}f_{17}^*+f_{8}f_{18}^* -
f_{12}f_{14}^*-f_{11}f_{13}^*), \\
\beta_{11}&=&\frac{\omega}{M}Im(f_{7}f_{5}^*+f_{8}f_{6}^* -
f_{12}f_{2}^*-f_{11}f_{1}^*), \\
\beta_{12}&=&\frac{\omega}{M}Re(f_{7}f_{17}^*+f_{8}f_{18}^* -
f_{12}f_{14}^*-f_{11}f_{13}^*), \\
\beta_{13}&=&\frac{\omega}{M}Re(f_{7}f_{5}^*+f_{8}f_{6}^* -
f_{12}f_{2}^*-f_{11}f_{1}^*), \\
\gamma _1&=&2\biggl[|f_{13}|^2+|f_{14}|^2-\frac{M^2}{\omega ^2}
(|f_{17}|^2+|f_{18}|^2)\biggr], \\
\gamma _2&=&2\biggl[|f_{1}|^2+|f_{2}|^2-\frac{M^2}{\omega ^2}
(|f_{5}|^2+|f_{6}|^2)\biggr], \\
\gamma _3&=&2\biggl[|f_{7}|^2+|f_{8}|^2-\frac{M^2}{\omega ^2}
(|f_{11}|^2+|f_{12}|^2)\biggr], \\
\gamma_4&=&2Re\biggl[f_{1}f_{13}^*+f_{2}f_{14}^* -\frac{M^2}{\omega ^2}
(f_{5}f_{17}^*+f_{6}f_{18}^*)\biggr], \\
\gamma_5&=&-2Im\biggl[f_{1}f_{13}^*+f_{2}f_{14}^* -\frac{M^2}{\omega ^2}
(f_{5}f_{17}^*+f_{6}f_{18}^*)\biggr], \\
\gamma _6&=&2\biggl[|f_{15}|^2+|f_{16}|^2-
|f_{17}|^2-|f_{18}|^2\biggr], \\
\gamma _7&=&2\biggl[|f_{3}|^2+|f_{4}|^2-
|f_{5}|^2-|f_{6}|^2\biggr], \\
\gamma _8&=&2\biggl[|f_{9}|^2+|f_{10}|^2-
|f_{11}|^2-|f_{12}|^2\biggr], \\  
\gamma_9&=&2Re(f_{3}f_{15}^*+f_{4}f_{16}^* -
f_{5}f_{17}^*-f_{6}f_{18}^*), \\
\gamma_{10}&=&2Im(f_{15}f_{3}^*+f_{16}f_{4}^* -
f_{17}f_{5}^*-f_{18}f_{6}^*), \\
\gamma_{11}&=&2Re(f_{13}f_{15}^*+f_{14}f_{16}^*), \\
\gamma_{12}&=&2Re(f_{1}f_{3}^*+f_{2}f_{4}^*), \\
\gamma_{13}&=&2Re(f_{7}f_{9}^*+f_{8}f_{10}^*), \\
\gamma_{14}&=&Re(f_{1}f_{15}^*+f_{3}f_{13}^* +
f_{16}f_{2}^*+f_{4}f_{14}^*), \\
\gamma_{15}&=&Im(f_{15}f_{1}^*+f_{13}f_{3}^* +
f_{16}^*f_{2}+f_{14}f_{4}^*), \\
\gamma_{16}&=&Re(f_{7}f_{17}^*+f_{8}f_{18}^* +
f_{12}f_{14}^*+f_{11}f_{13}^*), \\
\gamma_{17}&=&Re(f_{5}f_{7}^*+f_{6}f_{8}^* +
f_{2}f_{12}^*+f_{1}f_{11}^*), \\
\gamma_{18}&=&Im(f_{17}f_{7}^*+f_{18}f_{8}^* +
f_{14}f_{12}^*+f_{13}f_{11}^*), \\
\gamma_{19}&=&Im(f_{5}f_{7}^*+f_{6}f_{8}^* +
f_{2}f_{12}^*+f_{1}f_{11}^*), \\
\gamma_{20}&=&Re(f_{9}f_{17}^*+f_{10}f_{18}^* +
f_{12}f_{16}^*+f_{11}f_{15}^*), \\
\gamma_{21}&=&Re(f_{5}f_{9}^*+f_{6}f_{10}^* +
f_{4}f_{12}^*+f_{3}f_{11}^*), \\
\gamma_{22}&=&Im(f_{17}f_{9}^*+f_{18}f_{10}^* +
f_{16}f_{12}^*+f_{15}f_{11}^*), \\
\gamma_{23}&=&Im(f_{5}f_{9}^*+f_{6}f_{10}^* +
f_{4}f_{12}^*+f_{3}f_{11}^*), \\
\delta _1&=&|f_{15}|^2+|f_{16}|^2+
|f_{17}|^2+|f_{18}|^2, \\
\delta _2&=&|f_{3}|^2+|f_{4}|^2+  
|f_{5}|^2+|f_{6}|^2, \\
\delta _3&=&|f_{9}|^2+|f_{10}|^2+
|f_{11}|^2+|f_{12}|^2, \\
\delta_{4}&=&Re(f_{17}f_{5}^*+f_{18}f_{6}^* +
f_{16}f_{4}^*+f_{15}f_{3}^*), \\
\delta_{5}&=&Im(f_{17}f_{5}^*+f_{18}f_{6}^* +
f_{16}f_{4}^*+f_{15}f_{3}^*), \\
\delta_{6}&=&Im(f_{17}f_{9}^*+f_{18}f_{10}^* -
f_{16}f_{12}^*-f_{15}f_{11}^*), \\
\delta_{7}&=&Re(-f_{17}f_{9}^*-f_{18}f_{10}^* +
f_{16}f_{12}^*+f_{15}f_{11}^*), \\
\delta_{8}&=&Im(f_{5}f_{9}^*+f_{6}f_{10}^* -
f_{4}f_{12}^*-f_{3}f_{11}^*), \\
\delta_{9}&=&Re(-f_{5}f_{9}^*-f_{6}f_{10}^* +
f_{4}f_{12}^*+f_{3}f_{11}^*). 
\end{eqnarray*}
\section{Appendix 4: relations between the helicity amplitudes and 
the scalar amplitudes}

\begin{eqnarray*}
h_1&=&\displaystyle\frac{e}{2}[x(f_3+f_{11})+y(f_4+f_{12})+f_5-f_9],\\
h_2&=&\displaystyle\frac{e}{2}[x(f_3+f_{11})+y(f_4+f_{12})-f_5+f_9],\\
h_3&=&\displaystyle\frac{e}{\sqrt{2}}\displaystyle\frac{\omega }{M}(xf_1+yf_2-f_7),\\
h_4&=&\displaystyle\frac{e}{\sqrt{2}}\displaystyle\frac{\omega }{M}(xf_1+yf_2+f_7),  \\
h_5&=&\displaystyle\frac{e}{2}[x(f_{11}-f_{3})-y(f_4-f_{12})+f_5+f_9],\\h_6&=&\displaystyle\frac{e}{2}[x(f_{11}-f_{3})-y(f_4-f_{12})-f_5-f_9], \\
h_7&=&\displaystyle\frac{e}{2}[y(f_3+f_{11})-x(f_4+f_{12})+f_6-f_{10}],
\\
h_8&=&\displaystyle\frac{e}{2}[-y(f_3+f_{11})+x(f_4+f_{12})+f_6-f_{10}], \\
h_9&=&\displaystyle\frac{e}{\sqrt{2}}\displaystyle\frac{\omega }{M}(yf_1-xf_2-f_8),\\
h_{10}&=&\displaystyle\frac{e}{\sqrt{2}}\displaystyle\frac{\omega }{M}(-yf_1+xf_2-f_8),\\
h_{11}&=&\displaystyle\frac{e}{2}[-y(f_3-f_{11})+x(f_4-f_{12})+f_6+f_{10}]\\
h_{12}&=&\displaystyle\frac{e}{2}[y(f_3-f_{11})-x(f_4-f_{12})+f_6+f_{10}],\\
h_{13}&=&\displaystyle\frac{e}{\sqrt{2}}\displaystyle\frac{\sqrt{-k^2}}{k_0}(xf_{15}+yf_{16}+f_{17}),\\
h_{14}&=&e\displaystyle\frac{\omega }{M}\displaystyle\frac{\sqrt{-k^2}}{k_0}(xf_{13}+yf_{14}), \\
h_{15}&=&\displaystyle\frac{e}{\sqrt{2}}\displaystyle\frac{\sqrt{-k^2}}{k_0}(-xf_{15}-yf_{16}+f_{17}),\\
h_{16}&=&\displaystyle\frac{e}{\sqrt{2}}\displaystyle\frac{\sqrt{-k^2}}{k_0}(xf_{16}-yf_{15}+f_{18}), \\
h_{17}&=&e\displaystyle\frac{\omega }{M}\displaystyle\frac{\sqrt{-k^2}}{k_0}(xf_{14}-yf_{13}),~
\\h_{18}&=&\displaystyle\frac{e}{\sqrt{2}}\displaystyle\frac{\sqrt{-k^2}}{k_0}(yf_{15}-xf_{16}+f_{18}), 
\end{eqnarray*}
where $x=\cos\vartheta ,$ $y=\sin\vartheta $ ($\vartheta $ is the angle between 
the 3--momenta of the proton and virtual photon in the final--hadrons CMS), 
e is the proton charge, $k_0 (\omega )$ is the virtual--photon (deuteron) 
energy in the final--hadrons CMS.

\section{Appendix 5: expressions of SFs for  $ d(\vec e,e'\vec p) n$  
in terms of the scalar amplitudes}

Relations beween SFs $ P_i  $, $i=1-13$,  and
the scalar amplitudes  $f_i $, $i=1-18:$
\begin{eqnarray*}
P _1&=&-\frac{2}{3}Im\biggl[f_{17}f_{11}^*+f_{16}f_{10}^*-
f_{18}f_{12}^*-f_{15}f_{9}^*+
\frac{\omega ^2}{M^2}(f_{14}f_{8}^*-f_{13}f_{7}^*)\biggr], \\
P _2&=&-\frac{2}{3}Im\biggl[f_{5}f_{11}^*+f_{4}f_{10}^*-
f_{6}f_{12}^*-f_{3}f_{9}^*+
\frac{\omega ^2}{M^2}(f_{2}f_{8}^*-f_{1}f_{7}^*)\biggr], \\
P _3&=&\frac{2}{3}Re\biggl[f_{17}f_{11}^*+f_{16}f_{10}^*-
f_{18}f_{12}^*-f_{15}f_{9}^*+
\frac{\omega ^2}{M^2}(f_{14}f_{8}^*-f_{13}f_{7}^*)\biggr], \\
P _4&=&\frac{2}{3}Re\biggl[f_{5}f_{11}^*+f_{4}f_{10}^*-
f_{6}f_{12}^*-f_{3}f_{9}^*+
\frac{\omega ^2}{M^2}(f_{2}f_{8}^*-f_{1}f_{7}^*)\biggr], \\
P _5&=&-\frac{2}{3}Im\biggl[f_{17}f_{12}^*+f_{18}f_{11}^*-
f_{16}f_{9}^*-f_{15}f_{10}^*-
\frac{\omega ^2}{M^2}(f_{14}f_{7}^*+f_{13}f_{8}^*)\biggr], \\
P _6&=&-\frac{2}{3}Im\biggl[f_{5}f_{12}^*+f_{6}f_{11}^*-
f_{4}f_{9}^*-f_{3}f_{10}^*-
\frac{\omega ^2}{M^2}(f_{2}f_{7}^*+f_{1}f_{8}^*)\biggr], \\
P _7&=&\frac{2}{3}Re\biggl[f_{17}f_{12}^*+f_{18}f_{11}^*-
f_{16}f_{9}^*-f_{15}f_{10}^*-
\frac{\omega ^2}{M^2}(f_{14}f_{7}^*+f_{13}f_{8}^*)\biggr], \\
P _8&=&\frac{2}{3}Re\biggl[f_{5}f_{12}^*+f_{6}f_{11}^*-
f_{4}f_{9}^*-f_{3}f_{10}^*-
\frac{\omega ^2}{M^2}(f_{2}f_{7}^*+f_{1}f_{8}^*)\biggr], \\
P _9&=&-\frac{4}{3}Im\biggl(f_{15}f_{16}^*+f_{17}f_{18}^*+
\frac{\omega ^2}{M^2}f_{13}f_{14}^*\biggr), \\
P _{10}&=&-\frac{4}{3}Im\biggl(f_{3}f_{4}^*+f_{5}f_{6}^*+
\frac{\omega ^2}{M^2}f_{1}f_{2}^*\biggr), \\
P _{11}&=&-\frac{4}{3}Im\biggl(f_{9}f_{10}^*+f_{11}f_{12}^*+
\frac{\omega ^2}{M^2}f_{7}f_{8}^*\biggr), \\
P _{12}&=&\frac{2}{3}Im\biggl[f_{6}f_{17}^*+f_{4}f_{15}^*-
f_{5}f_{18}^*-f_{3}f_{16}^*+
\frac{\omega ^2}{M^2}(f_{2}f_{13}^*-f_{1}f_{14}^*)\biggr], \\
P _{13}&=&\frac{2}{3}Re\biggl[f_{6}f_{17}^*+f_{4}f_{15}^*-
f_{5}f_{18}^*-f_{3}f_{16}^*+
\frac{\omega ^2}{M^2}(f_{2}f_{13}^*-f_{1}f_{14}^*)\biggr], \\
\end{eqnarray*}
where $M$ is the deuteron mass and $\omega=(W^2+M^2+Q^2)/2W $ is the deuteron  energy in the 
$\gamma ^* +d\rightarrow n+p$ reaction CMS.

\section{Appendix 6: expressions of SFs for  $ d(\vec e,e'\vec n) p$ 
in terms of the scalar amplitudes}

Relations between SFs $ N_i $,~$i=1-13$, describing the tensor 
${\vec N}_{ij}$ Eq. (4.12) and the scalar amplitudes $f_i $,~$i=1-18$:
\begin{eqnarray*}
N _1&=&\frac{2}{3}Im\biggl[f_{17}f_{11}^*-f_{16}f_{10}^*+
f_{18}f_{12}^*-f_{15}f_{9}^*-
\frac{\omega ^2}{M^2}(f_{14}f_{8}^*+f_{13}f_{7}^*)\biggr], \\
N _2&=&\frac{2}{3}Im\biggl[f_{5}f_{11}^*-f_{4}f_{10}^*+
f_{6}f_{12}^*-f_{3}f_{9}^*-
\frac{\omega ^2}{M^2}(f_{2}f_{8}^*+f_{1}f_{7}^*)\biggr], \\
N _3&=&-\frac{2}{3}Re\biggl[f_{17}f_{11}^*-f_{16}f_{10}^*+
f_{18}f_{12}^*-f_{15}f_{9}^*-\frac{\omega ^2}{M^2}(f_{14}f_{8}^*+f_{13}f_{7}^*)\biggr], \\
N _4&=&-\frac{2}{3}Re\biggl[f_{5}f_{11}^*-f_{4}f_{10}^*+
f_{6}f_{12}^*-f_{3}f_{9}^*-
\frac{\omega ^2}{M^2}(f_{2}f_{8}^*+f_{1}f_{7}^*)\biggr], \\
N _5&=&\frac{2}{3}Im\biggl[f_{17}f_{12}^*-f_{18}f_{11}^*-
f_{16}f_{9}^*+f_{15}f_{10}^*-
\frac{\omega ^2}{M^2}(f_{14}f_{7}^*-f_{13}f_{8}^*)\biggr], \\
N _6&=&\frac{2}{3}Im\biggl[f_{5}f_{12}^*-f_{6}f_{11}^*-
f_{4}f_{9}^*+f_{3}f_{10}^*-
\frac{\omega ^2}{M^2}(f_{2}f_{7}^*-f_{1}f_{8}^*)\biggr], \\
N _7&=&-\frac{2}{3}Re\biggl[f_{17}f_{12}^*-f_{18}f_{11}^*-
f_{16}f_{9}^*+f_{15}f_{10}^*-
\frac{\omega ^2}{M^2}(f_{14}f_{7}^*-f_{13}f_{8}^*)\biggr], \\
N _8&=&-\frac{2}{3}Re\biggl[f_{5}f_{12}^*-f_{6}f_{11}^*-
f_{4}f_{9}^*+f_{3}f_{10}^*-
\frac{\omega ^2}{M^2}(f_{2}f_{7}^*-f_{1}f_{8}^*)\biggr], \\
N _9&=&\frac{4}{3}Im\biggl(-f_{15}f_{16}^*+f_{17}f_{18}^*-
\frac{\omega ^2}{M^2}f_{13}f_{14}^*\biggr), \\
N _{10}&=&\frac{4}{3}Im\biggl(-f_{3}f_{4}^*+f_{5}f_{6}^*-
\frac{\omega ^2}{M^2}f_{1}f_{2}^*\biggr), \\
N _{11}&=&\frac{4}{3}Im\biggl(f_{9}f_{10}^*-f_{11}f_{12}^*+
\frac{\omega ^2}{M^2}f_{7}f_{8}^*\biggr), \\
N _{12}&=&-\frac{2}{3}Im\biggl[f_{6}f_{17}^*-f_{4}f_{15}^*-
f_{5}f_{18}^*+f_{3}f_{16}^*-
\frac{\omega ^2}{M^2}(f_{2}f_{13}^*-f_{1}f_{14}^*)\biggr], \\
N _{13}&=&-\frac{2}{3}Re\biggl[f_{6}f_{17}^*-f_{4}f_{15}^*-
f_{5}f_{18}^*+f_{3}f_{16}^*-
\frac{\omega ^2}{M^2}(f_{2}f_{13}^*-f_{1}f_{14}^*)\biggr]. 
\end{eqnarray*}
\section{Appendix 7: structure functions for  $\vec d(e,e'\vec n)p$ }

Relations between SFs $ D_i$, $i=1-41,$ and the scalar amplitudes  $f_i$, $i=1-18$:
\begin{eqnarray*}
\\D_1&=&2\frac{\omega }{M}Re(f_{13}f_{17}^*+f_{14}f_{18}^*), \\
D_2&=&2\frac{\omega }{M}Re(f_{1}f_{5}^*+f_{2}f_{6}^*), \\
D_3&=&-2\frac{\omega }{M}Re(f_{7}f_{11}^*+f_{8}f_{12}^*), \\
D_4&=&\frac{\omega }{M}Re(f_{1}f_{17}^*+f_{2}f_{18}^*+
f_{5}f_{13}^*+f_{6}f_{14}^*), \\
D_5&=&-\frac{\omega }{M}Im(f_{1}f_{17}^*+f_{2}f_{18}^*+
f_{5}f_{13}^*+f_{6}f_{14}^*), \\
D_6&=&2\frac{\omega }{M}Re(f_{14}f_{17}^*-f_{13}f_{18}^*), \\
D_7&=&2\frac{\omega }{M}Re(f_{2}f_{5}^*-f_{1}f_{6}^*), \\
D_8&=&2\frac{\omega }{M}Re(f_{8}f_{11}^*-f_{7}f_{12}^*), \\
D_9&=&\frac{\omega }{M}Re(f_{2}f_{17}^*-f_{1}f_{18}^*-
f_{6}f_{13}^*+f_{5}f_{14}^*), \\
D_{10}&=&-\frac{\omega }{M}Im(f_{2}f_{17}^*-f_{1}f_{18}^*-
f_{6}f_{13}^*+f_{5}f_{14}^*), \\
D_{11}&=&\frac{\omega }{M}Re(f_{8}f_{17}^*-f_{7}f_{18}^*+
f_{12}f_{13}^*-f_{11}f_{14}^*), \\
D_{12}&=&\frac{\omega }{M}Re(f_{5}f_{8}^*+f_{1}f_{12}^*-
f_{2}f_{11}^*-f_{6}f_{7}^*), \\
D_{13}&=&-\frac{\omega }{M}Im(f_{8}f_{17}^*-f_{7}f_{18}^*+
f_{12}f_{13}^*-f_{11}f_{14}^*), \\
D_{14}&=&\frac{\omega }{M}Im(f_{5}f_{8}^*+f_{1}f_{12}^*-
f_{2}f_{11}^*-f_{6}f_{7}^*), \\
D_{15}&=&2Re(f_{15}f_{17}^*+f_{16}f_{18}^*), \\
D_{16}&=&2Re(f_{3}f_{5}^*+f_{4}f_{6}^*), \\
D_{17}&=&-2Re(f_{9}f_{11}^*+f_{10}f_{12}^*), \\
D_{18}&=&Re(f_{3}f_{17}^*+f_{5}f_{15}^*+
f_{4}f_{18}^*+f_{6}f_{16}^*), \\
D_{19}&=&-Im(f_{3}f_{17}^*+f_{5}f_{15}^*+
f_{4}f_{18}^*+f_{6}f_{16}^*), \\
D_{20}&=&2Re(f_{16}f_{17}^*-f_{15}f_{18}^*), \\
D_{21}&=&2Re(f_{4}f_{5}^*-f_{3}f_{6}^*), \\
D_{22}&=&2Re(f_{10}f_{11}^*-f_{9}f_{12}^*), \\
D_{23}&=&Re(f_{4}f_{17}^*-f_{3}f_{18}^*-
f_{6}f_{15}^*+f_{5}f_{16}^*), \\
D_{24}&=&-Im(f_{4}f_{17}^*-f_{3}f_{18}^*-
f_{6}f_{15}^*+f_{5}f_{16}^*), \\
D_{25}&=&Re(f_{10}f_{17}^*-f_{9}f_{18}^*+
f_{12}f_{15}^*-f_{11}f_{16}^*), \\
D_{26}&=&Re(f_{5}f_{10}^*+f_{3}f_{12}^*-
f_{4}f_{11}^*-f_{6}f_{9}^*), \\
D_{27}&=&-Im(f_{10}f_{17}^*-f_{9}f_{18}^*+
f_{12}f_{15}^*-f_{11}f_{16}^*), \\
D_{28}&=&Im(f_{5}f_{10}^*+f_{3}f_{12}^*-
f_{4}f_{11}^*-f_{6}f_{9}^*), \\
D_{29}&=&\frac{\omega }{M}Re(f_{15}f_{7}^*+f_{16}f_{8}^*-
f_{13}f_{9}^*-f_{14}f_{10}^*), \\
D_{30}&=&\frac{\omega }{M}Re(f_{3}f_{7}^*+f_{4}f_{8}^*-
f_{1}f_{9}^*-f_{2}f_{10}^*), \\
D_{31}&=&\frac{\omega }{M}Im(f_{15}f_{7}^*+f_{16}f_{8}^*-
f_{13}f_{9}^*-f_{14}f_{10}^*), \\
D_{32}&=&\frac{\omega }{M}Im(f_{3}f_{7}^*+f_{4}f_{8}^*-
f_{1}f_{9}^*-f_{2}f_{10}^*), \\
D_{33}&=&\frac{\omega }{M}Re(f_{16}f_{7}^*+f_{13}f_{10}^*-
f_{14}f_{9}^*-f_{15}f_{8}^*), \\
D_{34}&=&\frac{\omega }{M}Re(f_{1}f_{10}^*+f_{4}f_{7}^*-
f_{2}f_{9}^*-f_{3}f_{8}^*), \\
D_{35}&=&\frac{\omega }{M}Im(f_{16}f_{7}^*+f_{13}f_{10}^*-
f_{14}f_{9}^*-f_{15}f_{8}^*), \\
D_{36}&=&\frac{\omega }{M}Im(f_{1}f_{10}^*+f_{4}f_{7}^*-
f_{2}f_{9}^*-f_{3}f_{8}^*), \\
D_{37}&=&2\frac{\omega }{M}Re(f_{14}f_{15}^*-f_{13}f_{16}^*), \\
D_{38}&=&2\frac{\omega }{M}Re(f_{2}f_{3}^*-f_{1}f_{4}^*), \\
D_{39}&=&2\frac{\omega }{M}Re(f_{7}f_{10}^*-f_{8}f_{9}^*), \\
D_{40}&=&\frac{\omega }{M}Re(f_{14}f_{3}^*-f_{16}f_{1}^*-
f_{13}f_{4}^*+f_{15}f_{2}^*), \\
D_{41}&=&\frac{\omega }{M}Im(f_{14}f_{3}^*-f_{16}f_{1}^*-
f_{13}f_{4}^*+f_{15}f_{2}^*).
\end{eqnarray*}

\section{Appendix 8: structure functions for
${\vec d}(e, e'{\vec p})n$}

Relations between SFs $B _i, \ i=1 - 41,$ and the scalar amplitudes
$f_i, \ i=1 - 18:$
\begin{eqnarray*}
B_1&=&2\frac{\omega}{M}Re(f_{13}f_{17}^*- f_{14}f_{18}^*), \\
B_2&=&2\frac{\omega}{M}Re(f_{1}f_{5}^*- f_{2}f_{6}^*), \\
B_3&=&2\frac{\omega}{M}Re(f_{8}f_{12}^*- f_{7}f_{11}^*), \\
B_4&=&\frac{\omega}{M}Re(f_{1}f_{17}^*+f_{5}f_{13}^*
-f_{2}f_{18}^*-f_{6}f_{14}^*), \\
B_5&=&-\frac{\omega}{M}Im(f_{1}f_{17}^*+f_{5}f_{13}^*
-f_{2}f_{18}^*-f_{6}f_{14}^*), \\
B_6&=&2\frac{\omega}{M}Re(f_{13}f_{18}^*+ f_{14}f_{17}^*), \\
B_7&=&2\frac{\omega}{M}Re(f_{1}f_{6}^*+ f_{2}f_{5}^*), \\
B_8&=&-2\frac{\omega}{M}Re(f_{7}f_{12}^*+ f_{8}f_{11}^*), \\
B_9&=&\frac{\omega}{M}Re(f_{1}f_{18}^*+f_{6}f_{13}^*
+f_{2}f_{17}^*+f_{5}f_{14}^*), \\
B_{10}&=&-\frac{\omega}{M}Im(f_{1}f_{18}^*+f_{6}f_{13}^*
+f_{2}f_{17}^*+f_{5}f_{14}^*), \\
B_{11}&=&\frac{\omega}{M}Re(f_{14}f_{11}^*+f_{17}f_{8}^*
-f_{13}f_{12}^*-f_{18}f_{7}^*), \\
B_{12}&=&\frac{\omega}{M}Re(f_{8}f_{5}^*+f_{11}f_{2}^*
-f_{7}f_{6}^*-f_{12}f_{1}^*), \\
B_{13}&=&\frac{\omega}{M}Im(f_{14}f_{11}^*+f_{17}f_{8}^*
-f_{13}f_{12}^*-f_{18}f_{7}^*), \\
B_{14}&=&-\frac{\omega}{M}Im(f_{8}f_{5}^*+f_{11}f_{2}^*
-f_{7}f_{6}^*-f_{12}f_{1}^*), \\
B_{15}&=&2Re(f_{16}f_{18}^*- f_{15}f_{17}^*), \\
B_{16}&=&2Re(f_{4}f_{6}^*- f_{3}f_{5}^*), \\
B_{17}&=&2Re(f_{9}f_{11}^*- f_{10}f_{12}^*), \\
B_{18}&=&Re(f_{4}f_{18}^*+f_{6}f_{16}^*
-f_{5}f_{15}^*-f_{3}f_{17}^*), \\
B_{19}&=&-Im(f_{4}f_{18}^*+f_{6}f_{16}^*
-f_{5}f_{15}^*-f_{3}f_{17}^*), \\
B_{20}&=&-2Re(f_{16}f_{17}^*+ f_{15}f_{18}^*), \\
B_{21}&=&-2Re(f_{4}f_{5}^*+ f_{3}f_{6}^*), \\
B_{22}&=&2Re(f_{10}f_{11}^*+ f_{9}f_{12}^*), \\
B_{23}&=&-Re(f_{5}f_{16}^*+f_{6}f_{15}^*
+f_{4}f_{17}^*+f_{3}f_{18}^*), \\
B_{24}&=&Im(f_{5}f_{16}^*+f_{6}f_{15}^*
+f_{4}f_{17}^*+f_{3}f_{18}^*), \\
B_{25}&=&Re(f_{18}f_{9}^*+f_{15}f_{12}^*
-f_{17}f_{10}^*-f_{16}f_{11}^*), \\
B_{26}&=&Re(f_{9}f_{6}^*+f_{12}f_{3}^*
-f_{11}f_{4}^*-f_{10}f_{5}^*), \\
B_{27}&=&Im(f_{18}f_{9}^*+f_{15}f_{12}^*
-f_{17}f_{10}^*-f_{16}f_{11}^*), \\
B_{28}&=&-Im(f_{9}f_{6}^*+f_{12}f_{3}^*
-f_{11}f_{4}^*-f_{10}f_{5}^*), \\
B_{29}&=&\frac{\omega}{M}Re(f_{15}f_{7}^*+f_{14}f_{10}^*
-f_{13}f_{9}^*-f_{16}f_{8}^*), \\
B_{30}&=&\frac{\omega}{M}Re(f_{3}f_{7}^*+f_{2}f_{10}^*
-f_{1}f_{9}^*-f_{4}f_{8}^*), \\
B_{31}&=&\frac{\omega}{M}Im(f_{15}f_{7}^*+f_{14}f_{10}^*
-f_{13}f_{9}^*-f_{16}f_{8}^*), \\
B_{32}&=&\frac{\omega}{M}Im(f_{3}f_{7}^*+f_{2}f_{10}^*
-f_{1}f_{9}^*-f_{4}f_{8}^*), \\
B_{33}&=&\frac{\omega}{M}Re(f_{16}f_{7}^*+f_{15}f_{8}^*
-f_{14}f_{9}^*-f_{13}f_{10}^*), \\
B_{34}&=&\frac{\omega}{M}Re(f_{4}f_{7}^*+f_{3}f_{8}^*
-f_{2}f_{9}^*-f_{1}f_{10}^*), \\
B_{35}&=&\frac{\omega}{M}Im(f_{16}f_{7}^*+f_{15}f_{8}^*
-f_{14}f_{9}^*-f_{13}f_{10}^*), \\
B_{36}&=&\frac{\omega}{M}Im(f_{4}f_{7}^*+f_{3}f_{8}^*
-f_{2}f_{9}^*-f_{1}f_{10}^*), \\
B_{37}&=&2\frac{\omega}{M}Re(f_{13}f_{16}^*- f_{14}f_{15}^*), \\
B_{38}&=&2\frac{\omega}{M}Re(f_{1}f_{4}^*- f_{2}f_{3}^*), \\
B_{39}&=&2\frac{\omega}{M}Re(f_{7}f_{10}^*- f_{8}f_{9}^*), \\
B_{40}&=&\frac{\omega}{M}Re(f_{13}f_{4}^*+f_{16}f_{1}^*
-f_{15}f_{2}^*-f_{14}f_{3}^*), \\
B_{41}&=&\frac{\omega}{M}Im(f_{13}f_{4}^*+f_{16}f_{1}^*
-f_{15}f_{2}^*-f_{14}f_{3}^*). 
\end{eqnarray*}
\section{Appendix 9: the deuteron spin-density matrix}

\hspace{0.7cm}

In this Appendix we give general formulae describing the polarization of the deuteron. 

For the case of arbitrary polarization the deuteron is described by a general spin--density matrix (defined, in general case, by 8 
parameters) which in the coordinate representation has thefollowing  form:
\begin{equation}\label{C1}
\rho_{\mu\nu}=-\frac{1}{3}\bigl(g_{\mu\nu}-\frac{p_{\mu}p_{\nu}}{M^2}\bigr)
+\frac{i}{2M}\varepsilon_{\mu\nu\lambda\rho}s_{\lambda}p_{\rho}+ Q_{\mu\nu},
\ \  
\end{equation}
$$Q_{\mu\nu}=Q_{\nu\mu}, \ \ Q_{\mu\mu}=0\ , \ \ p_{\mu}Q_{\mu\nu}=0\ , $$
where $p_{\mu }$ (M) is the deuteron 4-momentum (mass), $s_{\mu}$ and $Q_{\mu\nu}$ are
the deuteron polarization 4-vector and quadrupole--polarization tensor. The deuteron, therefore,  is described, in the general case, by vector (three parameters) and tensor (five parameters) polarizations.

In the deuteron rest frame, the above formula is written as
\begin{equation}\label{C2}
\rho_{ij}=\frac{1}{3}\delta_{ij}-\frac{i}{2}\varepsilon
_{ijk}s_k+Q_{ij}, \ ij=x,y,z.
\end{equation}
This spin--density matrix can be written in the helicity representation
using the following relation
\begin{equation}\label{C3}
\rho_{\lambda\lambda'}=\rho_{ij}e_i^{(\lambda )*}e_j^{(\lambda')}, \
\lambda ,\lambda'=+,-,0,
\end{equation}
where $e_i^{(\lambda )}$ are the deuteron spin functions which have the spin projection $\lambda $ on the quantization axis ($z$ axis).
They are
\begin{equation}\label{C4}
e^{(\pm )}=\mp \frac{1}{\sqrt{2}}(1,\pm i,0), \
e^{(0)}=(0,0,1).
\end{equation}
The elements of the spin--density matrix in the helicity representation
are related to the ones in the coordinate representation by: 
\begin{equation}\label{C5}
\rho _{\pm\pm}=\frac{1}{3}\pm \frac{1}{2}s_z-\frac{1}{2}Q_{zz}, \
\rho_{00}=\frac{1}{3}+Q_{zz}, \
\rho_{+-}=-\frac{1}{2}(Q_{xx}-Q_{yy})+iQ_{xy}, \
\end{equation}
$$\rho_{+0}=\frac{1}{2\sqrt{2}}(s_x-is_y)-
\frac{1}{\sqrt{2}}(Q_{xz}-iQ_{yz}),
\rho_{-0}=\frac{1}{2\sqrt{2}}(s_x+is_y)+
\frac{1}{\sqrt{2}}(Q_{xz}+iQ_{yz}), \ $$
$$\rho_{\lambda\lambda'}=
(\rho_{\lambda'\lambda})^*, $$
with  $Q_{xx}+Q_{yy}+Q_{zz}=0.$

Polarized deuteron targets,  described by the population numbers
$n_+, \ $ $n_- $ and $n_0 $, are often used in  spin experiments. Here
$n_+, \ $ $n_- $ and $n_0 $ are the fractions of the atoms with
the nuclear spin projection on the quantization axis
$m=+1, \ $ $m=-1$ and $m=0,$
respectively. If the spin--density matrix is normalized to 1, i.e.,
$Tr\rho =1$, then we have $n_++n_-+n_0=1.$ Thus, the polarization state of
the deuteron target is defined in this case by two parameters: the so--called
V (vector) and T (tensor) polarizations
\begin{equation}\label{C6}
V=n_+-n_-, \ T=1-3n_0.
\end{equation}
Defining the quantities $n_{\pm ,0}$ as 
\begin{equation}\label{C7}
n_{\pm }=\rho_{ij}e_i^{(\pm )*}e_j^{(\pm )}, \
n_0=\rho_{ij}e_i^{(0)*}e_j^{(0)},
\end{equation}
we have the following relation between $V$ and $T$ parameters and parameters
of the spin--density matrix in the coordinate representation (in the case
when the quantization axis is directed along the $z$ axis)
\begin{equation}\label{C8}
n_0=\frac{1}{3}+Q_{zz}, \ n_{\pm }=\frac{1}{3}\pm \frac{1}{2}s_z-
\frac{1}{2}Q_{zz},
\end{equation}
or
\begin{equation}\label{C9}
T=-3Q_{zz}, \ V=s_z.
\end{equation}

Sometimes (for example, when calculating the radiative corrections to various
processes) it is convenient to parametrize the polarization states of the
particles in the considered reaction, in terms of the four-momenta of
the particles involved.
Therefore, first of all, we have to fix a  coordinate system and to express the polarization 
in a covariant form, in terms of the particle four --momenta.

Let us consider different (standard) sets of coordinates.

If we choose, in the laboratory system of the deuteron electrodisintegration
reaction, the longitudinal direction $\vec \ell$ along the electron beam and the
transverse one $\vec t$ in the plane $(\vec  k_1,\vec k_2)$ and perpendicular to
$\vec \ell$ , then
$$
S^{(\ell)}_{\mu}=\frac{2\eta k_{1\mu}-P_{\mu}}{M}, 
S^{(t)}_{\mu}=\frac{Vk_{2\mu}-(V-Vy-2Q^2\eta)k_{1\mu}-Q^2P_{\mu}}{Vd}, $$
\begin{equation}\label{1}
S^{(n)}_{\mu}=\frac{2\varepsilon_{\mu\lambda\rho\sigma} P_{\lambda} k_{1\rho}
k_{2\sigma}}{Vd}\ ,
\end{equation}
$$d=\sqrt{bQ^2}, \ \ b =1-y-\frac{Q^2}{V}\eta, \ \ \eta =M^2/V, $$
where $V=2k_1\cdot P$, $y=k\cdot P/k_1\cdot P$ and in the laboratory system 
we have $V=2ME$ and $y=1-E'/E.$


One can verify that the set of the 4-vectors $S_{\mu}^{(\ell,t,n)}$ satisfies
the following properties
\begin{equation}\label{2}
S_{\mu}^{(\alpha)}S_{\mu}^{(\beta)} = -\delta_{\alpha\beta}, \ \
S_{\mu}^{(\alpha)}P_{\mu} =0, \ \ \alpha,  \beta = \ell,t,n.
\end{equation}
One can make sure also that in the rest frame of the deuteron (the laboratory
system)
$$ S_{\mu}^{(\ell)}=(0, \vec \ell), \ \ S_{\mu}^{(t)}=(0,\vec  t), \ \
S_{\mu}^{(n)}=(0,\vec  n) \ , $$
\begin{equation}\label{3}
\vec  \ell = \vec n_1 , \ \ \vec  t = 
\displaystyle\frac{\vec n_2-(\vec n_1\cdot \vec n_2)\vec n_1}{\sqrt{1-(\vec n_1\cdot \vec n_2)^2}}, \ \ \vec  n=\displaystyle\frac{\vec n_1\times \vec n_2}{\sqrt{1-(\vec n_1\cdot \vec n_2)^2}}, \ \ 
\vec  n_{1,2}=\displaystyle\frac{\vec k_{1,2}}{|\vec k_{1,2}|}\ .
\end{equation}

Adding one more four--vector $S_{\mu}^{(0)}=P_{\mu}/M$ to this set,
we build a complete set of the orthogonal four--vectors with the following
properties
\begin{equation}\label{4}
S_{\mu}^{(m)}S_{\nu}^{(m)} = g_{\mu\nu}, \ \
S_{\mu}^{(m)}S_{\mu}^{(n)} = g_{mn}, \ \ m,n = 0,\ell,t,n.
\end{equation}
This allows to relate the deuteron quadrupole polarization tensor, given 
in the arbitrary system, to the one in the laboratory system
\begin{equation}\label{5}
Q_{\mu\nu} = S_{\mu}^{(m)}S_{\nu}^{(n)}R_{mn} \equiv S_{\mu}^{(\alpha)}
S_{\nu}^{(\beta)}R_{\alpha\beta}, \ \ R_{\alpha\beta}=R_{\beta\alpha}, \
R_{\alpha\alpha}=0
\end{equation}
because the components $R_{00},\ R_{0\alpha}$ and $R_{\alpha 0}$
identically equal to zero due to condition $Q_{\mu\nu}P_{\nu}=0.$ The tensor 
$R_{\alpha\beta}$ is the deuteron quadrupole polarization tensor in the 
laboratory system.

Consider just one more choice of the coordinate axes, commonly used, where the  
components of the deuteron polarization tensor are defined in the coordinate
system with the axes along directions $\vec L$, $\vec T$ and  $\vec N$ in the rest
frame of the deuteron, where
\begin{equation}
\label{7}
\vec L = \displaystyle\frac{\vec k_1-\vec k_2}{|\vec k_1-\vec k_2|},~
\vec  T = \frac{\vec n_1-(\vec n_1\cdot \vec L)\vec L
}{\sqrt{1-(\vec n_1\cdot \vec L)^2}},~ \vec  N = \vec n.
\end{equation}
The respective covariant form of this set reads
$$S_{\mu}^{(L)} =\frac{2\eta (k_1-k_2)_{\mu} -yP_{\mu}}{M\sqrt{yh}}$$
\begin{equation}\label{8}
S_{\mu}^{(T)} =\frac{(Vy+2\eta Q^2)k_{2\mu}-[Vy(1-y)-2\eta Q^2)k_{1\mu}
-Q^2(2-y)P_{\mu}}{V\sqrt{ybhQ^2}}\,
\end{equation}
$$S_{\mu}^{(N)} =S_{\mu}^{(n)}\ , \ \ h=y+4\frac{\eta Q^2}{yV}\ . $$

These two sets of the orthogonal 4-vectors are connected by means of
orthogonal matrix which describes the rotation in the plane perpendicular
to direction $\vec n = \vec N$
\begin{equation}\label{10}
S_{\mu}^{(L)} = \cos{\psi }S_{\mu}^{(l)}+\sin{\psi }S_{\mu}^{(t)}, \ \
S_{\mu}^{(T)} = -\sin{\psi }S_{\mu}^{(l)} +\cos{\psi }S_{\mu}^{(t)},
\end{equation} 
$$\cos{\psi}=\frac{Vy+2\eta Q^2}{V\sqrt{yh}}\ , \ \
\sin{\psi }= -2\sqrt{\frac{b\eta Q^2}{yhV}}\ . $$

\addcontentsline{toc}{chapter}{\protect\numberline{7} {Bibliography}}

\end{document}